\documentclass[amsmath,amssymb,aps,10pt,prd,twocolumn,letterpaper,nofootinbib,balancelastpage,superscriptaddress,floatfix,preprintnumbers]{revtex4-2}
\usepackage{graphicx}	
\usepackage{bm}		    
\usepackage{soul}
\usepackage[sort&compress]{natbib}	 	
	\setcitestyle{square,numbers,comma}	
\usepackage[colorlinks=true,urlcolor=blue,linkcolor=red,citecolor=red]{hyperref}	

\usepackage{setspace}
\usepackage{amsmath}
\usepackage{bm}
\usepackage{fancyhdr}
\usepackage{euscript}
\usepackage{enumerate}
\usepackage[shortlabels]{enumitem}
\usepackage{braket}
\usepackage{float,fancyvrb}
\usepackage[T1]{fontenc}
\usepackage{lmodern}
\usepackage{booktabs}
\usepackage{caption}
\usepackage{scrextend}
\usepackage{comment}
\usepackage{subcaption}
\usepackage{caption}
\usepackage{url}
\usepackage{amssymb}
\usepackage{xpatch}
\usepackage{helvet}
\usepackage[displaymath]{lineno}
\usepackage{color}
\usepackage{float}
\usepackage{placeins}

\graphicspath{{./Figures/}}

\newcommand{\uu}[1]{\ensuremath{\, \mathrm{#1}}} 

\newcommand{\V}[1]{\ensuremath{\bm{#1}}}

\begin{document}

\title{Spin Squeezing of Macroscopic Nuclear Spin Ensembles}

\author{Eric Boyers}
\affiliation{Department of Physics, Boston University, Boston, MA 02215, USA}
\author{Garry Goldstein}
\affiliation{garrygoldsteinwinnipeg@gmail.com}
\author{Alexander O. Sushkov}\email{asu@bu.edu}
\affiliation{Department of Physics, Boston University, Boston, MA 02215, USA}
\affiliation{Department of Electrical and Computer Engineering, Boston University, Boston, MA 02215, USA}
\affiliation{Photonics Center, Boston University, Boston, MA 02215, USA}


\begin{abstract}
	Spin squeezing has been explored in atomic systems as a tool for quantum sensing, improving experimental sensitivity beyond the spin standard quantum limit for certain measurements. To optimize absolute metrological sensitivity, it is beneficial to consider macroscopic spin ensembles, such as nuclear spins in solids and liquids. Coupling a macroscopic spin ensemble to a parametrically-modulated resonant circuit can create collective spin squeezing by generating spin correlations mediated by the circuit. We analyze the squeezing dynamics in the presence of decoherence and finite spin polarization, showing that achieving 7\,dB spin squeezing is feasible in several nuclear spin systems. The metrological benefit of squeezing a macroscopic spin ensemble lies in the suppression of technical noise sources in the spin detection system relative to the spin projection noise. This expands the experimental sensitivity bandwidth when searching for signals of unknown frequency and can improve the resonant signal-to-noise ratio. Squeezing macroscopic spin ensembles may prove to be a useful technique for fundamental physics experiments aimed at detecting spin interactions with oscillating background fields, such as ultralight dark matter. 
\end{abstract}

\maketitle



\clearpage

\section{Introduction}

The tools of quantum science have found applications in a variety of precision experiments, including the use of non-classical states of light in gravitational wave detectors, spin squeezed states in atomic clocks, squeezed microwave states in searches for axion dark matter and in electron paramagnetic resonance~\cite{aasi2013enhanced,Hosten2016,malnou2019squeezed, backes2021a,vasilakis2015generation,Bienfait2017b,Pedrozo-Penafiel2020}.
The present work focuses on experiments with ensembles of spin qubits.
Because the signals detected from such ensembles usually scale with the number of spins $N$, precision experiments often seek to maximize the ensemble size~\cite{Degen2017,Safronova2018,JacksonKimball2023}.
A fundamental limit to the sensitivity of such experiments is the spin projection noise, leading to the spin standard quantum limit (SQL) that scales as $1/\sqrt{N}$~\cite{Bloch1940}. 
Several experimental platforms have achieved the sensitivity sufficient to reach the spin projection noise limit, including ultracold atoms, atomic vapor cells, color centers in solids, and inductively-detected nuclear magnetic resonance (NMR)~\cite{Sleator1985,Crooker2004,Shah2010,Schlagnitweit2012}.


Several approaches have been suggested to surpass the SQL in spin-based measurements. 
The SQL is valid for an ensemble of uncorrelated spins and can be evaded by creating spin correlations. Examples include GHZ states that acquire a quantum phase faster than uncorrelated spins~\cite{Bollinger1996,Leibfried2004,Omran2019} and squeezed states that have reduced spin noise along a particular projection~\cite{Kitagawa1993, Wineland1992}. Notably, spin correlations do not have to be quantum: a ferromagnetic gyroscope, whose angular momentum is dominated by atomic spins, can evade the SQL, as well as the energy resolution limit~\cite{JacksonKimball2016,Vinante2021a}.

Squeezed spin states in particular have been explored in several platforms, especially cold atom and atomic vapor systems, where the spin projection noise has been squeezed by up to a factor of 100~\cite{Schleier-Smith2010,Leroux2010,Hosten2016,Vasilakis2015,Bao2020,Pedrozo-Penafiel2020,Colombo2022a}. 
However, atomic experiments have so far been limited to mesoscopic ensembles of up to $10^{10}$ atoms, with the notable exception of proposed experiments with $^3$He atoms in vapor cells~\cite{Serafin2021}. Such relatively small systems are important for applications such as high-spatial-resolution sensors and fundamental physics experiments that use radioactive atoms and molecules~\cite{Arrowsmith-Kron2024}. But their absolute metrological sensitivity, defined, for example, as the angular variance of the collective spin, is limited, even if improvement beyond the SQL is achieved. 
Experiments that use NMR of solids and liquids have considerable sensitivity potential, by virtue of making measurements on macroscopic spin ensembles, with $N$ on the order of a mole~\cite{Budker2006,Eckel2012,Budker2014}. Let us note that realizing this potential requires significant technical effort, since reaching the SQL-limited sensitivity necessitates suppression of thermal and technical noise below this level, which becomes more difficult for large $N$~\cite{Aybas2021b}. It is also important to achieve large spin ensemble polarization, which often requires hyperpolarization techniques~\cite{Eichhorn2022}.

In the present work we focus on macroscopic ensembles of nuclear spins and explore how spin squeezing can be used to maximize their metrological potential.
As a macroscopic NMR experiment approaches the SQL-limited sensitivity and exhausts the means of increasing the signal with classical techniques, we explore how the tools of quantum science can be leveraged to improve the experiment further. Previously, the NMR platform has been used to simulate spin squeezing using internal states of molecules and nuclei~\cite{Sinha2003,Auccaise2015,AksuKorkmaz2016}. Techniques for squeezing nuclear spins have been proposed in quantum dots~\cite{Rudner2011,Schuetz2013} and vapor cells filled with gaseous ${}^3\mathrm{He}$~\cite{Serafin2021,Sinatra2022}. 
We analyze the method of creating spin-squeezed states in a macroscopic ensemble of nuclear spins, inductively coupled to a resonant radiofrequency (RF) circuit.
Our approach is inspired by stroboscopic quantum non-demolition techniques that have achieved squeezing in atomic and optomechanical systems~\cite{Vasilakis2015,Groszkowski2020}. We determine the technical requirements to create significant squeezing and show that it is feasible to achieve $\approx 7\uu{dB}$ of squeezing for several nuclear spin systems.

For sensing schemes such as Ramsey spectroscopy, spin squeezing cannot significantly improve the sensitivity of an optimized experiment limited by Markovian spin decoherence~\cite{Auzinsh2004, Ma2011, Shah2010, Chin2012}. It is therefore necessary to determine what measurements can benefit from squeezing. 
We focus on steady-state continuous-wave (CW) NMR experiments sensing a weak oscillating field, such as that created by ultralight wave-like dark matter interacting with the spin ensemble~\cite{Sushkov2023d}.
We show how spin amplification can improve the sensitivity of such experiments limited by technical noise sources, and how squeezing can improve the sensitivity bandwidth of experiments even if they are limited by quantum spin projection noise.
Such approaches have been previously analyzed in the context of magnetometry and cavity QED~\cite{Hosten2016a,Davis2016,Nolan2017,Leroux2018,Guarrera2019,Koppenhofer2022,Koppenhofer2023}. For squeezed modes of the electromagnetic field, analogous approaches have been used in a search for the electromagnetic interaction of axion dark matter~\cite{Malnou2019,Backes2021}.
Our analysis is motivated by fundamental physics experiments aimed at detecting spin interactions with oscillating background fields, such as the Cosmic Axion Spin Precession Experiment (CASPEr)~\cite{Budker2014}. The CASPEr-e and CASPEr-g experiments use precision magnetic resonance with macroscopic nuclear spin ensembles in solids and liquids to search for the EDM and the gradient interactions of axion dark matter~\cite{Aybas2021a,Garcon2019}. Implementing spin squeezing in these experiments could significantly enhance their potential for discovery.

\section{Creating Squeezing in NMR}
\label{sec:2}

\begin{figure}[!t]
	\centering
	\includegraphics[width=7cm]{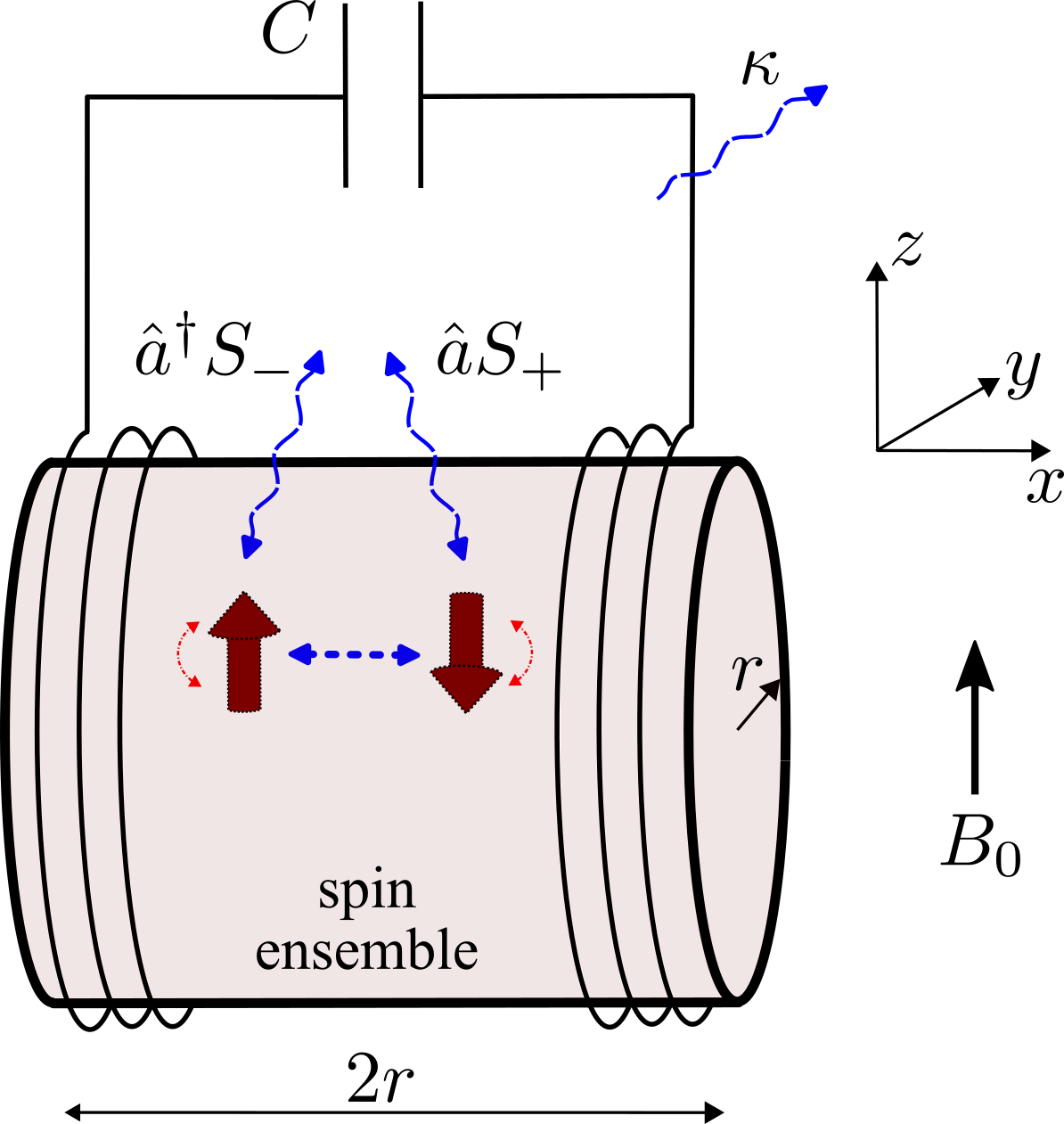}
	\caption{\label{fig:1} Setup schematic that shows a macroscopic spin ensemble, inductively coupled to a resonant circuit. Spin-spin correlations are created by photon emission and absorption terms $\hat{a}^{\dagger}S_-$ and $\hat{a}S_+$, indicated by wavy dashed blue lines.}
\end{figure}

Motivated by precision experiments with macroscopic nuclear spin ensembles in solids and liquids, we note that the relevant magnetic resonance frequencies are in the radiofrequency (RF) range. Therefore we consider a resonant RF circuit that consists of a capacitor $C$ and a solenoidal inductor $L$, with the number of turns $\eta$, radius $r$, and length $2r$, Fig.~\ref{fig:1}. The resonant angular frequency of the circuit is given by $\omega_c=1/\sqrt{LC}$ and the circuit quality factor is $Q=\omega_c/\kappa$, where $\kappa$ is the rate of energy dissipation due to effective circuit resistance.
The circuit electromagnetic field mode can be quantized in the usual way~\cite{Blais2021}, by writing the hamiltonian
\begin{align}
	H_c = LI^2/2+Q^2/2C,
	\label{eq:210}
\end{align}
where the circuit current $I$ and capacitor charge $C$ can be expressed in terms of photon creation and annihilation operators $\hat{a}^{\dagger}$ and $\hat{a}$:
\begin{align}
	I &= \sqrt{\frac{\hbar\omega_c}{2L}}(\hat{a}^{\dagger}+\hat{a}),\nonumber \\
	Q &= i\sqrt{\frac{\hbar\omega_cC}{2}}(\hat{a}^{\dagger}-\hat{a}).
	\label{eq:220}
\end{align}
The circuit hamiltonian then becomes
\begin{align}
	H_c = \hbar\omega_c\left(\hat{a}^{\dagger}\hat{a}+\frac{1}{2}\right).
	\label{eq:230}
\end{align}

The solenoidal coil is filled with the cylindrical sample that hosts a paramagnetic ensemble of $N$ nuclear spins, each with gyromagnetic ratio $\gamma$ and spin 1/2. 
The collective spin is given by $\V{S}=\sum_j \V{\sigma}_j$, where the sum is over all the spins $\V{\sigma}_j$ in the ensemble.
A bias magnetic field $B_0$, along the z-axis, sets the spin Larmor frequency $\omega_s=\gamma B_0$, which is tuned close to resonance with the circuit. The solenoid axis is aligned with the x-axis. A current $I$ in the circuit produces a magnetic field $B_x \approx \mu_0\eta I/(2r)$ inside the solenoid. This magnetic field produces the interaction between the circuit and the spin ensemble: 
\begin{align}
	H' = \hbar\gamma B_xS_x/2 = \hbar g(\hat{a}^{\dagger}+\hat{a})(S_+ + S_-),
	\label{eq:240}
\end{align}
where $S_+$ and $S_-$ are the collective spin raising and lowering operators, and we introduced the single-spin coupling constant $g$. We are also assuming spin interaction with the co-rotating component of the oscillating field $B_x$. The solenoid coil has equal length and diameter, thus we approximate its inductance as $L\approx \mu_0\eta^2 r$. Then the single-spin coupling constant is given by
\begin{align}
	\label{eq:g_bare_single_spin}
	g = \frac{\gamma}{4r} \sqrt{\frac{\mu_0 \hbar \omega_c }{2r}}.
\end{align}

Let us write out the total lab-frame system Hamiltonian of the coupled spin-circuit system:
\begin{align}
	\label{eq:H_Dicke_lab}
	H_\mathrm{lab}/\hbar &= \omega_s S_z + \omega_c \hat{a}^\dagger \hat{a} + g ( \hat{a}^\dagger + \hat{a} ) (S_+ + S_-)
\end{align}
This is the Dicke model, often used in cavity and circuit QED~\cite{Dicke1954}. Table~\ref{tab:1} lists numerical estimates for the single-spin coupling constant $g$ for three nuclear spin systems, assuming $r=3\,$mm and $\omega_c/2\pi=30\,$MHz. We see that the values are small and may appear negligible. Nevertheless, squeezing is a collective effect and the effective coupling strength is greatly enhanced.

\subsection{Spin squeezing by modulating the circuit coupling}
\label{sec:2a}

There are several approaches one can pursue to create spin-squeezed states using the coupled spin-circuit system described in the previous section. 
For example, squeezing can be created via feedback, by amplifying and feeding back the resonant circuit current to the spin ensemble~\cite{Schleier-Smith2010,Kohler2017}. In this work, we focus on the scheme that achieves spin squeezing by modulating the coupling between the circuit and the spin ensemble. In an experiment, this can be realized by using a broadband transformer between the resonant circuit and the spin sample, and modulating the mutual inductance coupling of this transformer by driving its core between superconducting and normal states. The transformer can have counter-wound coils, allowing a change of the sign of the effective coupling. The experimental details are beyond the scope of the present work, but experimental realizations are being explored in our laboratory.

Let us analyze the Hamiltonian in Eq.~(\ref{eq:H_Dicke_lab}), allowing the spin-circuit detuning $\Delta \equiv \omega_c - \omega_s $ and coupling $g$ to vary with time. We choose their time dependence to be periodic and write them in terms of the Fourier series: 
\begin{align}
	\Delta(t) &= \sum_{l=-\infty}^{\infty} \Delta_n e^{i l \omega_s t} \\
	g(t)      &= \sum_{l=-\infty}^{\infty} g_n e^{i l \omega_s t},
	\label{eq:Fourier}
\end{align}
where $\Delta_l$ and $g_l$ are the Fourier components.

We first consider the case where both $\Delta=\Delta_0$ and $g=g_0$ are constant. Transforming to the frame rotating at frequency $\omega_s$ and performing a high-frequency Magnus expansion in the dominant Larmor term with strength $\omega_s$, the effective rotating frame Hamiltonian is:
\begin{align}
	\label{eq:H_Dicke_rotating_eff}
	H_\mathrm{rot}/\hbar = \Delta_0 \hat{a}^\dagger \hat{a} + g_0 (\hat{a}^\dagger S_- + \hat{a} S_+) + ...,
\end{align}
where we have omitted higher-order terms which are suppressed by powers of the parameter $\Delta_0/\omega_s$. Since the parameters are constant, this is equivalent to the rotating wave approximation.

Consider the second term in Eq.~\eqref{eq:H_Dicke_rotating_eff}. It describes emission or absorption of a resonator photon, accompanied by a spin flip. Let us tune the system to the dispersive regime, where the detuning between the spin and the circuit is large compared to the collective coupling: $\Delta_0 \gg g_0\sqrt{N \bar{n}}$, where $\bar{n} = \braket{a^\dagger a}$ is the mean number of photons in the circuit. The spin energy shift is then given by second-order perturbation theory that describes the subsequent emission and absorption of a cavity photon. 
This energy shift can be estimated as follows.
Starting with the initial resonator photon population $n$, we have to add two terms. (1) A photon is first emitted into the resonator, and then re-absorbed; the spin energy shift is $\approx -g_0^2(n+1)S_+S_-/\Delta_0$. (2) A photon is first absorbed from the resonator, and then re-emitted; the spin energy shift is $\approx -g_0^2nS_-S_+/(-\Delta_0)$. Adding these two terms and dropping the term proportional to $S_z$, which can be absorbed into the detuning, we obtain the following rotating-frame effective spin Hamiltonian:
\begin{align}
	\label{eq:H_z_OAT}
	H_z/\hbar =- \frac{g_0^2}{\Delta_0} S_z^2.
\end{align}
Let us note that the number $n$ of photons in the resonator does not enter the effective coupling, due to the interference between the virtual photon emission and absorption.
The derivation of this result is given in Appendix~\ref{app:Sx2_derivation}.

This is the one-axis-twist (OAT) Hamiltonian that creates a squeezed state if the collective spin is initialized in the xy-plane. The resulting dynamics have been thoroughly studied theoretically and experimentally in cold-atom platforms~\cite{Kitagawa1993,Jin2009,Pezze2018,Colombo2022a}. Our squeezing mechanism is equivalent to that of vacuum squeezing in atomic systems~\cite{Hu2017}, which can be understood via the schematic in Fig.~\ref{fig:1}. Due to the detuning $\Delta$, exchanging a single excitation between the spin and the circuit via $a^\dagger S_-$ or $a S_+$ does not conserve energy and is suppressed. However, the process can maintain energy conservation if it occurs in pairs where the spins emit a photon to the circuit while the circuit simultaneously emits a photon back to the spins. Thus, in second-order perturbation theory, the circuit mediates effective spin-spin interactions which lead to squeezing. 

Let us estimate the numerical magnitude of the effective coupling for an ensemble of $^1$H nuclear spins. If we choose $\Delta_0/2\pi=3\,$MHz, then $g_0^2/\Delta_0 \approx 2 \pi \times 10^{-17} \uu{Hz}$. This appears to be hopelessly small.
However, because Hamiltonian in Eq.~\eqref{eq:H_z_OAT} contains the factor $S_z^2$, its strength is collectively enhanced. 
The exact details depend on the spin state, but the effective enhancement factor is on order of the number of spins in the ensemble, $N \approx 10^{22}$. Suppose the expectation value of the collective spin vector is in the transverse xy-plane, so that the only z-component is the spin projection noise $S_z\approx\sqrt{N}$. Then the magnitude of the spin Hamiltonian is $Ng_0^2/\Delta_0 \approx 2\pi\times 10^{5} \uu{Hz}$, which is larger than the relevant decoherence rates, as discussed below. Therefore spin squeezing can be created in this way.

In a macroscopic NMR experiment, the spin ensemble is initially prepared with the collective spin along the z-axis. In order to use the Hamiltonian in Eq.~\eqref{eq:H_z_OAT} to create spin squeezing, one has to apply a $\pi/2$ pulse to rotate the spins to the x-axis, let them evolve under the OAT Hamiltonian, and then apply additional pulses to reorient the spin ensemble for readout~\cite{Colombo2022a}.
However, there are technical reasons that make this scheme difficult to realize in practice. Pulse imperfections can create errors in the manipulation of the collective spin. Spin ensemble decoherence, while the collective spin is in the transverse plane, negates any advantage gained by squeezing, especially if spin polarization is lost and needs to be restored via slow thermal relaxation or hyperpolarization. Therefore in the next section we consider an alternative spin-squeezing scheme, that never tilts the collective spin vector into the transverse plane.

\subsection{Squeezing in the x-y plane}\label{sec:xy}

Several magnetic resonance-based searches for new fundamental physics, such as CASPEr, operate by preparing the collective spin along the z-axis and searching for small collective spin tilts due to new interactions~\cite{Budker2014}. To create spin squeezing in such a setup, we aim to create the $S_x^2$ squeezing Hamiltonian, which allows for squeezing of the collective spin without exposing the spin ensemble to pulses or decoherence in the transverse plane. We can create this Hamiltonian by modulating the coupling between the resonant circuit and the spin ensemble at twice the Larmor frequency. 

Let us consider the modulation scheme with the time-variation of the coupling $g(t)=g(1+2\cos2\omega_st)$, so that $g_0=g_2=g$ in Eq.~\eqref{eq:Fourier}. Once again we perform the Magnus expansion of Eq.~\eqref{eq:H_Dicke_lab} and neglect terms suppressed by the parameter $\Delta_0/\omega_s$. The rotating frame Hamiltonian becomes:
\begin{align}
	\label{eq:H_Dicke_rotating_modulated}
	H_\mathrm{rot}/\hbar &= \Delta_0 a^\dagger a + g (a^\dagger + a)  (S_+ + S_-) \\
	&= \Delta_0 a^\dagger a + 2g (a^\dagger + a) S_x.
\end{align}
Note that only the average detuning, $\Delta_0$, appears at the lowest order, so that effects due to detuning variation (which could occur as $g$ is modulated)  are suppressed.

We once again tune the system to the dispersive regime $\Delta_0\gg g\sqrt{N \bar{n}}$ and use second-order perturbation theory to obtain the rotating frame spin Hamiltonian:
\begin{align}
	\label{eq:H_modulated_Sx2}
	H_x/\hbar =  
	-\frac{4 g^2}{\Delta_0}S_x^2 = -g_eS_x^2,
\end{align}
where we defined the effective spin coupling $g_e = 4 g^2/\Delta_0$.
The detailed derivation of the effective Hamiltonian~\eqref{eq:H_modulated_Sx2}, based on the Schrieffer–Wolff transformation, can be found in Appendix~\ref{app:Sx2_derivation}.

The OAT Hamiltonian in Eq.~(\ref{eq:H_modulated_Sx2}) creates spin squeezing in the xy plane, for the collective spin state initially oriented along the z-axis. Let us note that the rotating-frame Hamiltonian in Eq.~(\ref{eq:H_Dicke_rotating_modulated}) has the same form as the lab-frame Hamiltonian in Eq.~(\ref{eq:H_Dicke_lab}). The difference is the Larmor term which, in the lab frame, rotates the collective spin much faster than squeezing can be generated and thus averages it away. By modulating the coupling at twice the Larmor frequency, we ensure that the spin squeezing is not averaged away, but can instead be generated in the rotating frame. Similar methods have been demonstrated in atomic systems~\cite{Vasilakis2015}.

In an experimental implementation, it may be difficult to achieve the exact coupling modulation function $g(t)=g(1 + 2 \cos 2\omega_s t)$. The modulation scheme that may be easier to implement consists of the square-wave modulation, with the coupling switching between two discrete values with a tunable duty cycle. 
This suppresses the effective coupling strength $g_e$ by a factor of order unity, 
as shown in appendix~\ref{app:square_modulation}.

\section{Squeezing dynamics}
\label{sec:3}

The OAT Hamiltonian in Eq.~(\ref{eq:H_modulated_Sx2}) can create spin squeezing for the collective spin oriented along the z-axis. The squeezing parameter $\xi^2$ quantifies the degree of squeezing by relating the minimum squeezed spin variance to the spin projection noise of the coherent spin state. Several squeezing parameters can be defined~\cite{Ma2011}. Let us use the Kitagawa and Ueda parameter
\begin{align}
	\xi^2 = \frac{\Delta^2 S_{\perp, \mathrm{min}}}{N/4},
	\label{eq:xi}
\end{align}
where $\Delta^2 S_{\perp, \mathrm{min}}$ is the minimal variance in the direction perpendicular to the collective spin~\cite{Kitagawa1993,Ma2011,Pezze2018}. 

Let us first consider an experiment that begins with the spin ensemble initialized in the $S_z=N/2$ eigenstate. The modulation $g(t)$ is switched on at time $t=0$, implementing the OAT Hamiltonian in Eq.~(\ref{eq:H_modulated_Sx2}). The evolution of the squeezing parameter with time is well-known~\cite{Kitagawa1993,Sorensen2001,Pezze2018}.
For an intuitive understanding of spin evolution, we can approximate it as if the spin experiences a magnetic field with magnitude on the order of $g_e\sqrt{N}$ along the x-direction. 
As usual with one-axis twisting, the spin vectors with initial $S_x$ components of opposite sign evolve in opposite directions in the z-y plane.
After time $t>1/g_eN$ the y-projection evolves by the angle with magnitude $\theta\approx g_e\sqrt{N}t$. The maximum transverse spin component is then $\Delta S_{max}\approx g_etN^{3/2}$. By the Heisnberg uncertainty principle, the minimum transverse component is $\Delta S_{min}\approx 1/g_et\sqrt{N}$. We substitute into Eq.~\eqref{eq:xi} to obtain the initial squeezing evolution term of order $ 1/(Ng_et)^2$. 
The full time evolution of $\xi^2$ under the OAT Hamiltonian for $t>1/g_eN$ is given by~\cite{Kitagawa1993}
\begin{align}
	\xi^2 \approx  \frac{1}{(Ng_et)^2} + \frac{1}{6} N^2 (g_et)^4.
\end{align}
The second term is the so-called over-squeezing term that is due to the spin state wrapping around the Bloch sphere.
In practice, noise and decoherence limit squeezing long before this oversqueezing term does, and we consider them in the following sections.

\begin{figure*}[!ht]
	\centering
	\includegraphics[width=\textwidth]{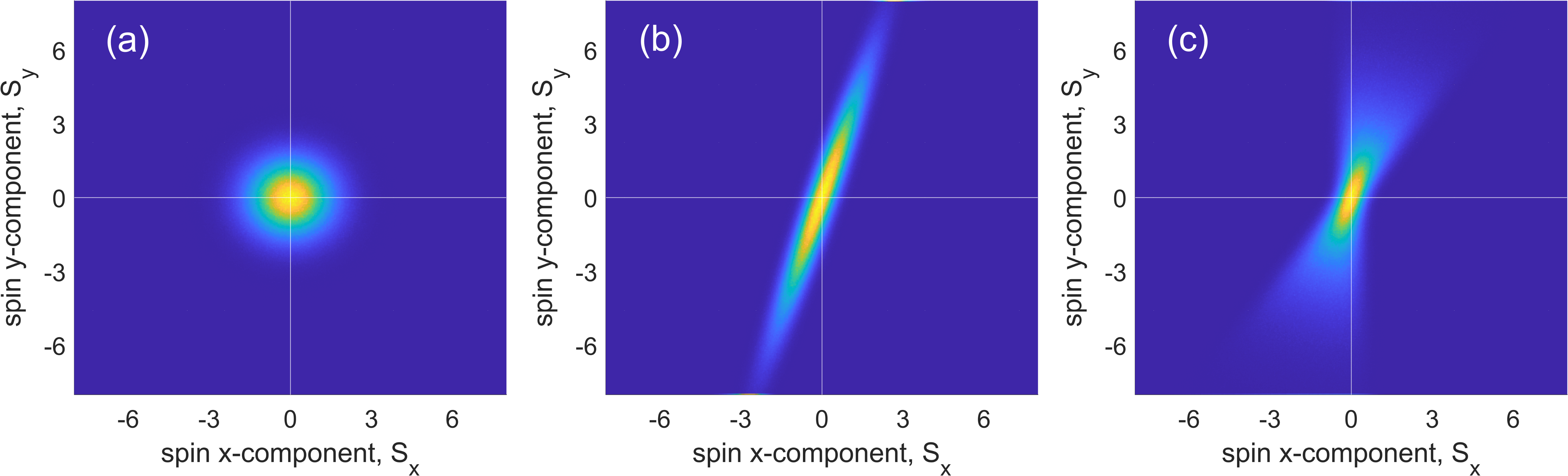}
	\caption{\label{fig:sq1} Projection of the collective spin on the transverse plane. (a) The collective spin is initialized along the z-axis. The spread in the transverse spin components is the spin projection noise. (b) Transverse collective spin projection, after evolution under the OAT Hamiltonian, to achieve 10\,dB of spin squeezing, with no spin decoherence. (c) Transverse collective spin projection, after evolution under the OAT Hamiltonian, in the presence of spin decoherence. The optimal evolution time is given by Eq.~\eqref{eq:optimal}, corresponding to 7\,dB of spin squeezing.}
\end{figure*}

\subsection{Spin polarization}

As we consider the technical requirements for achieving spin squeezing in a macroscopic NMR setup, let us first address the issue of spin polarization. The key advantage of atomic quantum optics experiments is the high-fidelity quantum state preparation of atomic ensembles, enabled by optical pumping. Such near-unity state preparation fidelities are not available in macroscopic ensembles of nuclear spin qubits. If nuclear spins are thermally polarized in an applied magnetic field, polarization of order 1\% is possible. Hyperpolarization techniques can improve this to tens of percent~\cite{Eichhorn2022,Bracker2005}. 

Let us consider the same experiment as in the previous section, but the spin ensemble is initialized in a mixed state with polarization fraction $p$. In Appendix~\ref{sec:Finite-polarization-effects} we show that the OAT evolution of the squeezing parameter is given by
\begin{align}
	\xi^2 \approx  \frac{1}{(pNg_et)^2} + \frac{1}{6} N^2 (g_et)^4.
\end{align}
We see that the buildup of squeezing is slowed down, as if instead of $N$ spins we have $pN$ spins. The spin projection noise, however, is independent of spin polarization. We note that using the Wineland squeezing parameter would add factors of $1/p^2$, however the metrological implications, discussed in Sec.~\ref{sec:5}, would be unchanged~\cite{Rudner2011}.

\subsection{Spin decoherence}
\label{sec:3b}

We consider the CW NMR scheme, with the collective spin staying close to the z-axis, which avoids having to apply error-prone pulses, as discussed in Sec.~\ref{sec:2a}. The tradeoff is the lack of possibility to refocus spin evolution due to fluctuating and inhomogeneous magnetic fields, or, possibly, dipolar interactions. The exact nature of spin decoherence depends on the environment of the chosen spin ensemble. Non-Markovian dephasing is less destructive for metrology with entangled states~\cite{Matsuzaki2011,Chin2012}. Nevertheless, here we conservatively assume the worst-case scenario of intrinsic spin dephasing caused by uncorrelated Markovian noise.

In Appendix~\ref{sec:Decay-of-squeezed} we show that, given an uncorrelated spin dephasing rate $\gamma_s$, the squeezing parameter degrades at the rate $2\gamma_s$. 
The intuition is that spin squeezing is a minimal form of entanglement, that consists of pairwise correlations between spins~\cite{Ma2011,Wang2010}. If the expectation of each spin operator decays at the rate $\gamma_s$, then pairwise spin correlations decay at the rate $2\gamma_s$~\cite{Hu2017,Lewis-Swan2018}.
This is in contrast to highly entangled states, such as the GHZ state involving $N$-qubit correlations, whose decay rate scales as $N \gamma_s$~\cite{Huelga1997}. 

\subsection{Circuit backaction}
\label{sec:3c}

The final source of dephasing that we consider is due to the backaction of the resonant circuit on the collective spin. The thermal Johnson-Nyquist noise, sourced by the dissipative elements in the resonant circuit, creates magnetic field fluctuations in the inductor that couples the circuit to the spin ensemble.
We assume that readout is performed by a weakly-coupled detector, such as a SQUID, so that the detector backaction is negligible, compared to the circuit thermal noise.

The interaction between the spin ensemble and the circuit is described by Eq.~\eqref{eq:H_Dicke_rotating_modulated}. If the mean thermal population of the circuit is 
\begin{align}
	\bar{n}\approx k_BT/\hbar\omega_c\gg 1,
\end{align}
then the spin ensemble experiences a fluctuating magnetic field with magnitude $g\sqrt{\bar{n}}$, correlation time $1/\kappa=Q/\omega_c$, and detuning $\Delta_0$ from the spin Larmor frequency. Here $T$ is the circuit temperature and $k_B$ is the Boltzmann constant.
The random phase acquired by the collective spin in a single correlation time $1/\kappa$ is of order $(g\sqrt{\bar{n}}/\kappa)(\kappa/\Delta_0)=(g\sqrt{\bar{n}}/\Delta_0)$. Therefore the spin dephasing rate $\gamma_c$ due to the circuit backaction is approximately given by $(g\sqrt{\bar{n}}/\Delta_0)\sqrt{\kappa/\gamma_c}\approx 1$, resulting in 
\begin{align}
	\gamma_c\approx \frac{g^2\bar{n}\kappa}{\Delta_0^2}.
	\label{eq:gammac}
\end{align}

A more rigorous theoretical treatment models the dissipation in the circuit at finite temperature as a generalized amplitude damping channel (GADC), which can be shown by adiabatic elimination of the cavity to lead to an effective GADC on the collective spin with rates $g^2\kappa (\bar{n}+1)/\Delta_0^2 $ and $ g^2\kappa \bar{n}/\Delta_0^2 $ for emission and absorption respectively~\cite{Agarwal1997}. Because the coupling is modulated, the exact behavior of the noise is complicated. However, since the modulation is periodic with a frequency much greater than the decay rate, $2\omega_s \gg \gamma_c$ we can replace the oscillating noise strength with an effectively constant value at its average, $\gamma_c \approx \bar{\gamma}_c $, so that the circuit noise rate is given approximately by $\bar{\gamma}_c \approx 0.5g^2\kappa \bar{n}/\Delta_0^2$. This agrees with Eq.~\eqref{eq:gammac} up to a numerical constant.

For short times $t \ll 1/\gamma_c$ we can use the collective spin random walk model to estimate the growth of the variance of the transverse spin projection: $\Delta^2S_{\perp}\approx(pNg\sqrt{\bar{n}}/\Delta_0)^2\kappa t\approx p^2N^2\gamma_ct$. Thus the squeezing parameter degrades at the rate $4p^2N\gamma_c$.

\subsection{Optimal squeezing}
\label{sec:3d}

\begin{table}[!t]
	\centering
	\begin{tabular}{|c|c|c|c|}
		\midrule
		Parameter & $^1$H & $^{207}$Pb & $^3$He \\
		\midrule
		$\gamma/2\pi$ (MHz/T) & 42.6 & 9.03 & -32.4 \\
		\midrule
		$N$ & $3\times10^{21}$ & $5\times10^{20}$ & $10^{19}$  \\
		\midrule
		$g/2\pi$ (Hz) & $5\times10^{-5}$ & $10^{-5}$ & $3\times10^{-5}$ \\
		\midrule
		$\Delta_0/2\pi$ (Hz) & $3\times10^{6}$ & $3\times10^{5}$ & $3\times10^{5}$ \\
		\midrule
		$pNg_e/2\pi$ (Hz) & $2\times10^{4}$ & $1500$ & $500$ \\
		\midrule
		$\gamma_s/2\pi$ (Hz) & $1000$ & $100$ & $10$ \\
		\midrule
		$4p^2N\gamma_c/2\pi$ (Hz) & $80$ & $50$ & $20$ \\
		\midrule
		$t_o$ ($\mu$s) & $15$ & $200$ & $700$ \\
		\midrule
		$\xi_o$ & $0.2$ & $0.2$ & $0.2$ \\
		\midrule
	\end{tabular}
	\caption{\fontsize{10}{11}\selectfont
		Numerical estimates for the optimal squeezing achievable in three macroscopic nuclear spin ensembles. We assume the RF circuit resonance frequency $\omega_c/2\pi = 30\,$MHz, spin polarization $p=0.1$, and circuit quality factor $Q=10^4$.}\label{tab:1}
\end{table}

Let us put together the dominant contributions to the squeezing dynamics under the OAT Hamiltonian:
\begin{align}
	\xi^2 \approx  \frac{1}{(pNg_et)^2} + (4p^2N\gamma_c+2\gamma_s)t,
\end{align}
where we neglected the over-squeezing term. 
Up to numerical factors of order unity, this is consistent with results for squeezing in atomic systems~\cite{Hu2017,Lewis-Swan2018}. At short times the collective spin is squeezed, but at long times the decoherence mechanisms introduce noise into the collective spin, and the squeezing is degraded. The optimal squeezing time $t_o$ minimizes the squeezing parameter at the minimum value $\xi_o$:
\begin{align}
	\xi_o &\approx \left(\frac{4p^2N\gamma_c+2\gamma_s}{pNg_e}\right)^{2/3}\\
	t_o &\approx \left(\frac{1}{pNg_e}\right)^{2/3}\left(\frac{1}{4p^2N\gamma_c+2\gamma_s}\right)^{1/3},
	\label{eq:optimal}
\end{align}
where we dropped factors of order unity.

Table~\ref{tab:1} shows experimentally-relevant parameters that were chosen for numerical estimates of the optimal squeezing achievable in three macroscopic nuclear spin ensembles. Estimates show that it is feasible to achieve 7\,dB of squeezing in several macroscopic NMR systems. Let us note that for solid samples with $^1$H and $^{207}$Pb nuclei, spin decoherence is dominated by the intrinsic relaxation $\gamma_s$, due, for example, to dipolar interactions. For the gaseous $^3$He sample, however, the intrinsic spin linewidth is much narrower, due to dipolar averaging, and the collective spin decoherence is dominated by the coupling to the RF circuit. 
Figure~\ref{fig:sq1} shows the results of a simulation of the evolution of the collective spin under the action of the OAT Hamiltonian in Eq.~\eqref{eq:H_modulated_Sx2}, with the parameters chosen so that the optimal squeezing is 7\,dB. Transverse spin relaxation is modeled as a uniformly-distributed stochastic rotation angle of each spin in the ensemble. Inhomogeneous broadening of spin couplings does not present an obstacle to squeezing, as shown in Appendix~\ref{app:inhomogeneous}. We also note that on these time scales squeezing does not reduce spin polarization fraction $p$.

\begin{figure*}[!t]
	\centering
	\includegraphics[width=\textwidth]{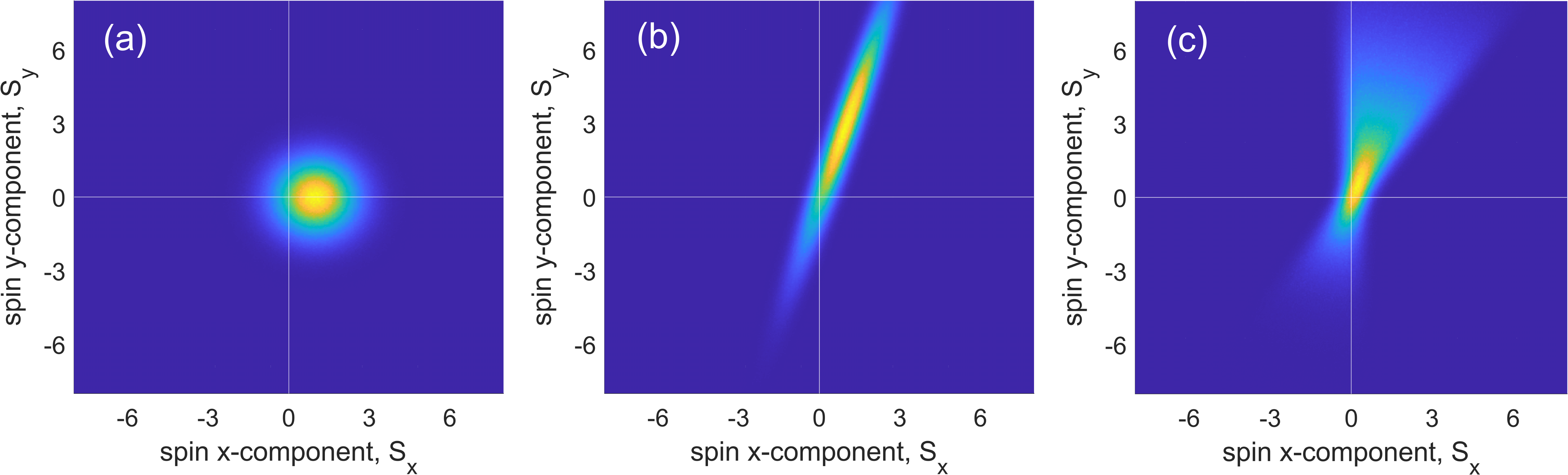}
	\caption{\label{fig:sq2} Evolution of the collective spin under the OAT Hamitonian and a tilt due to a hypothetical background field. (a) The collective spin is displaced from the z-axis by a small angle proportional to Rabi frequency $\Omega_1$. (b) Transverse collective spin projection, after evolution under the weak drive $\Omega_1$ and the OAT Hamiltonian, with no spin decoherence. Squeezing amplifies the transverse spin component and the spin projection noise. (c) Transverse collective spin projection, after evolution under the weak drive $\Omega_1$ and the OAT Hamiltonian, in the presence of spin decoherence. The evolution time is $t_o$, given by Eq.~\eqref{eq:optimal}, corresponding to 7\,dB of spin squeezing.}
\end{figure*}


\section{Metrological gain due to squeezing}
\label{sec:5}

We have argued that a squeezed state can be generated in a macroscopic ensemble of nuclear spins, but it does not necessarily follow that squeezing is metrologically useful for improving the sensitivity of an experiment. For example, it has been argued that spin squeezing does not improve the sensitivity of an optimized Ramsey interferometer or atomic magnetometer, limited by Markovian dephasing~\cite{Auzinsh2004,Ma2011,Pezze2018}. 
Nevertheless, we will show that, in certain cases, spin squeezing may offer an advantage in precision magnetic resonance experiments.

Let us return to the CASPEr CW NMR experimental scheme, searching for a torque on the collective spin, due to an interaction with a new background field, such as ultralight dark matter~\cite{Budker2014,Aybas2021a}. This interaction is quantified by a Rabi frequency $\Omega_1$. The experiment can search, for example, for a background field that oscillates at an unknown frequency and phase, with coherence time that is much longer than the $T_2$ of the spin ensemble.
The collective spin is initialized along the z-axis, and we start with the case of no squeezing. If the Larmor frequency is tuned near the oscillation frequency of the background field, then the collective spin is tilted by a small angle proportional to $\Omega_1$, Fig.~\ref{fig:sq2}(a). The observable is the oscillating transverse spin magnetization, measured, for example, by a SQUID sensor that is inductively coupled to a resonant RF circuit. For simplicity, we can assume that this is a separate, weakly-coupled, receiver circuit from the one that is used to create spin squeezing, Fig.~\ref{fig:1}.

\begin{figure*}[ht!]
	\centering
	\includegraphics[width=\textwidth]{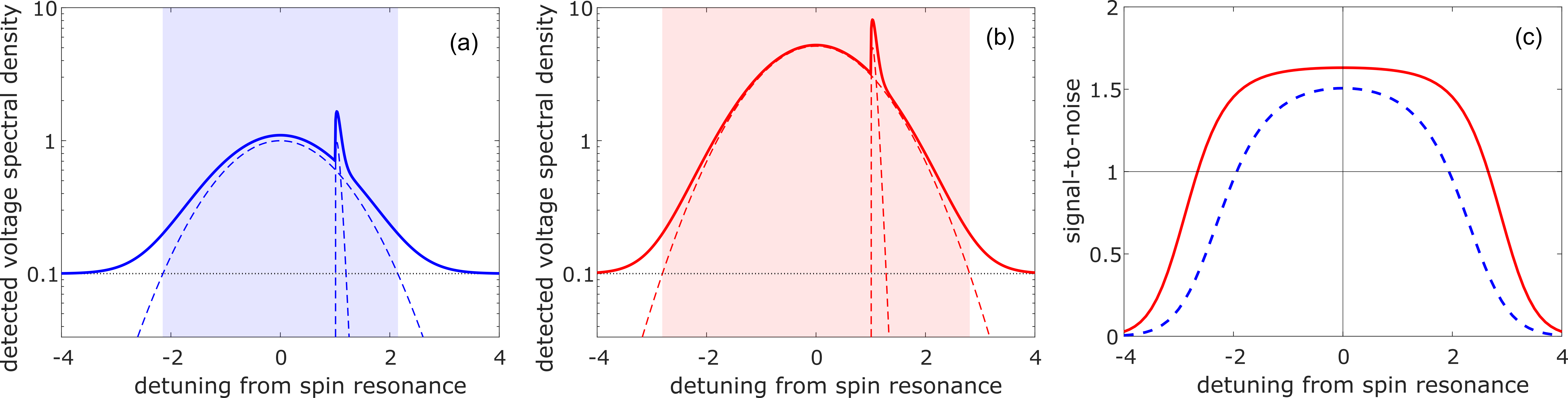}
	\caption{\label{fig:sq3} Detection of spin ensemble evolution in the presence of spin tilt due to a hypothetical background field. (a) Detected voltage power spectral density with no squeezing. The dotted horizontal line represents technical noise, such as the thermal Johnson-Nyquist circuit noise. The broad dashed Gaussian curve represents the quantum spin projection noise. The narrow dashed curve represents the collective spin tilt due to the hypothetical background field. The thick solid line is the sum of the three voltage contributions. (b) Detected voltage power spectral density with 7\,dB OAT squeezing. The technical noise is unchanged, while the collective spin transverse components are amplified. (c) The signal-to-noise ratio for detection without spin squeezing (dashed blue line) and with 7\,dB OAT squeezing (solid red line). Squeezing expands the sensitivity bandwidth and improves the on-resonance SNR.}
\end{figure*}

Let us consider the readout of the SQUID sensor, referred to the power spectral density of the voltage created in the receiver circuit. The first contribution $S_{VV}^{(th)}$ to this voltage is the thermal Johnson-Nyquist noise due to the dissipative elements in the circuit. We assume this white noise is the dominant technical noise in the experiment. It is indicated by the horizontal dotted line in Fig.~\ref{fig:sq3}(a). The voltage power spectral density created by the spin ensemble is proportional to $(S_x^2+S_y^2)\lambda(\delta)$, where $\lambda$ is the spin lineshape factor that depends on the detuning $\delta$ between the receiver circuit resonance and the Larmor frequency. Here we assume that $\lambda$ is Gaussian and re-scale the detuning by the spin linewidth $\gamma_s$. Near resonance, the spin projection noise acts as a voltage source, creating the second contribution $S_{VV}^{(spn)}$ to the voltage power spectral density, shown as the broad dashed line in Fig.~\ref{fig:sq3}(a). The final contribution is the collective spin tilt due to the interaction with the hypothetical background field. The corresponding voltage spectral density $S_{VV}^{(1)}$ is shown by the narrow dashed lineshape in Fig.~\ref{fig:sq3}(a). To illustrate a possible metrological arrangement, we have chosen the simulation parameters such that, on resonance, the experiment is limited by the quantum spin projection noise, with the circuit Johnson-Nyquist noise being a factor of 10 lower. We also set the oscillation frequency of the hypothetical background field at one spin linewidth away from spin resonance, and the linewidth of the background field to one-tenth of the spin linewidth. The blue-shaded region indicates the frequency band over which the spin projection noise dominates over the circuit Johnson-Nyquist noise.

Let us now consider the readout when we implement the OAT squeezing~\eqref{eq:H_modulated_Sx2}, as described in Sections \ref{sec:2},\ref{sec:3}. The OAT spin evolution is now combined with the spin tilt due to the new background field. The resulting transverse projection of the collective spin is shown in Fig.~\ref{fig:sq2}(b). We have chosen the same parameters as those in Fig.~\ref{fig:sq1}(b), resulting in 10\,dB of squeezing (in the absence of spin decoherence). In addition to the squeezing of spin projection noise, the mean collective spin tilt angle is amplified by the OAT evolution. Let us note, however, that the corresponding component of the spin projection noise is also amplified, along the direction from the origin to the mean collective spin. 
Transverse spin relaxation is again modeled as a uniformly-distributed stochastic rotation angle of each spin in the ensemble. This smears out the transverse components of the squeezed spin distribution, Fig.~\ref{fig:sq2}(c), reducing the squeezing parameter down to 7\,dB. Importantly, however, transverse spin relaxation only affects the value of the detected voltage power spectral density via the lineshape $\lambda(\delta)$, and does not affect the value of $(S_x^2+S_y^2)$.  

The voltage power spectral density, created by the spin ensemble, is amplified near spin resonance by a factor of $\approx 5$, Fig.~\ref{fig:sq3}(b). Importantly, the ratio of the voltage due to the collective spin tilt and the voltage due to the spin projection noise, $S_{VV}^{(1)}/S_{VV}^{(spn)}$, remains unchanged. However, this gain diminishes the relative importance of the technical noise sources, such as the circuit thermal noise $S_{VV}^{(th)}$, which is not amplified. It also expands the frequency band over which the spin projection noise dominates over the circuit Johnson-Nyquist noise. This band is indicated by the pink-shaded region in Fig.~\ref{fig:sq3}(b).

Let us define the ``signal-to-noise ratio'' (SNR) as the ratio of the background field-induced voltage spectral density to the sum of the voltage spectral densities due to the thermal noise and the spin projection noise, all evaluated at the background field oscillation frequency:
\begin{align}
	SNR = \frac{S_{VV}^{(1)}}{S_{VV}^{(spn)}+S_{VV}^{(th)}}.
	\label{eq:snr}
\end{align}
This is proportional, but not identical to the actual signal-to-noise ratio that would be observed in an optimized experiment, which would depend on other parameters, such as the averaging time. 
Focusing on the impact of squeezing, let us consider the following measurement protocol: (i) prepare the squeezed state for time $t_o$, (ii) measure the receiver circuit signal for time $T_2/2$. We choose the measurement time based on the assumption that the squeezing decay is dominated by the rate $2\gamma_s$ due to spin decoherence. Referring to the numerical estimates listed in Tab.~\ref{tab:1}, we also assume that $t_o\ll T_2/2$, so that the squeezing preparation has a negligible impact on the measurement duty cycle. In this case, the effective averaging time is the same with and without squeezing.
Therefore the quantity defined in Eq.~\eqref{eq:snr} captures the metrological implications of squeezing, Fig.~\ref{fig:sq3}(c). We observe that squeezing (red line) broadens the band of frequencies where SNR is above unity, compared to the measurement without squeezing (blue dashed line). This is related to the bandwidth gain in measurements that are performed with entangled spin systems~\cite{Huelga1997,Andre2004,Auzinsh2004,Pezze2018}. In the context of ultralight dark matter searches, the approach of electromagnetic mode squeezing has been used to accelerate a search for axion dark matter~\cite{Malnou2019,Backes2021}. Spin squeezing can be utilized in the same way: the broader sensitivity bandwidth implies that fewer spin resonance tuning steps are necessary to scan over a range of background field oscillation frequencies. Specifically, for the parameters chosen in Fig.~\ref{fig:sq3} where on resonance $S_{VV}^{(spn)}=10S_{VV}^{(th)}$, the sensitivity bandwidth is broader by a factor of 1.36, and therefore the scan rate can be improved by this factor. Let us emphasize that the exact magnitude of the scan rate improvement depends on the relative magnitudes of the thermal noise and the spin projection noise. If the thermal noise is relatively more substantial, say, for example, on resonance $S_{VV}^{(spn)}=2S_{VV}^{(th)}$, then the scan rate improvement is larger -- a factor of 2.5. The ultimate gain is determined by the experimental parameters of the relevant search.

In addition to the bandwidth gain, spin squeezing amplifies the collective spin tilt~\cite{Hosten2016a}.
The spin amplification can be understood by linearizing the spin equations of motion under the OAT Hamiltonian in Eq.~\eqref{eq:H_modulated_Sx2}. Assuming $\langle S_z\rangle\approx pN/2$, we obtain
\begin{align}
	\frac{dS_x}{dt} &= 0,\\
	\frac{dS_y}{dt} &= pNg_eS_x.
	\label{eq:eqsmo}
\end{align}
Thus the x-component of the collective spin is mapped to the y-component with a magnification factor on the order of $pNg_et$. The evolution time $t$ is limited by spin decoherence, as discussed in Sec.~\ref{sec:3d}. OAT squeezing amplifies both the spin projection noise and the collective spin tilt due to the background field.
This reduces the impact of thermal and technical noise on the experimental sensitivity, improving the SNR even when the hypothetical background field frequency is near spin resonance, Fig.~\ref{fig:sq3}(c). This improvement depends on the level of the broadband thermal noise $S_{VV}^{(th)}$, relative to the spin projection noise $S_{VV}^{(spn)}$. If $S_{VV}^{(spn)}\gg S_{VV}^{(th)}$, then there is no SNR gain near spin resonance. On the other hand, if $S_{VV}^{(spn)}< S_{VV}^{(th)}$, then the SNR improvement can be substantial.

\section{Outlook}

We have shown how coupling a macroscopic spin ensemble to a parametrically-modulated resonant circuit can create collective spin squeezing, by creating spin correlations, mediated by the circuit. These correlations can persist, despite noise and decoherence, with the potential to create squeezing on the order of 7\,dB in several nuclear spin systems. This approach is equivalent to vacuum squeezing, explored in experiments with atomic systems~\cite{Hu2017}. We are now studying experimental implementations of this and related squeezing schemes, such as a driven resonant circuit and measurement-based squeezing. These techniques have the potential to improve the sensitivity and bandwidth of magnetic resonance experiments, reducing the role of technical and thermal noise sources that are present in the spin detection system.


\begin{thebibliography}{70}%
	\makeatletter
	\providecommand \@ifxundefined [1]{%
		\@ifx{#1\undefined}
	}%
	\providecommand \@ifnum [1]{%
		\ifnum #1\expandafter \@firstoftwo
		\else \expandafter \@secondoftwo
		\fi
	}%
	\providecommand \@ifx [1]{%
		\ifx #1\expandafter \@firstoftwo
		\else \expandafter \@secondoftwo
		\fi
	}%
	\providecommand \natexlab [1]{#1}%
	\providecommand \enquote  [1]{``#1''}%
	\providecommand \bibnamefont  [1]{#1}%
	\providecommand \bibfnamefont [1]{#1}%
	\providecommand \citenamefont [1]{#1}%
	\providecommand \href@noop [0]{\@secondoftwo}%
	\providecommand \href [0]{\begingroup \@sanitize@url \@href}%
	\providecommand \@href[1]{\@@startlink{#1}\@@href}%
	\providecommand \@@href[1]{\endgroup#1\@@endlink}%
	\providecommand \@sanitize@url [0]{\catcode `\\12\catcode `\$12\catcode
		`\&12\catcode `\#12\catcode `\^12\catcode `\_12\catcode `\%12\relax}%
	\providecommand \@@startlink[1]{}%
	\providecommand \@@endlink[0]{}%
	\providecommand \url  [0]{\begingroup\@sanitize@url \@url }%
	\providecommand \@url [1]{\endgroup\@href {#1}{\urlprefix }}%
	\providecommand \urlprefix  [0]{URL }%
	\providecommand \Eprint [0]{\href }%
	\providecommand \doibase [0]{https://doi.org/}%
	\providecommand \selectlanguage [0]{\@gobble}%
	\providecommand \bibinfo  [0]{\@secondoftwo}%
	\providecommand \bibfield  [0]{\@secondoftwo}%
	\providecommand \translation [1]{[#1]}%
	\providecommand \BibitemOpen [0]{}%
	\providecommand \bibitemStop [0]{}%
	\providecommand \bibitemNoStop [0]{.\EOS\space}%
	\providecommand \EOS [0]{\spacefactor3000\relax}%
	\providecommand \BibitemShut  [1]{\csname bibitem#1\endcsname}%
	\let\auto@bib@innerbib\@empty
	\bibitem [{\citenamefont {Aasi}\ \emph {et~al.}(2013)\citenamefont {Aasi},
		\citenamefont {Abadie}, \citenamefont {Abbott}, \citenamefont {Abbott},
		\citenamefont {Abbott}, \citenamefont {Abernathy}, \citenamefont {Adams},
		\citenamefont {Adams}, \citenamefont {Addesso}, \citenamefont {Adhikari},
		\citenamefont {Affeldt}, \citenamefont {Aguiar}, \citenamefont {Ajith},
		\citenamefont {Allen}, \citenamefont {Ceron}, \citenamefont {Amariutei},
		\citenamefont {Anderson}, \citenamefont {Anderson}, \citenamefont {Arai},
		\citenamefont {Araya}, \citenamefont {Arceneaux}, \citenamefont {Ast},
		\citenamefont {Aston}, \citenamefont {Atkinson}, \citenamefont {Aufmuth},
		\citenamefont {Aulbert}, \citenamefont {Austin}, \citenamefont {Aylott},
		\citenamefont {Babak}, \citenamefont {Baker}, \citenamefont {Ballmer},
		\citenamefont {Bao}, \citenamefont {Barayoga}, \citenamefont {Barker},
		\citenamefont {Barr}, \citenamefont {Barsotti}, \citenamefont {Barton},
		\citenamefont {Bartos}, \citenamefont {Bassiri}, \citenamefont {Batch},
		\citenamefont {Bauchrowitz}, \citenamefont {Behnke}, \citenamefont {Bell},
		\citenamefont {Bell}, \citenamefont {Bergmann}, \citenamefont {Berliner},
		\citenamefont {Bertolini}, \citenamefont {Betzwieser}, \citenamefont
		{Beveridge}, \citenamefont {Beyersdorf}, \citenamefont {Bhadbhade},
		\citenamefont {Bilenko}, \citenamefont {Billingsley}, \citenamefont {Birch},
		\citenamefont {Biscans}, \citenamefont {Black}, \citenamefont {Blackburn},
		\citenamefont {Blackburn}, \citenamefont {Blair}, \citenamefont {Bland},
		\citenamefont {Bock}, \citenamefont {Bodiya}, \citenamefont {Bogan},
		\citenamefont {Bond}, \citenamefont {Bork}, \citenamefont {Born},
		\citenamefont {Bose}, \citenamefont {Bowers}, \citenamefont {Brady},
		\citenamefont {Braginsky}, \citenamefont {Brau}, \citenamefont {Breyer},
		\citenamefont {Bridges}, \citenamefont {Brinkmann}, \citenamefont {Britzger},
		\citenamefont {Brooks}, \citenamefont {Brown}, \citenamefont {Brown},
		\citenamefont {Buckland}, \citenamefont {Br{\"{u}}ckner}, \citenamefont
		{Buchler}, \citenamefont {Buonanno}, \citenamefont {Burguet-Castell},
		\citenamefont {Byer}, \citenamefont {Cadonati}, \citenamefont {Camp},
		\citenamefont {Campsie}, \citenamefont {Cannon}, \citenamefont {Cao},
		\citenamefont {Capano}, \citenamefont {Carbone}, \citenamefont {Caride},
		\citenamefont {Castiglia}, \citenamefont {Caudill}, \citenamefont
		{Cavagli{\`{a}}}, \citenamefont {Cepeda}, \citenamefont {Chalermsongsak},
		\citenamefont {Chao}, \citenamefont {Charlton}, \citenamefont {Chen},
		\citenamefont {Chen}, \citenamefont {Cho}, \citenamefont {Chow},
		\citenamefont {Christensen}, \citenamefont {Chu}, \citenamefont {Chua},
		\citenamefont {Chung}, \citenamefont {Ciani}, \citenamefont {Clara},
		\citenamefont {Clark}, \citenamefont {Clark}, \citenamefont {Constancio},
		\citenamefont {Cook}, \citenamefont {Corbitt}, \citenamefont {Cordier},
		\citenamefont {Cornish}, \citenamefont {Corsi}, \citenamefont {Costa},
		\citenamefont {Coughlin}, \citenamefont {Countryman}, \citenamefont
		{Couvares}, \citenamefont {Coward}, \citenamefont {Cowart}, \citenamefont
		{Coyne}, \citenamefont {Craig}, \citenamefont {Creighton}, \citenamefont
		{Creighton}, \citenamefont {Cumming}, \citenamefont {Cunningham},
		\citenamefont {Dahl}, \citenamefont {Damjanic}, \citenamefont {Danilishin},
		\citenamefont {Danzmann}, \citenamefont {Daudert}, \citenamefont {Daveloza},
		\citenamefont {Davies}, \citenamefont {Daw}, \citenamefont {Dayanga},
		\citenamefont {Deleeuw}, \citenamefont {Denker}, \citenamefont {Dent},
		\citenamefont {Dergachev}, \citenamefont {DeRosa}, \citenamefont {DeSalvo},
		\citenamefont {Dhurandhar}, \citenamefont {{Di Palma}}, \citenamefont
		{D{\'{i}}az}, \citenamefont {Dietz}, \citenamefont {Donovan}, \citenamefont
		{Dooley}, \citenamefont {Doravari}, \citenamefont {Drasco}, \citenamefont
		{Drever}, \citenamefont {Driggers}, \citenamefont {Du}, \citenamefont
		{Dumas}, \citenamefont {Dwyer}, \citenamefont {Eberle}, \citenamefont
		{Edwards}, \citenamefont {Effler}, \citenamefont {Ehrens}, \citenamefont
		{Eikenberry}, \citenamefont {Engel}, \citenamefont {Essick}, \citenamefont
		{Etzel}, \citenamefont {Evans}, \citenamefont {Evans}, \citenamefont {Evans},
		\citenamefont {Factourovich}, \citenamefont {Fairhurst}, \citenamefont
		{Fang}, \citenamefont {Farr}, \citenamefont {Farr}, \citenamefont {Favata},
		\citenamefont {Fazi}, \citenamefont {Fehrmann}, \citenamefont {Feldbaum},
		\citenamefont {Finn}, \citenamefont {Fisher}, \citenamefont {Foley},
		\citenamefont {Forsi}, \citenamefont {Fotopoulos}, \citenamefont {Frede},
		\citenamefont {Frei}, \citenamefont {Frei}, \citenamefont {Freise},
		\citenamefont {Frey}, \citenamefont {Fricke}, \citenamefont {Friedrich},
		\citenamefont {Fritschel}, \citenamefont {Frolov}, \citenamefont {Fujimoto},
		\citenamefont {Fulda}, \citenamefont {Fyffe}, \citenamefont {Gair},
		\citenamefont {Garcia}, \citenamefont {Gehrels}, \citenamefont {Gelencser},
		\citenamefont {Gergely}, \citenamefont {Ghosh}, \citenamefont {Giaime},
		\citenamefont {Giampanis}, \citenamefont {Giardina}, \citenamefont
		{Gil-Casanova}, \citenamefont {Gill}, \citenamefont {Gleason}, \citenamefont
		{Goetz}, \citenamefont {Gonz{\'{a}}lez}, \citenamefont {Gordon},
		\citenamefont {Gorodetsky}, \citenamefont {Gossan}, \citenamefont
		{Go{\ss}ler}, \citenamefont {Graef}, \citenamefont {Graff}, \citenamefont
		{Grant}, \citenamefont {Gras}, \citenamefont {Gray}, \citenamefont
		{Greenhalgh}, \citenamefont {Gretarsson}, \citenamefont {Griffo},
		\citenamefont {Grote}, \citenamefont {Grover}, \citenamefont {Grunewald},
		\citenamefont {Guido}, \citenamefont {Gustafson}, \citenamefont {Gustafson},
		\citenamefont {Hammer}, \citenamefont {Hammond}, \citenamefont {Hanks},
		\citenamefont {Hanna}, \citenamefont {Hanson}, \citenamefont {Haris},
		\citenamefont {Harms}, \citenamefont {Harry}, \citenamefont {Harry},
		\citenamefont {Harstad}, \citenamefont {Hartman}, \citenamefont {Haughian},
		\citenamefont {Hayama}, \citenamefont {Heefner}, \citenamefont {Heintze},
		\citenamefont {Hendry}, \citenamefont {Heng}, \citenamefont {Heptonstall},
		\citenamefont {Heurs}, \citenamefont {Hewitson}, \citenamefont {Hild},
		\citenamefont {Hoak}, \citenamefont {Hodge}, \citenamefont {Holt},
		\citenamefont {Holtrop}, \citenamefont {Hong}, \citenamefont {Hooper},
		\citenamefont {Hough}, \citenamefont {Howell}, \citenamefont {Huang},
		\citenamefont {Huerta}, \citenamefont {Hughey}, \citenamefont {Huttner},
		\citenamefont {Huynh}, \citenamefont {Huynh-Dinh}, \citenamefont {Ingram},
		\citenamefont {Inta}, \citenamefont {Isogai}, \citenamefont {Ivanov},
		\citenamefont {Iyer}, \citenamefont {Izumi}, \citenamefont {Jacobson},
		\citenamefont {James}, \citenamefont {Jang}, \citenamefont {Jang},
		\citenamefont {Jesse}, \citenamefont {Johnson}, \citenamefont {Jones},
		\citenamefont {Jones}, \citenamefont {Jones}, \citenamefont {Ju},
		\citenamefont {Kalmus}, \citenamefont {Kalogera}, \citenamefont {Kandhasamy},
		\citenamefont {Kang}, \citenamefont {Kanner}, \citenamefont {Kasturi},
		\citenamefont {Katsavounidis}, \citenamefont {Katzman}, \citenamefont
		{Kaufer}, \citenamefont {Kawabe}, \citenamefont {Kawamura}, \citenamefont
		{Kawazoe}, \citenamefont {Keitel}, \citenamefont {Kelley}, \citenamefont
		{Kells}, \citenamefont {Keppel}, \citenamefont {Khalaidovski}, \citenamefont
		{Khalili}, \citenamefont {Khazanov}, \citenamefont {Kim}, \citenamefont
		{Kim}, \citenamefont {Kim}, \citenamefont {Kim}, \citenamefont {Kim},
		\citenamefont {King}, \citenamefont {Kinzel}, \citenamefont {Kissel},
		\citenamefont {Klimenko}, \citenamefont {Kline}, \citenamefont {Kokeyama},
		\citenamefont {Kondrashov}, \citenamefont {Koranda}, \citenamefont {Korth},
		\citenamefont {Kozak}, \citenamefont {Kozameh}, \citenamefont {Kremin},
		\citenamefont {Kringel}, \citenamefont {Krishnan}, \citenamefont
		{Kucharczyk}, \citenamefont {Kuehn}, \citenamefont {Kumar}, \citenamefont
		{Kumar}, \citenamefont {Kuper}, \citenamefont {Kurdyumov}, \citenamefont
		{Kwee}, \citenamefont {Lam}, \citenamefont {Landry}, \citenamefont {Lantz},
		\citenamefont {Lasky}, \citenamefont {Lawrie}, \citenamefont {Lazzarini},
		\citenamefont {{Le Roux}}, \citenamefont {Leaci}, \citenamefont {Lee},
		\citenamefont {Lee}, \citenamefont {Lee}, \citenamefont {Lee}, \citenamefont
		{Leong}, \citenamefont {Levine}, \citenamefont {Lhuillier}, \citenamefont
		{Lin}, \citenamefont {Litvine}, \citenamefont {Liu}, \citenamefont {Liu},
		\citenamefont {Lockerbie}, \citenamefont {Lodhia}, \citenamefont {Loew},
		\citenamefont {Logue}, \citenamefont {Lombardi}, \citenamefont {Lormand},
		\citenamefont {Lough}, \citenamefont {Lubinski}, \citenamefont {L{\"{u}}ck},
		\citenamefont {Lundgren}, \citenamefont {Macarthur}, \citenamefont
		{Macdonald}, \citenamefont {Machenschalk}, \citenamefont {MacInnis},
		\citenamefont {Macleod}, \citenamefont {Maga{\~{n}}a-Sandoval}, \citenamefont
		{Mageswaran}, \citenamefont {Mailand}, \citenamefont {Manca}, \citenamefont
		{Mandel}, \citenamefont {Mandic}, \citenamefont {M{\'{a}}rka}, \citenamefont
		{M{\'{a}}rka}, \citenamefont {Markosyan}, \citenamefont {Maros},
		\citenamefont {Martin}, \citenamefont {Martin}, \citenamefont {Martinov},
		\citenamefont {Marx}, \citenamefont {Mason}, \citenamefont {Matichard},
		\citenamefont {Matone}, \citenamefont {Matzner}, \citenamefont {Mavalvala},
		\citenamefont {May}, \citenamefont {Mazzolo}, \citenamefont {McAuley},
		\citenamefont {McCarthy}, \citenamefont {McClelland}, \citenamefont
		{McGuire}, \citenamefont {McIntyre}, \citenamefont {McIver}, \citenamefont
		{Meadors}, \citenamefont {Mehmet}, \citenamefont {Meier}, \citenamefont
		{Melatos}, \citenamefont {Mendell}, \citenamefont {Mercer}, \citenamefont
		{Meshkov}, \citenamefont {Messenger}, \citenamefont {Meyer}, \citenamefont
		{Miao}, \citenamefont {Miller}, \citenamefont {Mingarelli}, \citenamefont
		{Mitra}, \citenamefont {Mitrofanov}, \citenamefont {Mitselmakher},
		\citenamefont {Mittleman}, \citenamefont {Moe}, \citenamefont {Mokler},
		\citenamefont {Mohapatra}, \citenamefont {Moraru}, \citenamefont {Moreno},
		\citenamefont {Mori}, \citenamefont {Morriss}, \citenamefont {Mossavi},
		\citenamefont {Mow-Lowry}, \citenamefont {Mueller}, \citenamefont {Mueller},
		\citenamefont {Mukherjee}, \citenamefont {Mullavey}, \citenamefont {Munch},
		\citenamefont {Murphy}, \citenamefont {Murray}, \citenamefont {Mytidis},
		\citenamefont {Kumar}, \citenamefont {Nash}, \citenamefont {Nayak},
		\citenamefont {Necula}, \citenamefont {Newton}, \citenamefont {Nguyen},
		\citenamefont {Nishida}, \citenamefont {Nishizawa}, \citenamefont {Nitz},
		\citenamefont {Nolting}, \citenamefont {Normandin}, \citenamefont {Nuttall},
		\citenamefont {O'Dell}, \citenamefont {O'Reilly}, \citenamefont
		{O'Shaughnessy}, \citenamefont {Ochsner}, \citenamefont {Oelker},
		\citenamefont {Ogin}, \citenamefont {Oh}, \citenamefont {Oh}, \citenamefont
		{Ohme}, \citenamefont {Oppermann}, \citenamefont {Osthelder}, \citenamefont
		{Ott}, \citenamefont {Ottaway}, \citenamefont {Ottens}, \citenamefont {Ou},
		\citenamefont {Overmier}, \citenamefont {Owen}, \citenamefont {Padilla},
		\citenamefont {Pai}, \citenamefont {Pan}, \citenamefont {Pankow},
		\citenamefont {Papa}, \citenamefont {Paris}, \citenamefont {Parkinson},
		\citenamefont {Pedraza}, \citenamefont {Penn}, \citenamefont {Peralta},
		\citenamefont {Perreca}, \citenamefont {Phelps}, \citenamefont {Pickenpack},
		\citenamefont {Pierro}, \citenamefont {Pinto}, \citenamefont {Pitkin},
		\citenamefont {Pletsch}, \citenamefont {P{\"{o}}ld}, \citenamefont
		{Postiglione}, \citenamefont {Poux}, \citenamefont {Predoi}, \citenamefont
		{Prestegard}, \citenamefont {Price}, \citenamefont {Prijatelj}, \citenamefont
		{Privitera}, \citenamefont {Prokhorov}, \citenamefont {Puncken},
		\citenamefont {Quetschke}, \citenamefont {Quintero}, \citenamefont
		{Quitzow-James}, \citenamefont {Raab}, \citenamefont {Radkins}, \citenamefont
		{Raffai}, \citenamefont {Raja}, \citenamefont {Rakhmanov}, \citenamefont
		{Ramet}, \citenamefont {Raymond}, \citenamefont {Reed}, \citenamefont {Reed},
		\citenamefont {Reid}, \citenamefont {Reitze}, \citenamefont {Riesen},
		\citenamefont {Riles}, \citenamefont {Roberts}, \citenamefont {Robertson},
		\citenamefont {Robinson}, \citenamefont {Roddy}, \citenamefont {Rodriguez},
		\citenamefont {Rodriguez}, \citenamefont {Rodruck}, \citenamefont {Rollins},
		\citenamefont {Romie}, \citenamefont {R{\"{o}}ver}, \citenamefont {Rowan},
		\citenamefont {R{\"{u}}diger}, \citenamefont {Ryan}, \citenamefont {Salemi},
		\citenamefont {Sammut}, \citenamefont {Sandberg}, \citenamefont {Sanders},
		\citenamefont {Sankar}, \citenamefont {Sannibale}, \citenamefont
		{Santamar{\'{i}}a}, \citenamefont {Santiago-Prieto}, \citenamefont
		{Santostasi}, \citenamefont {Sathyaprakash}, \citenamefont {Saulson},
		\citenamefont {Savage}, \citenamefont {Schilling}, \citenamefont {Schnabel},
		\citenamefont {Schofield}, \citenamefont {Schuette}, \citenamefont {Schulz},
		\citenamefont {Schutz}, \citenamefont {Schwinberg}, \citenamefont {Scott},
		\citenamefont {Scott}, \citenamefont {Seifert}, \citenamefont {Sellers},
		\citenamefont {Sengupta}, \citenamefont {Sergeev}, \citenamefont {Shaddock},
		\citenamefont {Shahriar}, \citenamefont {Shaltev}, \citenamefont {Shao},
		\citenamefont {Shapiro}, \citenamefont {Shawhan}, \citenamefont {Shoemaker},
		\citenamefont {Sidery}, \citenamefont {Siemens}, \citenamefont {Sigg},
		\citenamefont {Simakov}, \citenamefont {Singer}, \citenamefont {Singer},
		\citenamefont {Sintes}, \citenamefont {Skelton}, \citenamefont {Slagmolen},
		\citenamefont {Slutsky}, \citenamefont {Smith}, \citenamefont {Smith},
		\citenamefont {Smith}, \citenamefont {Smith-Lefebvre}, \citenamefont {Son},
		\citenamefont {Sorazu}, \citenamefont {Souradeep}, \citenamefont {Stefszky},
		\citenamefont {Steinert}, \citenamefont {Steinlechner}, \citenamefont
		{Steinlechner}, \citenamefont {Steplewski}, \citenamefont {Stevens},
		\citenamefont {Stochino}, \citenamefont {Stone}, \citenamefont {Strain},
		\citenamefont {Strigin}, \citenamefont {Stroeer}, \citenamefont {Stuver},
		\citenamefont {Summerscales}, \citenamefont {Susmithan}, \citenamefont
		{Sutton}, \citenamefont {Szeifert}, \citenamefont {Talukder}, \citenamefont
		{Tanner}, \citenamefont {Tarabrin}, \citenamefont {Taylor}, \citenamefont
		{Thomas}, \citenamefont {Thomas}, \citenamefont {Thorne}, \citenamefont
		{Thorne}, \citenamefont {Thrane}, \citenamefont {Tiwari}, \citenamefont
		{Tokmakov}, \citenamefont {Tomlinson}, \citenamefont {Torres}, \citenamefont
		{Torrie}, \citenamefont {Traylor}, \citenamefont {Tse}, \citenamefont
		{Ugolini}, \citenamefont {Unnikrishnan}, \citenamefont {Vahlbruch},
		\citenamefont {Vallisneri}, \citenamefont {{Van Der Sluys}}, \citenamefont
		{{Van Veggel}}, \citenamefont {Vass}, \citenamefont {Vaulin}, \citenamefont
		{Vecchio}, \citenamefont {Veitch}, \citenamefont {Veitch}, \citenamefont
		{Venkateswara}, \citenamefont {Verma}, \citenamefont {Vincent-Finley},
		\citenamefont {Vitale}, \citenamefont {Vo}, \citenamefont {Vorvick},
		\citenamefont {Vousden}, \citenamefont {Vyatchanin}, \citenamefont {Wade},
		\citenamefont {Wade}, \citenamefont {Wade}, \citenamefont {Waldman},
		\citenamefont {Wallace}, \citenamefont {Wan}, \citenamefont {Wang},
		\citenamefont {Wang}, \citenamefont {Wang}, \citenamefont {Wanner},
		\citenamefont {Ward}, \citenamefont {Was}, \citenamefont {Weinert},
		\citenamefont {Weinstein}, \citenamefont {Weiss}, \citenamefont {Welborn},
		\citenamefont {Wen}, \citenamefont {Wessels}, \citenamefont {West},
		\citenamefont {Westphal}, \citenamefont {Wette}, \citenamefont {Whelan},
		\citenamefont {Whitcomb}, \citenamefont {Wiseman}, \citenamefont {White},
		\citenamefont {Whiting}, \citenamefont {Wiesner}, \citenamefont {Wilkinson},
		\citenamefont {Willems}, \citenamefont {Williams}, \citenamefont {Williams},
		\citenamefont {Williams}, \citenamefont {Willis}, \citenamefont {Willke},
		\citenamefont {Wimmer}, \citenamefont {Winkelmann}, \citenamefont {Winkler},
		\citenamefont {Wipf}, \citenamefont {Wittel}, \citenamefont {Woan},
		\citenamefont {Wooley}, \citenamefont {Worden}, \citenamefont {Yablon},
		\citenamefont {Yakushin}, \citenamefont {Yamamoto}, \citenamefont {Yancey},
		\citenamefont {Yang}, \citenamefont {Yeaton-Massey}, \citenamefont {Yoshida},
		\citenamefont {Yum}, \citenamefont {Zanolin}, \citenamefont {Zhang},
		\citenamefont {Zhang}, \citenamefont {Zhao}, \citenamefont {Zhu},
		\citenamefont {Zhu}, \citenamefont {Zotov}, \citenamefont {Zucker},\ and\
		\citenamefont {Zweizig}}]{aasi2013enhanced}%
	\BibitemOpen
	\bibfield  {author} {\bibinfo {author} {\bibfnamefont {J.}~\bibnamefont
			{Aasi}}, \bibinfo {author} {\bibfnamefont {J.}~\bibnamefont {Abadie}},
		\bibinfo {author} {\bibfnamefont {B.~P.}\ \bibnamefont {Abbott}}, \bibinfo
		{author} {\bibfnamefont {R.}~\bibnamefont {Abbott}}, \bibinfo {author}
		{\bibfnamefont {T.~D.}\ \bibnamefont {Abbott}}, \bibinfo {author}
		{\bibfnamefont {M.~R.}\ \bibnamefont {Abernathy}}, \bibinfo {author}
		{\bibfnamefont {C.}~\bibnamefont {Adams}}, \bibinfo {author} {\bibfnamefont
			{T.}~\bibnamefont {Adams}}, \bibinfo {author} {\bibfnamefont
			{P.}~\bibnamefont {Addesso}}, \bibinfo {author} {\bibfnamefont {R.~X.}\
			\bibnamefont {Adhikari}}, \bibinfo {author} {\bibfnamefont {C.}~\bibnamefont
			{Affeldt}}, \bibinfo {author} {\bibfnamefont {O.~D.}\ \bibnamefont {Aguiar}},
		\bibinfo {author} {\bibfnamefont {P.}~\bibnamefont {Ajith}}, \bibinfo
		{author} {\bibfnamefont {B.}~\bibnamefont {Allen}}, \bibinfo {author}
		{\bibfnamefont {E.~A.}\ \bibnamefont {Ceron}}, \bibinfo {author}
		{\bibfnamefont {D.}~\bibnamefont {Amariutei}}, \bibinfo {author}
		{\bibfnamefont {S.~B.}\ \bibnamefont {Anderson}}, \bibinfo {author}
		{\bibfnamefont {W.~G.}\ \bibnamefont {Anderson}}, \bibinfo {author}
		{\bibfnamefont {K.}~\bibnamefont {Arai}}, \bibinfo {author} {\bibfnamefont
			{M.~C.}\ \bibnamefont {Araya}}, \bibinfo {author} {\bibfnamefont
			{C.}~\bibnamefont {Arceneaux}}, \bibinfo {author} {\bibfnamefont
			{S.}~\bibnamefont {Ast}}, \bibinfo {author} {\bibfnamefont {S.~M.}\
			\bibnamefont {Aston}}, \bibinfo {author} {\bibfnamefont {D.}~\bibnamefont
			{Atkinson}}, \bibinfo {author} {\bibfnamefont {P.}~\bibnamefont {Aufmuth}},
		\bibinfo {author} {\bibfnamefont {C.}~\bibnamefont {Aulbert}}, \bibinfo
		{author} {\bibfnamefont {L.}~\bibnamefont {Austin}}, \bibinfo {author}
		{\bibfnamefont {B.~E.}\ \bibnamefont {Aylott}}, \bibinfo {author}
		{\bibfnamefont {S.}~\bibnamefont {Babak}}, \bibinfo {author} {\bibfnamefont
			{P.~T.}\ \bibnamefont {Baker}}, \bibinfo {author} {\bibfnamefont
			{S.}~\bibnamefont {Ballmer}}, \bibinfo {author} {\bibfnamefont
			{Y.}~\bibnamefont {Bao}}, \bibinfo {author} {\bibfnamefont {J.~C.}\
			\bibnamefont {Barayoga}}, \bibinfo {author} {\bibfnamefont {D.}~\bibnamefont
			{Barker}}, \bibinfo {author} {\bibfnamefont {B.}~\bibnamefont {Barr}},
		\bibinfo {author} {\bibfnamefont {L.}~\bibnamefont {Barsotti}}, \bibinfo
		{author} {\bibfnamefont {M.~A.}\ \bibnamefont {Barton}}, \bibinfo {author}
		{\bibfnamefont {I.}~\bibnamefont {Bartos}}, \bibinfo {author} {\bibfnamefont
			{R.}~\bibnamefont {Bassiri}}, \bibinfo {author} {\bibfnamefont
			{J.}~\bibnamefont {Batch}}, \bibinfo {author} {\bibfnamefont
			{J.}~\bibnamefont {Bauchrowitz}}, \bibinfo {author} {\bibfnamefont
			{B.}~\bibnamefont {Behnke}}, \bibinfo {author} {\bibfnamefont {A.~S.}\
			\bibnamefont {Bell}}, \bibinfo {author} {\bibfnamefont {C.}~\bibnamefont
			{Bell}}, \bibinfo {author} {\bibfnamefont {G.}~\bibnamefont {Bergmann}},
		\bibinfo {author} {\bibfnamefont {J.~M.}\ \bibnamefont {Berliner}}, \bibinfo
		{author} {\bibfnamefont {A.}~\bibnamefont {Bertolini}}, \bibinfo {author}
		{\bibfnamefont {J.}~\bibnamefont {Betzwieser}}, \bibinfo {author}
		{\bibfnamefont {N.}~\bibnamefont {Beveridge}}, \bibinfo {author}
		{\bibfnamefont {P.~T.}\ \bibnamefont {Beyersdorf}}, \bibinfo {author}
		{\bibfnamefont {T.}~\bibnamefont {Bhadbhade}}, \bibinfo {author}
		{\bibfnamefont {I.~A.}\ \bibnamefont {Bilenko}}, \bibinfo {author}
		{\bibfnamefont {G.}~\bibnamefont {Billingsley}}, \bibinfo {author}
		{\bibfnamefont {J.}~\bibnamefont {Birch}}, \bibinfo {author} {\bibfnamefont
			{S.}~\bibnamefont {Biscans}}, \bibinfo {author} {\bibfnamefont
			{E.}~\bibnamefont {Black}}, \bibinfo {author} {\bibfnamefont {J.~K.}\
			\bibnamefont {Blackburn}}, \bibinfo {author} {\bibfnamefont {L.}~\bibnamefont
			{Blackburn}}, \bibinfo {author} {\bibfnamefont {D.}~\bibnamefont {Blair}},
		\bibinfo {author} {\bibfnamefont {B.}~\bibnamefont {Bland}}, \bibinfo
		{author} {\bibfnamefont {O.}~\bibnamefont {Bock}}, \bibinfo {author}
		{\bibfnamefont {T.~P.}\ \bibnamefont {Bodiya}}, \bibinfo {author}
		{\bibfnamefont {C.}~\bibnamefont {Bogan}}, \bibinfo {author} {\bibfnamefont
			{C.}~\bibnamefont {Bond}}, \bibinfo {author} {\bibfnamefont {R.}~\bibnamefont
			{Bork}}, \bibinfo {author} {\bibfnamefont {M.}~\bibnamefont {Born}}, \bibinfo
		{author} {\bibfnamefont {S.}~\bibnamefont {Bose}}, \bibinfo {author}
		{\bibfnamefont {J.}~\bibnamefont {Bowers}}, \bibinfo {author} {\bibfnamefont
			{P.~R.}\ \bibnamefont {Brady}}, \bibinfo {author} {\bibfnamefont {V.~B.}\
			\bibnamefont {Braginsky}}, \bibinfo {author} {\bibfnamefont {J.~E.}\
			\bibnamefont {Brau}}, \bibinfo {author} {\bibfnamefont {J.}~\bibnamefont
			{Breyer}}, \bibinfo {author} {\bibfnamefont {D.~O.}\ \bibnamefont {Bridges}},
		\bibinfo {author} {\bibfnamefont {M.}~\bibnamefont {Brinkmann}}, \bibinfo
		{author} {\bibfnamefont {M.}~\bibnamefont {Britzger}}, \bibinfo {author}
		{\bibfnamefont {A.~F.}\ \bibnamefont {Brooks}}, \bibinfo {author}
		{\bibfnamefont {D.~A.}\ \bibnamefont {Brown}}, \bibinfo {author}
		{\bibfnamefont {D.~D.}\ \bibnamefont {Brown}}, \bibinfo {author}
		{\bibfnamefont {K.}~\bibnamefont {Buckland}}, \bibinfo {author}
		{\bibfnamefont {F.}~\bibnamefont {Br{\"{u}}ckner}}, \bibinfo {author}
		{\bibfnamefont {B.~C.}\ \bibnamefont {Buchler}}, \bibinfo {author}
		{\bibfnamefont {A.}~\bibnamefont {Buonanno}}, \bibinfo {author}
		{\bibfnamefont {J.}~\bibnamefont {Burguet-Castell}}, \bibinfo {author}
		{\bibfnamefont {R.~L.}\ \bibnamefont {Byer}}, \bibinfo {author}
		{\bibfnamefont {L.}~\bibnamefont {Cadonati}}, \bibinfo {author}
		{\bibfnamefont {J.~B.}\ \bibnamefont {Camp}}, \bibinfo {author}
		{\bibfnamefont {P.}~\bibnamefont {Campsie}}, \bibinfo {author} {\bibfnamefont
			{K.}~\bibnamefont {Cannon}}, \bibinfo {author} {\bibfnamefont
			{J.}~\bibnamefont {Cao}}, \bibinfo {author} {\bibfnamefont {C.~D.}\
			\bibnamefont {Capano}}, \bibinfo {author} {\bibfnamefont {L.}~\bibnamefont
			{Carbone}}, \bibinfo {author} {\bibfnamefont {S.}~\bibnamefont {Caride}},
		\bibinfo {author} {\bibfnamefont {A.~D.}\ \bibnamefont {Castiglia}}, \bibinfo
		{author} {\bibfnamefont {S.}~\bibnamefont {Caudill}}, \bibinfo {author}
		{\bibfnamefont {M.}~\bibnamefont {Cavagli{\`{a}}}}, \bibinfo {author}
		{\bibfnamefont {C.}~\bibnamefont {Cepeda}}, \bibinfo {author} {\bibfnamefont
			{T.}~\bibnamefont {Chalermsongsak}}, \bibinfo {author} {\bibfnamefont
			{S.}~\bibnamefont {Chao}}, \bibinfo {author} {\bibfnamefont {P.}~\bibnamefont
			{Charlton}}, \bibinfo {author} {\bibfnamefont {X.}~\bibnamefont {Chen}},
		\bibinfo {author} {\bibfnamefont {Y.}~\bibnamefont {Chen}}, \bibinfo {author}
		{\bibfnamefont {H.~S.}\ \bibnamefont {Cho}}, \bibinfo {author} {\bibfnamefont
			{J.~H.}\ \bibnamefont {Chow}}, \bibinfo {author} {\bibfnamefont
			{N.}~\bibnamefont {Christensen}}, \bibinfo {author} {\bibfnamefont
			{Q.}~\bibnamefont {Chu}}, \bibinfo {author} {\bibfnamefont {S.~S.}\
			\bibnamefont {Chua}}, \bibinfo {author} {\bibfnamefont {C.~T.}\ \bibnamefont
			{Chung}}, \bibinfo {author} {\bibfnamefont {G.}~\bibnamefont {Ciani}},
		\bibinfo {author} {\bibfnamefont {F.}~\bibnamefont {Clara}}, \bibinfo
		{author} {\bibfnamefont {D.~E.}\ \bibnamefont {Clark}}, \bibinfo {author}
		{\bibfnamefont {J.~A.}\ \bibnamefont {Clark}}, \bibinfo {author}
		{\bibfnamefont {M.}~\bibnamefont {Constancio}}, \bibinfo {author}
		{\bibfnamefont {D.}~\bibnamefont {Cook}}, \bibinfo {author} {\bibfnamefont
			{T.~R.}\ \bibnamefont {Corbitt}}, \bibinfo {author} {\bibfnamefont
			{M.}~\bibnamefont {Cordier}}, \bibinfo {author} {\bibfnamefont
			{N.}~\bibnamefont {Cornish}}, \bibinfo {author} {\bibfnamefont
			{A.}~\bibnamefont {Corsi}}, \bibinfo {author} {\bibfnamefont {C.~A.}\
			\bibnamefont {Costa}}, \bibinfo {author} {\bibfnamefont {M.~W.}\ \bibnamefont
			{Coughlin}}, \bibinfo {author} {\bibfnamefont {S.}~\bibnamefont
			{Countryman}}, \bibinfo {author} {\bibfnamefont {P.}~\bibnamefont
			{Couvares}}, \bibinfo {author} {\bibfnamefont {D.~M.}\ \bibnamefont
			{Coward}}, \bibinfo {author} {\bibfnamefont {M.}~\bibnamefont {Cowart}},
		\bibinfo {author} {\bibfnamefont {D.~C.}\ \bibnamefont {Coyne}}, \bibinfo
		{author} {\bibfnamefont {K.}~\bibnamefont {Craig}}, \bibinfo {author}
		{\bibfnamefont {J.~D.}\ \bibnamefont {Creighton}}, \bibinfo {author}
		{\bibfnamefont {T.~D.}\ \bibnamefont {Creighton}}, \bibinfo {author}
		{\bibfnamefont {A.}~\bibnamefont {Cumming}}, \bibinfo {author} {\bibfnamefont
			{L.}~\bibnamefont {Cunningham}}, \bibinfo {author} {\bibfnamefont
			{K.}~\bibnamefont {Dahl}}, \bibinfo {author} {\bibfnamefont {M.}~\bibnamefont
			{Damjanic}}, \bibinfo {author} {\bibfnamefont {S.~L.}\ \bibnamefont
			{Danilishin}}, \bibinfo {author} {\bibfnamefont {K.}~\bibnamefont
			{Danzmann}}, \bibinfo {author} {\bibfnamefont {B.}~\bibnamefont {Daudert}},
		\bibinfo {author} {\bibfnamefont {H.}~\bibnamefont {Daveloza}}, \bibinfo
		{author} {\bibfnamefont {G.~S.}\ \bibnamefont {Davies}}, \bibinfo {author}
		{\bibfnamefont {E.~J.}\ \bibnamefont {Daw}}, \bibinfo {author} {\bibfnamefont
			{T.}~\bibnamefont {Dayanga}}, \bibinfo {author} {\bibfnamefont
			{E.}~\bibnamefont {Deleeuw}}, \bibinfo {author} {\bibfnamefont
			{T.}~\bibnamefont {Denker}}, \bibinfo {author} {\bibfnamefont
			{T.}~\bibnamefont {Dent}}, \bibinfo {author} {\bibfnamefont {V.}~\bibnamefont
			{Dergachev}}, \bibinfo {author} {\bibfnamefont {R.}~\bibnamefont {DeRosa}},
		\bibinfo {author} {\bibfnamefont {R.}~\bibnamefont {DeSalvo}}, \bibinfo
		{author} {\bibfnamefont {S.}~\bibnamefont {Dhurandhar}}, \bibinfo {author}
		{\bibfnamefont {I.}~\bibnamefont {{Di Palma}}}, \bibinfo {author}
		{\bibfnamefont {M.}~\bibnamefont {D{\'{i}}az}}, \bibinfo {author}
		{\bibfnamefont {A.}~\bibnamefont {Dietz}}, \bibinfo {author} {\bibfnamefont
			{F.}~\bibnamefont {Donovan}}, \bibinfo {author} {\bibfnamefont {K.~L.}\
			\bibnamefont {Dooley}}, \bibinfo {author} {\bibfnamefont {S.}~\bibnamefont
			{Doravari}}, \bibinfo {author} {\bibfnamefont {S.}~\bibnamefont {Drasco}},
		\bibinfo {author} {\bibfnamefont {R.~W.}\ \bibnamefont {Drever}}, \bibinfo
		{author} {\bibfnamefont {J.~C.}\ \bibnamefont {Driggers}}, \bibinfo {author}
		{\bibfnamefont {Z.}~\bibnamefont {Du}}, \bibinfo {author} {\bibfnamefont
			{J.~C.}\ \bibnamefont {Dumas}}, \bibinfo {author} {\bibfnamefont
			{S.}~\bibnamefont {Dwyer}}, \bibinfo {author} {\bibfnamefont
			{T.}~\bibnamefont {Eberle}}, \bibinfo {author} {\bibfnamefont
			{M.}~\bibnamefont {Edwards}}, \bibinfo {author} {\bibfnamefont
			{A.}~\bibnamefont {Effler}}, \bibinfo {author} {\bibfnamefont
			{P.}~\bibnamefont {Ehrens}}, \bibinfo {author} {\bibfnamefont {S.~S.}\
			\bibnamefont {Eikenberry}}, \bibinfo {author} {\bibfnamefont
			{R.}~\bibnamefont {Engel}}, \bibinfo {author} {\bibfnamefont
			{R.}~\bibnamefont {Essick}}, \bibinfo {author} {\bibfnamefont
			{T.}~\bibnamefont {Etzel}}, \bibinfo {author} {\bibfnamefont
			{K.}~\bibnamefont {Evans}}, \bibinfo {author} {\bibfnamefont
			{M.}~\bibnamefont {Evans}}, \bibinfo {author} {\bibfnamefont
			{T.}~\bibnamefont {Evans}}, \bibinfo {author} {\bibfnamefont
			{M.}~\bibnamefont {Factourovich}}, \bibinfo {author} {\bibfnamefont
			{S.}~\bibnamefont {Fairhurst}}, \bibinfo {author} {\bibfnamefont
			{Q.}~\bibnamefont {Fang}}, \bibinfo {author} {\bibfnamefont {B.~F.}\
			\bibnamefont {Farr}}, \bibinfo {author} {\bibfnamefont {W.}~\bibnamefont
			{Farr}}, \bibinfo {author} {\bibfnamefont {M.}~\bibnamefont {Favata}},
		\bibinfo {author} {\bibfnamefont {D.}~\bibnamefont {Fazi}}, \bibinfo {author}
		{\bibfnamefont {H.}~\bibnamefont {Fehrmann}}, \bibinfo {author}
		{\bibfnamefont {D.}~\bibnamefont {Feldbaum}}, \bibinfo {author}
		{\bibfnamefont {L.~S.}\ \bibnamefont {Finn}}, \bibinfo {author}
		{\bibfnamefont {R.~P.}\ \bibnamefont {Fisher}}, \bibinfo {author}
		{\bibfnamefont {S.}~\bibnamefont {Foley}}, \bibinfo {author} {\bibfnamefont
			{E.}~\bibnamefont {Forsi}}, \bibinfo {author} {\bibfnamefont
			{N.}~\bibnamefont {Fotopoulos}}, \bibinfo {author} {\bibfnamefont
			{M.}~\bibnamefont {Frede}}, \bibinfo {author} {\bibfnamefont {M.~A.}\
			\bibnamefont {Frei}}, \bibinfo {author} {\bibfnamefont {Z.}~\bibnamefont
			{Frei}}, \bibinfo {author} {\bibfnamefont {A.}~\bibnamefont {Freise}},
		\bibinfo {author} {\bibfnamefont {R.}~\bibnamefont {Frey}}, \bibinfo {author}
		{\bibfnamefont {T.~T.}\ \bibnamefont {Fricke}}, \bibinfo {author}
		{\bibfnamefont {D.}~\bibnamefont {Friedrich}}, \bibinfo {author}
		{\bibfnamefont {P.}~\bibnamefont {Fritschel}}, \bibinfo {author}
		{\bibfnamefont {V.~V.}\ \bibnamefont {Frolov}}, \bibinfo {author}
		{\bibfnamefont {M.~K.}\ \bibnamefont {Fujimoto}}, \bibinfo {author}
		{\bibfnamefont {P.~J.}\ \bibnamefont {Fulda}}, \bibinfo {author}
		{\bibfnamefont {M.}~\bibnamefont {Fyffe}}, \bibinfo {author} {\bibfnamefont
			{J.}~\bibnamefont {Gair}}, \bibinfo {author} {\bibfnamefont {J.}~\bibnamefont
			{Garcia}}, \bibinfo {author} {\bibfnamefont {N.}~\bibnamefont {Gehrels}},
		\bibinfo {author} {\bibfnamefont {G.}~\bibnamefont {Gelencser}}, \bibinfo
		{author} {\bibfnamefont {L.~{\'{A}}.}\ \bibnamefont {Gergely}}, \bibinfo
		{author} {\bibfnamefont {S.}~\bibnamefont {Ghosh}}, \bibinfo {author}
		{\bibfnamefont {J.~A.}\ \bibnamefont {Giaime}}, \bibinfo {author}
		{\bibfnamefont {S.}~\bibnamefont {Giampanis}}, \bibinfo {author}
		{\bibfnamefont {K.~D.}\ \bibnamefont {Giardina}}, \bibinfo {author}
		{\bibfnamefont {S.}~\bibnamefont {Gil-Casanova}}, \bibinfo {author}
		{\bibfnamefont {C.}~\bibnamefont {Gill}}, \bibinfo {author} {\bibfnamefont
			{J.}~\bibnamefont {Gleason}}, \bibinfo {author} {\bibfnamefont
			{E.}~\bibnamefont {Goetz}}, \bibinfo {author} {\bibfnamefont
			{G.}~\bibnamefont {Gonz{\'{a}}lez}}, \bibinfo {author} {\bibfnamefont
			{N.}~\bibnamefont {Gordon}}, \bibinfo {author} {\bibfnamefont {M.~L.}\
			\bibnamefont {Gorodetsky}}, \bibinfo {author} {\bibfnamefont
			{S.}~\bibnamefont {Gossan}}, \bibinfo {author} {\bibfnamefont
			{S.}~\bibnamefont {Go{\ss}ler}}, \bibinfo {author} {\bibfnamefont
			{C.}~\bibnamefont {Graef}}, \bibinfo {author} {\bibfnamefont {P.~B.}\
			\bibnamefont {Graff}}, \bibinfo {author} {\bibfnamefont {A.}~\bibnamefont
			{Grant}}, \bibinfo {author} {\bibfnamefont {S.}~\bibnamefont {Gras}},
		\bibinfo {author} {\bibfnamefont {C.}~\bibnamefont {Gray}}, \bibinfo {author}
		{\bibfnamefont {R.~J.}\ \bibnamefont {Greenhalgh}}, \bibinfo {author}
		{\bibfnamefont {A.~M.}\ \bibnamefont {Gretarsson}}, \bibinfo {author}
		{\bibfnamefont {C.}~\bibnamefont {Griffo}}, \bibinfo {author} {\bibfnamefont
			{H.}~\bibnamefont {Grote}}, \bibinfo {author} {\bibfnamefont
			{K.}~\bibnamefont {Grover}}, \bibinfo {author} {\bibfnamefont
			{S.}~\bibnamefont {Grunewald}}, \bibinfo {author} {\bibfnamefont
			{C.}~\bibnamefont {Guido}}, \bibinfo {author} {\bibfnamefont {E.~K.}\
			\bibnamefont {Gustafson}}, \bibinfo {author} {\bibfnamefont {R.}~\bibnamefont
			{Gustafson}}, \bibinfo {author} {\bibfnamefont {D.}~\bibnamefont {Hammer}},
		\bibinfo {author} {\bibfnamefont {G.}~\bibnamefont {Hammond}}, \bibinfo
		{author} {\bibfnamefont {J.}~\bibnamefont {Hanks}}, \bibinfo {author}
		{\bibfnamefont {C.}~\bibnamefont {Hanna}}, \bibinfo {author} {\bibfnamefont
			{J.}~\bibnamefont {Hanson}}, \bibinfo {author} {\bibfnamefont
			{K.}~\bibnamefont {Haris}}, \bibinfo {author} {\bibfnamefont
			{J.}~\bibnamefont {Harms}}, \bibinfo {author} {\bibfnamefont {G.~M.}\
			\bibnamefont {Harry}}, \bibinfo {author} {\bibfnamefont {I.~W.}\ \bibnamefont
			{Harry}}, \bibinfo {author} {\bibfnamefont {E.~D.}\ \bibnamefont {Harstad}},
		\bibinfo {author} {\bibfnamefont {M.~T.}\ \bibnamefont {Hartman}}, \bibinfo
		{author} {\bibfnamefont {K.}~\bibnamefont {Haughian}}, \bibinfo {author}
		{\bibfnamefont {K.}~\bibnamefont {Hayama}}, \bibinfo {author} {\bibfnamefont
			{J.}~\bibnamefont {Heefner}}, \bibinfo {author} {\bibfnamefont {M.~C.}\
			\bibnamefont {Heintze}}, \bibinfo {author} {\bibfnamefont {M.~A.}\
			\bibnamefont {Hendry}}, \bibinfo {author} {\bibfnamefont {I.~S.}\
			\bibnamefont {Heng}}, \bibinfo {author} {\bibfnamefont {A.~W.}\ \bibnamefont
			{Heptonstall}}, \bibinfo {author} {\bibfnamefont {M.}~\bibnamefont {Heurs}},
		\bibinfo {author} {\bibfnamefont {M.}~\bibnamefont {Hewitson}}, \bibinfo
		{author} {\bibfnamefont {S.}~\bibnamefont {Hild}}, \bibinfo {author}
		{\bibfnamefont {D.}~\bibnamefont {Hoak}}, \bibinfo {author} {\bibfnamefont
			{K.~A.}\ \bibnamefont {Hodge}}, \bibinfo {author} {\bibfnamefont
			{K.}~\bibnamefont {Holt}}, \bibinfo {author} {\bibfnamefont {M.}~\bibnamefont
			{Holtrop}}, \bibinfo {author} {\bibfnamefont {T.}~\bibnamefont {Hong}},
		\bibinfo {author} {\bibfnamefont {S.}~\bibnamefont {Hooper}}, \bibinfo
		{author} {\bibfnamefont {J.}~\bibnamefont {Hough}}, \bibinfo {author}
		{\bibfnamefont {E.~J.}\ \bibnamefont {Howell}}, \bibinfo {author}
		{\bibfnamefont {V.}~\bibnamefont {Huang}}, \bibinfo {author} {\bibfnamefont
			{E.~A.}\ \bibnamefont {Huerta}}, \bibinfo {author} {\bibfnamefont
			{B.}~\bibnamefont {Hughey}}, \bibinfo {author} {\bibfnamefont {S.~H.}\
			\bibnamefont {Huttner}}, \bibinfo {author} {\bibfnamefont {M.}~\bibnamefont
			{Huynh}}, \bibinfo {author} {\bibfnamefont {T.}~\bibnamefont {Huynh-Dinh}},
		\bibinfo {author} {\bibfnamefont {D.~R.}\ \bibnamefont {Ingram}}, \bibinfo
		{author} {\bibfnamefont {R.}~\bibnamefont {Inta}}, \bibinfo {author}
		{\bibfnamefont {T.}~\bibnamefont {Isogai}}, \bibinfo {author} {\bibfnamefont
			{A.}~\bibnamefont {Ivanov}}, \bibinfo {author} {\bibfnamefont {B.~R.}\
			\bibnamefont {Iyer}}, \bibinfo {author} {\bibfnamefont {K.}~\bibnamefont
			{Izumi}}, \bibinfo {author} {\bibfnamefont {M.}~\bibnamefont {Jacobson}},
		\bibinfo {author} {\bibfnamefont {E.}~\bibnamefont {James}}, \bibinfo
		{author} {\bibfnamefont {H.}~\bibnamefont {Jang}}, \bibinfo {author}
		{\bibfnamefont {Y.~J.}\ \bibnamefont {Jang}}, \bibinfo {author}
		{\bibfnamefont {E.}~\bibnamefont {Jesse}}, \bibinfo {author} {\bibfnamefont
			{W.~W.}\ \bibnamefont {Johnson}}, \bibinfo {author} {\bibfnamefont
			{D.}~\bibnamefont {Jones}}, \bibinfo {author} {\bibfnamefont {D.~I.}\
			\bibnamefont {Jones}}, \bibinfo {author} {\bibfnamefont {R.}~\bibnamefont
			{Jones}}, \bibinfo {author} {\bibfnamefont {L.}~\bibnamefont {Ju}}, \bibinfo
		{author} {\bibfnamefont {P.}~\bibnamefont {Kalmus}}, \bibinfo {author}
		{\bibfnamefont {V.}~\bibnamefont {Kalogera}}, \bibinfo {author}
		{\bibfnamefont {S.}~\bibnamefont {Kandhasamy}}, \bibinfo {author}
		{\bibfnamefont {G.}~\bibnamefont {Kang}}, \bibinfo {author} {\bibfnamefont
			{J.~B.}\ \bibnamefont {Kanner}}, \bibinfo {author} {\bibfnamefont
			{R.}~\bibnamefont {Kasturi}}, \bibinfo {author} {\bibfnamefont
			{E.}~\bibnamefont {Katsavounidis}}, \bibinfo {author} {\bibfnamefont
			{W.}~\bibnamefont {Katzman}}, \bibinfo {author} {\bibfnamefont
			{H.}~\bibnamefont {Kaufer}}, \bibinfo {author} {\bibfnamefont
			{K.}~\bibnamefont {Kawabe}}, \bibinfo {author} {\bibfnamefont
			{S.}~\bibnamefont {Kawamura}}, \bibinfo {author} {\bibfnamefont
			{F.}~\bibnamefont {Kawazoe}}, \bibinfo {author} {\bibfnamefont
			{D.}~\bibnamefont {Keitel}}, \bibinfo {author} {\bibfnamefont {D.~B.}\
			\bibnamefont {Kelley}}, \bibinfo {author} {\bibfnamefont {W.}~\bibnamefont
			{Kells}}, \bibinfo {author} {\bibfnamefont {D.~G.}\ \bibnamefont {Keppel}},
		\bibinfo {author} {\bibfnamefont {A.}~\bibnamefont {Khalaidovski}}, \bibinfo
		{author} {\bibfnamefont {F.~Y.}\ \bibnamefont {Khalili}}, \bibinfo {author}
		{\bibfnamefont {E.~A.}\ \bibnamefont {Khazanov}}, \bibinfo {author}
		{\bibfnamefont {B.~K.}\ \bibnamefont {Kim}}, \bibinfo {author} {\bibfnamefont
			{C.}~\bibnamefont {Kim}}, \bibinfo {author} {\bibfnamefont {K.}~\bibnamefont
			{Kim}}, \bibinfo {author} {\bibfnamefont {N.}~\bibnamefont {Kim}}, \bibinfo
		{author} {\bibfnamefont {Y.~M.}\ \bibnamefont {Kim}}, \bibinfo {author}
		{\bibfnamefont {P.~J.}\ \bibnamefont {King}}, \bibinfo {author}
		{\bibfnamefont {D.~L.}\ \bibnamefont {Kinzel}}, \bibinfo {author}
		{\bibfnamefont {J.~S.}\ \bibnamefont {Kissel}}, \bibinfo {author}
		{\bibfnamefont {S.}~\bibnamefont {Klimenko}}, \bibinfo {author}
		{\bibfnamefont {J.}~\bibnamefont {Kline}}, \bibinfo {author} {\bibfnamefont
			{K.}~\bibnamefont {Kokeyama}}, \bibinfo {author} {\bibfnamefont
			{V.}~\bibnamefont {Kondrashov}}, \bibinfo {author} {\bibfnamefont
			{S.}~\bibnamefont {Koranda}}, \bibinfo {author} {\bibfnamefont {W.~Z.}\
			\bibnamefont {Korth}}, \bibinfo {author} {\bibfnamefont {D.}~\bibnamefont
			{Kozak}}, \bibinfo {author} {\bibfnamefont {C.}~\bibnamefont {Kozameh}},
		\bibinfo {author} {\bibfnamefont {A.}~\bibnamefont {Kremin}}, \bibinfo
		{author} {\bibfnamefont {V.}~\bibnamefont {Kringel}}, \bibinfo {author}
		{\bibfnamefont {B.}~\bibnamefont {Krishnan}}, \bibinfo {author}
		{\bibfnamefont {C.}~\bibnamefont {Kucharczyk}}, \bibinfo {author}
		{\bibfnamefont {G.}~\bibnamefont {Kuehn}}, \bibinfo {author} {\bibfnamefont
			{P.}~\bibnamefont {Kumar}}, \bibinfo {author} {\bibfnamefont
			{R.}~\bibnamefont {Kumar}}, \bibinfo {author} {\bibfnamefont {B.~J.}\
			\bibnamefont {Kuper}}, \bibinfo {author} {\bibfnamefont {R.}~\bibnamefont
			{Kurdyumov}}, \bibinfo {author} {\bibfnamefont {P.}~\bibnamefont {Kwee}},
		\bibinfo {author} {\bibfnamefont {P.~K.}\ \bibnamefont {Lam}}, \bibinfo
		{author} {\bibfnamefont {M.}~\bibnamefont {Landry}}, \bibinfo {author}
		{\bibfnamefont {B.}~\bibnamefont {Lantz}}, \bibinfo {author} {\bibfnamefont
			{P.~D.}\ \bibnamefont {Lasky}}, \bibinfo {author} {\bibfnamefont
			{C.}~\bibnamefont {Lawrie}}, \bibinfo {author} {\bibfnamefont
			{A.}~\bibnamefont {Lazzarini}}, \bibinfo {author} {\bibfnamefont
			{A.}~\bibnamefont {{Le Roux}}}, \bibinfo {author} {\bibfnamefont
			{P.}~\bibnamefont {Leaci}}, \bibinfo {author} {\bibfnamefont {C.~H.}\
			\bibnamefont {Lee}}, \bibinfo {author} {\bibfnamefont {H.~K.}\ \bibnamefont
			{Lee}}, \bibinfo {author} {\bibfnamefont {H.~M.}\ \bibnamefont {Lee}},
		\bibinfo {author} {\bibfnamefont {J.}~\bibnamefont {Lee}}, \bibinfo {author}
		{\bibfnamefont {J.~R.}\ \bibnamefont {Leong}}, \bibinfo {author}
		{\bibfnamefont {B.}~\bibnamefont {Levine}}, \bibinfo {author} {\bibfnamefont
			{V.}~\bibnamefont {Lhuillier}}, \bibinfo {author} {\bibfnamefont {A.~C.}\
			\bibnamefont {Lin}}, \bibinfo {author} {\bibfnamefont {V.}~\bibnamefont
			{Litvine}}, \bibinfo {author} {\bibfnamefont {Y.}~\bibnamefont {Liu}},
		\bibinfo {author} {\bibfnamefont {Z.}~\bibnamefont {Liu}}, \bibinfo {author}
		{\bibfnamefont {N.~A.}\ \bibnamefont {Lockerbie}}, \bibinfo {author}
		{\bibfnamefont {D.}~\bibnamefont {Lodhia}}, \bibinfo {author} {\bibfnamefont
			{K.}~\bibnamefont {Loew}}, \bibinfo {author} {\bibfnamefont {J.}~\bibnamefont
			{Logue}}, \bibinfo {author} {\bibfnamefont {A.~L.}\ \bibnamefont {Lombardi}},
		\bibinfo {author} {\bibfnamefont {M.}~\bibnamefont {Lormand}}, \bibinfo
		{author} {\bibfnamefont {J.}~\bibnamefont {Lough}}, \bibinfo {author}
		{\bibfnamefont {M.}~\bibnamefont {Lubinski}}, \bibinfo {author}
		{\bibfnamefont {H.}~\bibnamefont {L{\"{u}}ck}}, \bibinfo {author}
		{\bibfnamefont {A.~P.}\ \bibnamefont {Lundgren}}, \bibinfo {author}
		{\bibfnamefont {J.}~\bibnamefont {Macarthur}}, \bibinfo {author}
		{\bibfnamefont {E.}~\bibnamefont {Macdonald}}, \bibinfo {author}
		{\bibfnamefont {B.}~\bibnamefont {Machenschalk}}, \bibinfo {author}
		{\bibfnamefont {M.}~\bibnamefont {MacInnis}}, \bibinfo {author}
		{\bibfnamefont {D.~M.}\ \bibnamefont {Macleod}}, \bibinfo {author}
		{\bibfnamefont {F.}~\bibnamefont {Maga{\~{n}}a-Sandoval}}, \bibinfo {author}
		{\bibfnamefont {M.}~\bibnamefont {Mageswaran}}, \bibinfo {author}
		{\bibfnamefont {K.}~\bibnamefont {Mailand}}, \bibinfo {author} {\bibfnamefont
			{G.}~\bibnamefont {Manca}}, \bibinfo {author} {\bibfnamefont
			{I.}~\bibnamefont {Mandel}}, \bibinfo {author} {\bibfnamefont
			{V.}~\bibnamefont {Mandic}}, \bibinfo {author} {\bibfnamefont
			{S.}~\bibnamefont {M{\'{a}}rka}}, \bibinfo {author} {\bibfnamefont
			{Z.}~\bibnamefont {M{\'{a}}rka}}, \bibinfo {author} {\bibfnamefont {A.~S.}\
			\bibnamefont {Markosyan}}, \bibinfo {author} {\bibfnamefont {E.}~\bibnamefont
			{Maros}}, \bibinfo {author} {\bibfnamefont {I.~W.}\ \bibnamefont {Martin}},
		\bibinfo {author} {\bibfnamefont {R.~M.}\ \bibnamefont {Martin}}, \bibinfo
		{author} {\bibfnamefont {D.}~\bibnamefont {Martinov}}, \bibinfo {author}
		{\bibfnamefont {J.~N.}\ \bibnamefont {Marx}}, \bibinfo {author}
		{\bibfnamefont {K.}~\bibnamefont {Mason}}, \bibinfo {author} {\bibfnamefont
			{F.}~\bibnamefont {Matichard}}, \bibinfo {author} {\bibfnamefont
			{L.}~\bibnamefont {Matone}}, \bibinfo {author} {\bibfnamefont {R.~A.}\
			\bibnamefont {Matzner}}, \bibinfo {author} {\bibfnamefont {N.}~\bibnamefont
			{Mavalvala}}, \bibinfo {author} {\bibfnamefont {G.}~\bibnamefont {May}},
		\bibinfo {author} {\bibfnamefont {G.}~\bibnamefont {Mazzolo}}, \bibinfo
		{author} {\bibfnamefont {K.}~\bibnamefont {McAuley}}, \bibinfo {author}
		{\bibfnamefont {R.}~\bibnamefont {McCarthy}}, \bibinfo {author}
		{\bibfnamefont {D.~E.}\ \bibnamefont {McClelland}}, \bibinfo {author}
		{\bibfnamefont {S.~C.}\ \bibnamefont {McGuire}}, \bibinfo {author}
		{\bibfnamefont {G.}~\bibnamefont {McIntyre}}, \bibinfo {author}
		{\bibfnamefont {J.}~\bibnamefont {McIver}}, \bibinfo {author} {\bibfnamefont
			{G.~D.}\ \bibnamefont {Meadors}}, \bibinfo {author} {\bibfnamefont
			{M.}~\bibnamefont {Mehmet}}, \bibinfo {author} {\bibfnamefont
			{T.}~\bibnamefont {Meier}}, \bibinfo {author} {\bibfnamefont
			{A.}~\bibnamefont {Melatos}}, \bibinfo {author} {\bibfnamefont
			{G.}~\bibnamefont {Mendell}}, \bibinfo {author} {\bibfnamefont {R.~A.}\
			\bibnamefont {Mercer}}, \bibinfo {author} {\bibfnamefont {S.}~\bibnamefont
			{Meshkov}}, \bibinfo {author} {\bibfnamefont {C.}~\bibnamefont {Messenger}},
		\bibinfo {author} {\bibfnamefont {M.~S.}\ \bibnamefont {Meyer}}, \bibinfo
		{author} {\bibfnamefont {H.}~\bibnamefont {Miao}}, \bibinfo {author}
		{\bibfnamefont {J.}~\bibnamefont {Miller}}, \bibinfo {author} {\bibfnamefont
			{C.~M.}\ \bibnamefont {Mingarelli}}, \bibinfo {author} {\bibfnamefont
			{S.}~\bibnamefont {Mitra}}, \bibinfo {author} {\bibfnamefont {V.~P.}\
			\bibnamefont {Mitrofanov}}, \bibinfo {author} {\bibfnamefont
			{G.}~\bibnamefont {Mitselmakher}}, \bibinfo {author} {\bibfnamefont
			{R.}~\bibnamefont {Mittleman}}, \bibinfo {author} {\bibfnamefont
			{B.}~\bibnamefont {Moe}}, \bibinfo {author} {\bibfnamefont {F.}~\bibnamefont
			{Mokler}}, \bibinfo {author} {\bibfnamefont {S.~R.}\ \bibnamefont
			{Mohapatra}}, \bibinfo {author} {\bibfnamefont {D.}~\bibnamefont {Moraru}},
		\bibinfo {author} {\bibfnamefont {G.}~\bibnamefont {Moreno}}, \bibinfo
		{author} {\bibfnamefont {T.}~\bibnamefont {Mori}}, \bibinfo {author}
		{\bibfnamefont {S.~R.}\ \bibnamefont {Morriss}}, \bibinfo {author}
		{\bibfnamefont {K.}~\bibnamefont {Mossavi}}, \bibinfo {author} {\bibfnamefont
			{C.~M.}\ \bibnamefont {Mow-Lowry}}, \bibinfo {author} {\bibfnamefont {C.~L.}\
			\bibnamefont {Mueller}}, \bibinfo {author} {\bibfnamefont {G.}~\bibnamefont
			{Mueller}}, \bibinfo {author} {\bibfnamefont {S.}~\bibnamefont {Mukherjee}},
		\bibinfo {author} {\bibfnamefont {A.}~\bibnamefont {Mullavey}}, \bibinfo
		{author} {\bibfnamefont {J.}~\bibnamefont {Munch}}, \bibinfo {author}
		{\bibfnamefont {D.}~\bibnamefont {Murphy}}, \bibinfo {author} {\bibfnamefont
			{P.~G.}\ \bibnamefont {Murray}}, \bibinfo {author} {\bibfnamefont
			{A.}~\bibnamefont {Mytidis}}, \bibinfo {author} {\bibfnamefont {D.~N.}\
			\bibnamefont {Kumar}}, \bibinfo {author} {\bibfnamefont {T.}~\bibnamefont
			{Nash}}, \bibinfo {author} {\bibfnamefont {R.}~\bibnamefont {Nayak}},
		\bibinfo {author} {\bibfnamefont {V.}~\bibnamefont {Necula}}, \bibinfo
		{author} {\bibfnamefont {G.}~\bibnamefont {Newton}}, \bibinfo {author}
		{\bibfnamefont {T.}~\bibnamefont {Nguyen}}, \bibinfo {author} {\bibfnamefont
			{E.}~\bibnamefont {Nishida}}, \bibinfo {author} {\bibfnamefont
			{A.}~\bibnamefont {Nishizawa}}, \bibinfo {author} {\bibfnamefont
			{A.}~\bibnamefont {Nitz}}, \bibinfo {author} {\bibfnamefont {D.}~\bibnamefont
			{Nolting}}, \bibinfo {author} {\bibfnamefont {M.~E.}\ \bibnamefont
			{Normandin}}, \bibinfo {author} {\bibfnamefont {L.~K.}\ \bibnamefont
			{Nuttall}}, \bibinfo {author} {\bibfnamefont {J.}~\bibnamefont {O'Dell}},
		\bibinfo {author} {\bibfnamefont {B.}~\bibnamefont {O'Reilly}}, \bibinfo
		{author} {\bibfnamefont {R.}~\bibnamefont {O'Shaughnessy}}, \bibinfo {author}
		{\bibfnamefont {E.}~\bibnamefont {Ochsner}}, \bibinfo {author} {\bibfnamefont
			{E.}~\bibnamefont {Oelker}}, \bibinfo {author} {\bibfnamefont {G.~H.}\
			\bibnamefont {Ogin}}, \bibinfo {author} {\bibfnamefont {J.~J.}\ \bibnamefont
			{Oh}}, \bibinfo {author} {\bibfnamefont {S.~H.}\ \bibnamefont {Oh}}, \bibinfo
		{author} {\bibfnamefont {F.}~\bibnamefont {Ohme}}, \bibinfo {author}
		{\bibfnamefont {P.}~\bibnamefont {Oppermann}}, \bibinfo {author}
		{\bibfnamefont {C.}~\bibnamefont {Osthelder}}, \bibinfo {author}
		{\bibfnamefont {C.~D.}\ \bibnamefont {Ott}}, \bibinfo {author} {\bibfnamefont
			{D.~J.}\ \bibnamefont {Ottaway}}, \bibinfo {author} {\bibfnamefont {R.~S.}\
			\bibnamefont {Ottens}}, \bibinfo {author} {\bibfnamefont {J.}~\bibnamefont
			{Ou}}, \bibinfo {author} {\bibfnamefont {H.}~\bibnamefont {Overmier}},
		\bibinfo {author} {\bibfnamefont {B.~J.}\ \bibnamefont {Owen}}, \bibinfo
		{author} {\bibfnamefont {C.}~\bibnamefont {Padilla}}, \bibinfo {author}
		{\bibfnamefont {A.}~\bibnamefont {Pai}}, \bibinfo {author} {\bibfnamefont
			{Y.}~\bibnamefont {Pan}}, \bibinfo {author} {\bibfnamefont {C.}~\bibnamefont
			{Pankow}}, \bibinfo {author} {\bibfnamefont {M.~A.}\ \bibnamefont {Papa}},
		\bibinfo {author} {\bibfnamefont {H.}~\bibnamefont {Paris}}, \bibinfo
		{author} {\bibfnamefont {W.}~\bibnamefont {Parkinson}}, \bibinfo {author}
		{\bibfnamefont {M.}~\bibnamefont {Pedraza}}, \bibinfo {author} {\bibfnamefont
			{S.}~\bibnamefont {Penn}}, \bibinfo {author} {\bibfnamefont {C.}~\bibnamefont
			{Peralta}}, \bibinfo {author} {\bibfnamefont {A.}~\bibnamefont {Perreca}},
		\bibinfo {author} {\bibfnamefont {M.}~\bibnamefont {Phelps}}, \bibinfo
		{author} {\bibfnamefont {M.}~\bibnamefont {Pickenpack}}, \bibinfo {author}
		{\bibfnamefont {V.}~\bibnamefont {Pierro}}, \bibinfo {author} {\bibfnamefont
			{I.~M.}\ \bibnamefont {Pinto}}, \bibinfo {author} {\bibfnamefont
			{M.}~\bibnamefont {Pitkin}}, \bibinfo {author} {\bibfnamefont {H.~J.}\
			\bibnamefont {Pletsch}}, \bibinfo {author} {\bibfnamefont {J.}~\bibnamefont
			{P{\"{o}}ld}}, \bibinfo {author} {\bibfnamefont {F.}~\bibnamefont
			{Postiglione}}, \bibinfo {author} {\bibfnamefont {C.}~\bibnamefont {Poux}},
		\bibinfo {author} {\bibfnamefont {V.}~\bibnamefont {Predoi}}, \bibinfo
		{author} {\bibfnamefont {T.}~\bibnamefont {Prestegard}}, \bibinfo {author}
		{\bibfnamefont {L.~R.}\ \bibnamefont {Price}}, \bibinfo {author}
		{\bibfnamefont {M.}~\bibnamefont {Prijatelj}}, \bibinfo {author}
		{\bibfnamefont {S.}~\bibnamefont {Privitera}}, \bibinfo {author}
		{\bibfnamefont {L.~G.}\ \bibnamefont {Prokhorov}}, \bibinfo {author}
		{\bibfnamefont {O.}~\bibnamefont {Puncken}}, \bibinfo {author} {\bibfnamefont
			{V.}~\bibnamefont {Quetschke}}, \bibinfo {author} {\bibfnamefont
			{E.}~\bibnamefont {Quintero}}, \bibinfo {author} {\bibfnamefont
			{R.}~\bibnamefont {Quitzow-James}}, \bibinfo {author} {\bibfnamefont {F.~J.}\
			\bibnamefont {Raab}}, \bibinfo {author} {\bibfnamefont {H.}~\bibnamefont
			{Radkins}}, \bibinfo {author} {\bibfnamefont {P.}~\bibnamefont {Raffai}},
		\bibinfo {author} {\bibfnamefont {S.}~\bibnamefont {Raja}}, \bibinfo {author}
		{\bibfnamefont {M.}~\bibnamefont {Rakhmanov}}, \bibinfo {author}
		{\bibfnamefont {C.}~\bibnamefont {Ramet}}, \bibinfo {author} {\bibfnamefont
			{V.}~\bibnamefont {Raymond}}, \bibinfo {author} {\bibfnamefont {C.~M.}\
			\bibnamefont {Reed}}, \bibinfo {author} {\bibfnamefont {T.}~\bibnamefont
			{Reed}}, \bibinfo {author} {\bibfnamefont {S.}~\bibnamefont {Reid}}, \bibinfo
		{author} {\bibfnamefont {D.~H.}\ \bibnamefont {Reitze}}, \bibinfo {author}
		{\bibfnamefont {R.}~\bibnamefont {Riesen}}, \bibinfo {author} {\bibfnamefont
			{K.}~\bibnamefont {Riles}}, \bibinfo {author} {\bibfnamefont
			{M.}~\bibnamefont {Roberts}}, \bibinfo {author} {\bibfnamefont {N.~A.}\
			\bibnamefont {Robertson}}, \bibinfo {author} {\bibfnamefont {E.~L.}\
			\bibnamefont {Robinson}}, \bibinfo {author} {\bibfnamefont {S.}~\bibnamefont
			{Roddy}}, \bibinfo {author} {\bibfnamefont {C.}~\bibnamefont {Rodriguez}},
		\bibinfo {author} {\bibfnamefont {L.}~\bibnamefont {Rodriguez}}, \bibinfo
		{author} {\bibfnamefont {M.}~\bibnamefont {Rodruck}}, \bibinfo {author}
		{\bibfnamefont {J.~G.}\ \bibnamefont {Rollins}}, \bibinfo {author}
		{\bibfnamefont {J.~H.}\ \bibnamefont {Romie}}, \bibinfo {author}
		{\bibfnamefont {C.}~\bibnamefont {R{\"{o}}ver}}, \bibinfo {author}
		{\bibfnamefont {S.}~\bibnamefont {Rowan}}, \bibinfo {author} {\bibfnamefont
			{A.}~\bibnamefont {R{\"{u}}diger}}, \bibinfo {author} {\bibfnamefont
			{K.}~\bibnamefont {Ryan}}, \bibinfo {author} {\bibfnamefont {F.}~\bibnamefont
			{Salemi}}, \bibinfo {author} {\bibfnamefont {L.}~\bibnamefont {Sammut}},
		\bibinfo {author} {\bibfnamefont {V.}~\bibnamefont {Sandberg}}, \bibinfo
		{author} {\bibfnamefont {J.}~\bibnamefont {Sanders}}, \bibinfo {author}
		{\bibfnamefont {S.}~\bibnamefont {Sankar}}, \bibinfo {author} {\bibfnamefont
			{V.}~\bibnamefont {Sannibale}}, \bibinfo {author} {\bibfnamefont
			{L.}~\bibnamefont {Santamar{\'{i}}a}}, \bibinfo {author} {\bibfnamefont
			{I.}~\bibnamefont {Santiago-Prieto}}, \bibinfo {author} {\bibfnamefont
			{G.}~\bibnamefont {Santostasi}}, \bibinfo {author} {\bibfnamefont {B.~S.}\
			\bibnamefont {Sathyaprakash}}, \bibinfo {author} {\bibfnamefont {P.~R.}\
			\bibnamefont {Saulson}}, \bibinfo {author} {\bibfnamefont {R.~L.}\
			\bibnamefont {Savage}}, \bibinfo {author} {\bibfnamefont {R.}~\bibnamefont
			{Schilling}}, \bibinfo {author} {\bibfnamefont {R.}~\bibnamefont {Schnabel}},
		\bibinfo {author} {\bibfnamefont {R.~M.}\ \bibnamefont {Schofield}}, \bibinfo
		{author} {\bibfnamefont {D.}~\bibnamefont {Schuette}}, \bibinfo {author}
		{\bibfnamefont {B.}~\bibnamefont {Schulz}}, \bibinfo {author} {\bibfnamefont
			{B.~F.}\ \bibnamefont {Schutz}}, \bibinfo {author} {\bibfnamefont
			{P.}~\bibnamefont {Schwinberg}}, \bibinfo {author} {\bibfnamefont
			{J.}~\bibnamefont {Scott}}, \bibinfo {author} {\bibfnamefont {S.~M.}\
			\bibnamefont {Scott}}, \bibinfo {author} {\bibfnamefont {F.}~\bibnamefont
			{Seifert}}, \bibinfo {author} {\bibfnamefont {D.}~\bibnamefont {Sellers}},
		\bibinfo {author} {\bibfnamefont {A.~S.}\ \bibnamefont {Sengupta}}, \bibinfo
		{author} {\bibfnamefont {A.}~\bibnamefont {Sergeev}}, \bibinfo {author}
		{\bibfnamefont {D.~A.}\ \bibnamefont {Shaddock}}, \bibinfo {author}
		{\bibfnamefont {M.~S.}\ \bibnamefont {Shahriar}}, \bibinfo {author}
		{\bibfnamefont {M.}~\bibnamefont {Shaltev}}, \bibinfo {author} {\bibfnamefont
			{Z.}~\bibnamefont {Shao}}, \bibinfo {author} {\bibfnamefont {B.}~\bibnamefont
			{Shapiro}}, \bibinfo {author} {\bibfnamefont {P.}~\bibnamefont {Shawhan}},
		\bibinfo {author} {\bibfnamefont {D.~H.}\ \bibnamefont {Shoemaker}}, \bibinfo
		{author} {\bibfnamefont {T.~L.}\ \bibnamefont {Sidery}}, \bibinfo {author}
		{\bibfnamefont {X.}~\bibnamefont {Siemens}}, \bibinfo {author} {\bibfnamefont
			{D.}~\bibnamefont {Sigg}}, \bibinfo {author} {\bibfnamefont {D.}~\bibnamefont
			{Simakov}}, \bibinfo {author} {\bibfnamefont {A.}~\bibnamefont {Singer}},
		\bibinfo {author} {\bibfnamefont {L.}~\bibnamefont {Singer}}, \bibinfo
		{author} {\bibfnamefont {A.~M.}\ \bibnamefont {Sintes}}, \bibinfo {author}
		{\bibfnamefont {G.~R.}\ \bibnamefont {Skelton}}, \bibinfo {author}
		{\bibfnamefont {B.~J.}\ \bibnamefont {Slagmolen}}, \bibinfo {author}
		{\bibfnamefont {J.}~\bibnamefont {Slutsky}}, \bibinfo {author} {\bibfnamefont
			{J.~R.}\ \bibnamefont {Smith}}, \bibinfo {author} {\bibfnamefont {M.~R.}\
			\bibnamefont {Smith}}, \bibinfo {author} {\bibfnamefont {R.~J.}\ \bibnamefont
			{Smith}}, \bibinfo {author} {\bibfnamefont {N.~D.}\ \bibnamefont
			{Smith-Lefebvre}}, \bibinfo {author} {\bibfnamefont {E.~J.}\ \bibnamefont
			{Son}}, \bibinfo {author} {\bibfnamefont {B.}~\bibnamefont {Sorazu}},
		\bibinfo {author} {\bibfnamefont {T.}~\bibnamefont {Souradeep}}, \bibinfo
		{author} {\bibfnamefont {M.}~\bibnamefont {Stefszky}}, \bibinfo {author}
		{\bibfnamefont {E.}~\bibnamefont {Steinert}}, \bibinfo {author}
		{\bibfnamefont {J.}~\bibnamefont {Steinlechner}}, \bibinfo {author}
		{\bibfnamefont {S.}~\bibnamefont {Steinlechner}}, \bibinfo {author}
		{\bibfnamefont {S.}~\bibnamefont {Steplewski}}, \bibinfo {author}
		{\bibfnamefont {D.}~\bibnamefont {Stevens}}, \bibinfo {author} {\bibfnamefont
			{A.}~\bibnamefont {Stochino}}, \bibinfo {author} {\bibfnamefont
			{R.}~\bibnamefont {Stone}}, \bibinfo {author} {\bibfnamefont {K.~A.}\
			\bibnamefont {Strain}}, \bibinfo {author} {\bibfnamefont {S.~E.}\
			\bibnamefont {Strigin}}, \bibinfo {author} {\bibfnamefont {A.~S.}\
			\bibnamefont {Stroeer}}, \bibinfo {author} {\bibfnamefont {A.~L.}\
			\bibnamefont {Stuver}}, \bibinfo {author} {\bibfnamefont {T.~Z.}\
			\bibnamefont {Summerscales}}, \bibinfo {author} {\bibfnamefont
			{S.}~\bibnamefont {Susmithan}}, \bibinfo {author} {\bibfnamefont {P.~J.}\
			\bibnamefont {Sutton}}, \bibinfo {author} {\bibfnamefont {G.}~\bibnamefont
			{Szeifert}}, \bibinfo {author} {\bibfnamefont {D.}~\bibnamefont {Talukder}},
		\bibinfo {author} {\bibfnamefont {D.~B.}\ \bibnamefont {Tanner}}, \bibinfo
		{author} {\bibfnamefont {S.~P.}\ \bibnamefont {Tarabrin}}, \bibinfo {author}
		{\bibfnamefont {R.}~\bibnamefont {Taylor}}, \bibinfo {author} {\bibfnamefont
			{M.}~\bibnamefont {Thomas}}, \bibinfo {author} {\bibfnamefont
			{P.}~\bibnamefont {Thomas}}, \bibinfo {author} {\bibfnamefont {K.~A.}\
			\bibnamefont {Thorne}}, \bibinfo {author} {\bibfnamefont {K.~S.}\
			\bibnamefont {Thorne}}, \bibinfo {author} {\bibfnamefont {E.}~\bibnamefont
			{Thrane}}, \bibinfo {author} {\bibfnamefont {V.}~\bibnamefont {Tiwari}},
		\bibinfo {author} {\bibfnamefont {K.~V.}\ \bibnamefont {Tokmakov}}, \bibinfo
		{author} {\bibfnamefont {C.}~\bibnamefont {Tomlinson}}, \bibinfo {author}
		{\bibfnamefont {C.~V.}\ \bibnamefont {Torres}}, \bibinfo {author}
		{\bibfnamefont {C.~I.}\ \bibnamefont {Torrie}}, \bibinfo {author}
		{\bibfnamefont {G.}~\bibnamefont {Traylor}}, \bibinfo {author} {\bibfnamefont
			{M.}~\bibnamefont {Tse}}, \bibinfo {author} {\bibfnamefont {D.}~\bibnamefont
			{Ugolini}}, \bibinfo {author} {\bibfnamefont {C.~S.}\ \bibnamefont
			{Unnikrishnan}}, \bibinfo {author} {\bibfnamefont {H.}~\bibnamefont
			{Vahlbruch}}, \bibinfo {author} {\bibfnamefont {M.}~\bibnamefont
			{Vallisneri}}, \bibinfo {author} {\bibfnamefont {M.~V.}\ \bibnamefont {{Van
					Der Sluys}}}, \bibinfo {author} {\bibfnamefont {A.~A.}\ \bibnamefont {{Van
					Veggel}}}, \bibinfo {author} {\bibfnamefont {S.}~\bibnamefont {Vass}},
		\bibinfo {author} {\bibfnamefont {R.}~\bibnamefont {Vaulin}}, \bibinfo
		{author} {\bibfnamefont {A.}~\bibnamefont {Vecchio}}, \bibinfo {author}
		{\bibfnamefont {P.~J.}\ \bibnamefont {Veitch}}, \bibinfo {author}
		{\bibfnamefont {J.}~\bibnamefont {Veitch}}, \bibinfo {author} {\bibfnamefont
			{K.}~\bibnamefont {Venkateswara}}, \bibinfo {author} {\bibfnamefont
			{S.}~\bibnamefont {Verma}}, \bibinfo {author} {\bibfnamefont
			{R.}~\bibnamefont {Vincent-Finley}}, \bibinfo {author} {\bibfnamefont
			{S.}~\bibnamefont {Vitale}}, \bibinfo {author} {\bibfnamefont
			{T.}~\bibnamefont {Vo}}, \bibinfo {author} {\bibfnamefont {C.}~\bibnamefont
			{Vorvick}}, \bibinfo {author} {\bibfnamefont {W.~D.}\ \bibnamefont
			{Vousden}}, \bibinfo {author} {\bibfnamefont {S.~P.}\ \bibnamefont
			{Vyatchanin}}, \bibinfo {author} {\bibfnamefont {A.}~\bibnamefont {Wade}},
		\bibinfo {author} {\bibfnamefont {L.}~\bibnamefont {Wade}}, \bibinfo {author}
		{\bibfnamefont {M.}~\bibnamefont {Wade}}, \bibinfo {author} {\bibfnamefont
			{S.~J.}\ \bibnamefont {Waldman}}, \bibinfo {author} {\bibfnamefont
			{L.}~\bibnamefont {Wallace}}, \bibinfo {author} {\bibfnamefont
			{Y.}~\bibnamefont {Wan}}, \bibinfo {author} {\bibfnamefont {M.}~\bibnamefont
			{Wang}}, \bibinfo {author} {\bibfnamefont {J.}~\bibnamefont {Wang}}, \bibinfo
		{author} {\bibfnamefont {X.}~\bibnamefont {Wang}}, \bibinfo {author}
		{\bibfnamefont {A.}~\bibnamefont {Wanner}}, \bibinfo {author} {\bibfnamefont
			{R.~L.}\ \bibnamefont {Ward}}, \bibinfo {author} {\bibfnamefont
			{M.}~\bibnamefont {Was}}, \bibinfo {author} {\bibfnamefont {M.}~\bibnamefont
			{Weinert}}, \bibinfo {author} {\bibfnamefont {A.~J.}\ \bibnamefont
			{Weinstein}}, \bibinfo {author} {\bibfnamefont {R.}~\bibnamefont {Weiss}},
		\bibinfo {author} {\bibfnamefont {T.}~\bibnamefont {Welborn}}, \bibinfo
		{author} {\bibfnamefont {L.}~\bibnamefont {Wen}}, \bibinfo {author}
		{\bibfnamefont {P.}~\bibnamefont {Wessels}}, \bibinfo {author} {\bibfnamefont
			{M.}~\bibnamefont {West}}, \bibinfo {author} {\bibfnamefont {T.}~\bibnamefont
			{Westphal}}, \bibinfo {author} {\bibfnamefont {K.}~\bibnamefont {Wette}},
		\bibinfo {author} {\bibfnamefont {J.~T.}\ \bibnamefont {Whelan}}, \bibinfo
		{author} {\bibfnamefont {S.~E.}\ \bibnamefont {Whitcomb}}, \bibinfo {author}
		{\bibfnamefont {A.~G.}\ \bibnamefont {Wiseman}}, \bibinfo {author}
		{\bibfnamefont {D.~J.}\ \bibnamefont {White}}, \bibinfo {author}
		{\bibfnamefont {B.~F.}\ \bibnamefont {Whiting}}, \bibinfo {author}
		{\bibfnamefont {K.}~\bibnamefont {Wiesner}}, \bibinfo {author} {\bibfnamefont
			{C.}~\bibnamefont {Wilkinson}}, \bibinfo {author} {\bibfnamefont {P.~A.}\
			\bibnamefont {Willems}}, \bibinfo {author} {\bibfnamefont {L.}~\bibnamefont
			{Williams}}, \bibinfo {author} {\bibfnamefont {R.}~\bibnamefont {Williams}},
		\bibinfo {author} {\bibfnamefont {T.}~\bibnamefont {Williams}}, \bibinfo
		{author} {\bibfnamefont {J.~L.}\ \bibnamefont {Willis}}, \bibinfo {author}
		{\bibfnamefont {B.}~\bibnamefont {Willke}}, \bibinfo {author} {\bibfnamefont
			{M.}~\bibnamefont {Wimmer}}, \bibinfo {author} {\bibfnamefont
			{L.}~\bibnamefont {Winkelmann}}, \bibinfo {author} {\bibfnamefont
			{W.}~\bibnamefont {Winkler}}, \bibinfo {author} {\bibfnamefont {C.~C.}\
			\bibnamefont {Wipf}}, \bibinfo {author} {\bibfnamefont {H.}~\bibnamefont
			{Wittel}}, \bibinfo {author} {\bibfnamefont {G.}~\bibnamefont {Woan}},
		\bibinfo {author} {\bibfnamefont {R.}~\bibnamefont {Wooley}}, \bibinfo
		{author} {\bibfnamefont {J.}~\bibnamefont {Worden}}, \bibinfo {author}
		{\bibfnamefont {J.}~\bibnamefont {Yablon}}, \bibinfo {author} {\bibfnamefont
			{I.}~\bibnamefont {Yakushin}}, \bibinfo {author} {\bibfnamefont
			{H.}~\bibnamefont {Yamamoto}}, \bibinfo {author} {\bibfnamefont {C.~C.}\
			\bibnamefont {Yancey}}, \bibinfo {author} {\bibfnamefont {H.}~\bibnamefont
			{Yang}}, \bibinfo {author} {\bibfnamefont {D.}~\bibnamefont {Yeaton-Massey}},
		\bibinfo {author} {\bibfnamefont {S.}~\bibnamefont {Yoshida}}, \bibinfo
		{author} {\bibfnamefont {H.}~\bibnamefont {Yum}}, \bibinfo {author}
		{\bibfnamefont {M.}~\bibnamefont {Zanolin}}, \bibinfo {author} {\bibfnamefont
			{F.}~\bibnamefont {Zhang}}, \bibinfo {author} {\bibfnamefont
			{L.}~\bibnamefont {Zhang}}, \bibinfo {author} {\bibfnamefont
			{C.}~\bibnamefont {Zhao}}, \bibinfo {author} {\bibfnamefont {H.}~\bibnamefont
			{Zhu}}, \bibinfo {author} {\bibfnamefont {X.~J.}\ \bibnamefont {Zhu}},
		\bibinfo {author} {\bibfnamefont {N.}~\bibnamefont {Zotov}}, \bibinfo
		{author} {\bibfnamefont {M.~E.}\ \bibnamefont {Zucker}},\ and\ \bibinfo
		{author} {\bibfnamefont {J.}~\bibnamefont {Zweizig}},\ }\bibfield  {title}
	{\bibinfo {title} {{Enhanced sensitivity of the LIGO gravitational wave
				detector by using squeezed states of light}},\ }\href
	{https://doi.org/10.1038/nphoton.2013.177} {\bibfield  {journal} {\bibinfo
			{journal} {Nature Photonics}\ }\textbf {\bibinfo {volume} {7}},\ \bibinfo
		{pages} {613} (\bibinfo {year} {2013})},\ \Eprint
	{https://arxiv.org/abs/1310.0383} {arXiv:1310.0383} \BibitemShut {NoStop}%
	\bibitem [{\citenamefont {Hosten}\ \emph
		{et~al.}(2016{\natexlab{a}})\citenamefont {Hosten}, \citenamefont {Engelsen},
		\citenamefont {Krishnakumar},\ and\ \citenamefont {Kasevich}}]{Hosten2016}%
	\BibitemOpen
	\bibfield  {author} {\bibinfo {author} {\bibfnamefont {O.}~\bibnamefont
			{Hosten}}, \bibinfo {author} {\bibfnamefont {N.~J.}\ \bibnamefont
			{Engelsen}}, \bibinfo {author} {\bibfnamefont {R.}~\bibnamefont
			{Krishnakumar}},\ and\ \bibinfo {author} {\bibfnamefont {M.~A.}\ \bibnamefont
			{Kasevich}},\ }\bibfield  {title} {\bibinfo {title} {Measurement noise 100
			times lower than the quantum-projection limit using entangled atoms},\ }\href
	{https://doi.org/10.1038/nature16176} {\bibfield  {journal} {\bibinfo
			{journal} {Nature}\ }\textbf {\bibinfo {volume} {529}},\ \bibinfo {pages}
		{505} (\bibinfo {year} {2016}{\natexlab{a}})}\BibitemShut {NoStop}%
	\bibitem [{\citenamefont {Malnou}\ \emph
		{et~al.}(2019{\natexlab{a}})\citenamefont {Malnou}, \citenamefont {Palken},
		\citenamefont {Brubaker}, \citenamefont {Vale}, \citenamefont {Hilton},\ and\
		\citenamefont {Lehnert}}]{malnou2019squeezed}%
	\BibitemOpen
	\bibfield  {author} {\bibinfo {author} {\bibfnamefont {M.}~\bibnamefont
			{Malnou}}, \bibinfo {author} {\bibfnamefont {D.~A.}\ \bibnamefont {Palken}},
		\bibinfo {author} {\bibfnamefont {B.~M.}\ \bibnamefont {Brubaker}}, \bibinfo
		{author} {\bibfnamefont {L.~R.}\ \bibnamefont {Vale}}, \bibinfo {author}
		{\bibfnamefont {G.~C.}\ \bibnamefont {Hilton}},\ and\ \bibinfo {author}
		{\bibfnamefont {K.~W.}\ \bibnamefont {Lehnert}},\ }\bibfield  {title}
	{\bibinfo {title} {{Squeezed Vacuum Used to Accelerate the Search for a Weak
				Classical Signal}},\ }\href {https://doi.org/10.1103/PhysRevX.9.021023}
	{\bibfield  {journal} {\bibinfo  {journal} {Physical Review X}\ }\textbf
		{\bibinfo {volume} {9}},\ \bibinfo {pages} {21023} (\bibinfo {year}
		{2019}{\natexlab{a}})},\ \Eprint {https://arxiv.org/abs/1809.06470}
	{arXiv:1809.06470} \BibitemShut {NoStop}%
	\bibitem [{\citenamefont {Backes}\ \emph
		{et~al.}(2021{\natexlab{a}})\citenamefont {Backes}, \citenamefont {Palken},
		\citenamefont {Kenany}, \citenamefont {Brubaker}, \citenamefont {Cahn},
		\citenamefont {Droster}, \citenamefont {Hilton}, \citenamefont {Ghosh},
		\citenamefont {Jackson}, \citenamefont {Lamoreaux}, \citenamefont {Leder},
		\citenamefont {Lehnert}, \citenamefont {Lewis}, \citenamefont {Malnou},
		\citenamefont {Maruyama}, \citenamefont {Rapidis}, \citenamefont
		{Simanovskaia}, \citenamefont {Singh}, \citenamefont {Speller}, \citenamefont
		{Urdinaran}, \citenamefont {Vale}, \citenamefont {van Assendelft},
		\citenamefont {van Bibber},\ and\ \citenamefont {Wang}}]{backes2021a}%
	\BibitemOpen
	\bibfield  {author} {\bibinfo {author} {\bibfnamefont {K.~M.}\ \bibnamefont
			{Backes}}, \bibinfo {author} {\bibfnamefont {D.~A.}\ \bibnamefont {Palken}},
		\bibinfo {author} {\bibfnamefont {S.~A.}\ \bibnamefont {Kenany}}, \bibinfo
		{author} {\bibfnamefont {B.~M.}\ \bibnamefont {Brubaker}}, \bibinfo {author}
		{\bibfnamefont {S.~B.}\ \bibnamefont {Cahn}}, \bibinfo {author}
		{\bibfnamefont {A.}~\bibnamefont {Droster}}, \bibinfo {author} {\bibfnamefont
			{G.~C.}\ \bibnamefont {Hilton}}, \bibinfo {author} {\bibfnamefont
			{S.}~\bibnamefont {Ghosh}}, \bibinfo {author} {\bibfnamefont
			{H.}~\bibnamefont {Jackson}}, \bibinfo {author} {\bibfnamefont {S.~K.}\
			\bibnamefont {Lamoreaux}}, \bibinfo {author} {\bibfnamefont {A.~F.}\
			\bibnamefont {Leder}}, \bibinfo {author} {\bibfnamefont {K.~W.}\ \bibnamefont
			{Lehnert}}, \bibinfo {author} {\bibfnamefont {S.~M.}\ \bibnamefont {Lewis}},
		\bibinfo {author} {\bibfnamefont {M.}~\bibnamefont {Malnou}}, \bibinfo
		{author} {\bibfnamefont {R.~H.}\ \bibnamefont {Maruyama}}, \bibinfo {author}
		{\bibfnamefont {N.~M.}\ \bibnamefont {Rapidis}}, \bibinfo {author}
		{\bibfnamefont {M.}~\bibnamefont {Simanovskaia}}, \bibinfo {author}
		{\bibfnamefont {S.}~\bibnamefont {Singh}}, \bibinfo {author} {\bibfnamefont
			{D.~H.}\ \bibnamefont {Speller}}, \bibinfo {author} {\bibfnamefont
			{I.}~\bibnamefont {Urdinaran}}, \bibinfo {author} {\bibfnamefont {L.~R.}\
			\bibnamefont {Vale}}, \bibinfo {author} {\bibfnamefont {E.~C.}\ \bibnamefont
			{van Assendelft}}, \bibinfo {author} {\bibfnamefont {K.}~\bibnamefont {van
				Bibber}},\ and\ \bibinfo {author} {\bibfnamefont {H.}~\bibnamefont {Wang}},\
	}\bibfield  {title} {\bibinfo {title} {{A quantum enhanced search for dark
				matter axions}},\ }\href {https://doi.org/10.1038/s41586-021-03226-7}
	{\bibfield  {journal} {\bibinfo  {journal} {Nature}\ }\textbf {\bibinfo
			{volume} {590}},\ \bibinfo {pages} {238} (\bibinfo {year}
		{2021}{\natexlab{a}})},\ \Eprint {https://arxiv.org/abs/2008.01853}
	{arXiv:2008.01853} \BibitemShut {NoStop}%
	\bibitem [{\citenamefont {Vasilakis}\ \emph
		{et~al.}(2015{\natexlab{a}})\citenamefont {Vasilakis}, \citenamefont {Shen},
		\citenamefont {Jensen}, \citenamefont {Balabas}, \citenamefont {Salart},
		\citenamefont {Chen},\ and\ \citenamefont
		{Polzik}}]{vasilakis2015generation}%
	\BibitemOpen
	\bibfield  {author} {\bibinfo {author} {\bibfnamefont {G.}~\bibnamefont
			{Vasilakis}}, \bibinfo {author} {\bibfnamefont {H.}~\bibnamefont {Shen}},
		\bibinfo {author} {\bibfnamefont {K.}~\bibnamefont {Jensen}}, \bibinfo
		{author} {\bibfnamefont {M.}~\bibnamefont {Balabas}}, \bibinfo {author}
		{\bibfnamefont {D.}~\bibnamefont {Salart}}, \bibinfo {author} {\bibfnamefont
			{B.}~\bibnamefont {Chen}},\ and\ \bibinfo {author} {\bibfnamefont {E.~S.}\
			\bibnamefont {Polzik}},\ }\bibfield  {title} {\bibinfo {title} {{Generation
				of a squeezed state of an oscillator by stroboscopic back-action-evading
				measurement}},\ }\href {https://doi.org/10.1038/nphys3280} {\bibfield
		{journal} {\bibinfo  {journal} {Nature Physics}\ }\textbf {\bibinfo {volume}
			{11}},\ \bibinfo {pages} {389} (\bibinfo {year}
		{2015}{\natexlab{a}})}\BibitemShut {NoStop}%
	\bibitem [{\citenamefont {Bienfait}\ \emph {et~al.}(2017)\citenamefont
		{Bienfait}, \citenamefont {{Campagne-Ibarcq}}, \citenamefont {Kiilerich},
		\citenamefont {Zhou}, \citenamefont {Probst}, \citenamefont {Pla},
		\citenamefont {Schenkel}, \citenamefont {Vion}, \citenamefont {Esteve},
		\citenamefont {Morton}, \citenamefont {Moelmer},\ and\ \citenamefont
		{Bertet}}]{Bienfait2017b}%
	\BibitemOpen
	\bibfield  {author} {\bibinfo {author} {\bibfnamefont {A.}~\bibnamefont
			{Bienfait}}, \bibinfo {author} {\bibfnamefont {P.}~\bibnamefont
			{{Campagne-Ibarcq}}}, \bibinfo {author} {\bibfnamefont {A.~H.}\ \bibnamefont
			{Kiilerich}}, \bibinfo {author} {\bibfnamefont {X.}~\bibnamefont {Zhou}},
		\bibinfo {author} {\bibfnamefont {S.}~\bibnamefont {Probst}}, \bibinfo
		{author} {\bibfnamefont {J.~J.}\ \bibnamefont {Pla}}, \bibinfo {author}
		{\bibfnamefont {T.}~\bibnamefont {Schenkel}}, \bibinfo {author}
		{\bibfnamefont {D.}~\bibnamefont {Vion}}, \bibinfo {author} {\bibfnamefont
			{D.}~\bibnamefont {Esteve}}, \bibinfo {author} {\bibfnamefont {J.~J.}\
			\bibnamefont {Morton}}, \bibinfo {author} {\bibfnamefont {K.}~\bibnamefont
			{Moelmer}},\ and\ \bibinfo {author} {\bibfnamefont {P.}~\bibnamefont
			{Bertet}},\ }\bibfield  {title} {\bibinfo {title} {Magnetic resonance with
			squeezed microwaves},\ }\href {https://doi.org/10.1103/PhysRevX.7.041011}
	{\bibfield  {journal} {\bibinfo  {journal} {Physical Review X}\ }\textbf
		{\bibinfo {volume} {7}},\ \bibinfo {pages} {041011} (\bibinfo {year}
		{2017})},\ \Eprint {https://arxiv.org/abs/1610.03329} {arXiv:1610.03329}
	\BibitemShut {NoStop}%
	\bibitem [{\citenamefont {{Pedrozo-Pe{\~n}afiel}}\ \emph
		{et~al.}(2020)\citenamefont {{Pedrozo-Pe{\~n}afiel}}, \citenamefont
		{Colombo}, \citenamefont {Shu}, \citenamefont {Adiyatullin}, \citenamefont
		{Li}, \citenamefont {Mendez}, \citenamefont {Braverman}, \citenamefont
		{Kawasaki}, \citenamefont {Akamatsu}, \citenamefont {Xiao},\ and\
		\citenamefont {Vuleti{\'c}}}]{Pedrozo-Penafiel2020}%
	\BibitemOpen
	\bibfield  {author} {\bibinfo {author} {\bibfnamefont {E.}~\bibnamefont
			{{Pedrozo-Pe{\~n}afiel}}}, \bibinfo {author} {\bibfnamefont {S.}~\bibnamefont
			{Colombo}}, \bibinfo {author} {\bibfnamefont {C.}~\bibnamefont {Shu}},
		\bibinfo {author} {\bibfnamefont {A.~F.}\ \bibnamefont {Adiyatullin}},
		\bibinfo {author} {\bibfnamefont {Z.}~\bibnamefont {Li}}, \bibinfo {author}
		{\bibfnamefont {E.}~\bibnamefont {Mendez}}, \bibinfo {author} {\bibfnamefont
			{B.}~\bibnamefont {Braverman}}, \bibinfo {author} {\bibfnamefont
			{A.}~\bibnamefont {Kawasaki}}, \bibinfo {author} {\bibfnamefont
			{D.}~\bibnamefont {Akamatsu}}, \bibinfo {author} {\bibfnamefont
			{Y.}~\bibnamefont {Xiao}},\ and\ \bibinfo {author} {\bibfnamefont
			{V.}~\bibnamefont {Vuleti{\'c}}},\ }\bibfield  {title} {\bibinfo {title}
		{Entanglement on an optical atomic-clock transition},\ }\href
	{https://doi.org/10.1038/s41586-020-3006-1} {\bibfield  {journal} {\bibinfo
			{journal} {Nature}\ }\textbf {\bibinfo {volume} {588}},\ \bibinfo {pages}
		{414} (\bibinfo {year} {2020})}\BibitemShut {NoStop}%
	\bibitem [{\citenamefont {Degen}\ \emph {et~al.}(2017)\citenamefont {Degen},
		\citenamefont {Reinhard},\ and\ \citenamefont {Cappellaro}}]{Degen2017}%
	\BibitemOpen
	\bibfield  {author} {\bibinfo {author} {\bibfnamefont {C.~L.}\ \bibnamefont
			{Degen}}, \bibinfo {author} {\bibfnamefont {F.}~\bibnamefont {Reinhard}},\
		and\ \bibinfo {author} {\bibfnamefont {P.}~\bibnamefont {Cappellaro}},\
	}\bibfield  {title} {\bibinfo {title} {Quantum sensing},\ }\href
	{https://doi.org/10.1103/RevModPhys.89.035002} {\bibfield  {journal}
		{\bibinfo  {journal} {Reviews of Modern Physics}\ }\textbf {\bibinfo {volume}
			{89}},\ \bibinfo {pages} {035002} (\bibinfo {year} {2017})},\ \Eprint
	{https://arxiv.org/abs/1611.02427} {arXiv:1611.02427} \BibitemShut {NoStop}%
	\bibitem [{\citenamefont {Safronova}\ \emph {et~al.}(2018)\citenamefont
		{Safronova}, \citenamefont {Budker}, \citenamefont {Demille}, \citenamefont
		{Kimball}, \citenamefont {Derevianko},\ and\ \citenamefont
		{Clark}}]{Safronova2018}%
	\BibitemOpen
	\bibfield  {author} {\bibinfo {author} {\bibfnamefont {M.~S.}\ \bibnamefont
			{Safronova}}, \bibinfo {author} {\bibfnamefont {D.}~\bibnamefont {Budker}},
		\bibinfo {author} {\bibfnamefont {D.}~\bibnamefont {Demille}}, \bibinfo
		{author} {\bibfnamefont {D.~F.}\ \bibnamefont {Kimball}}, \bibinfo {author}
		{\bibfnamefont {A.}~\bibnamefont {Derevianko}},\ and\ \bibinfo {author}
		{\bibfnamefont {C.~W.}\ \bibnamefont {Clark}},\ }\bibfield  {title} {\bibinfo
		{title} {Search for new physics with atoms and molecules},\ }\href
	{https://doi.org/10.1103/RevModPhys.90.025008} {\bibfield  {journal}
		{\bibinfo  {journal} {Reviews of Modern Physics}\ }\textbf {\bibinfo {volume}
			{90}},\ \bibinfo {pages} {025008} (\bibinfo {year} {2018})},\ \Eprint
	{https://arxiv.org/abs/1710.01833} {arXiv:1710.01833} \BibitemShut {NoStop}%
	\bibitem [{\citenamefont {Jackson~Kimball}\ \emph {et~al.}(2023)\citenamefont
		{Jackson~Kimball}, \citenamefont {Budker}, \citenamefont {Chupp},
		\citenamefont {Geraci}, \citenamefont {Kolkowitz}, \citenamefont {Singh},\
		and\ \citenamefont {Sushkov}}]{JacksonKimball2023}%
	\BibitemOpen
	\bibfield  {author} {\bibinfo {author} {\bibfnamefont {D.~F.}\ \bibnamefont
			{Jackson~Kimball}}, \bibinfo {author} {\bibfnamefont {D.}~\bibnamefont
			{Budker}}, \bibinfo {author} {\bibfnamefont {T.~E.}\ \bibnamefont {Chupp}},
		\bibinfo {author} {\bibfnamefont {A.~A.}\ \bibnamefont {Geraci}}, \bibinfo
		{author} {\bibfnamefont {S.}~\bibnamefont {Kolkowitz}}, \bibinfo {author}
		{\bibfnamefont {J.~T.}\ \bibnamefont {Singh}},\ and\ \bibinfo {author}
		{\bibfnamefont {A.~O.}\ \bibnamefont {Sushkov}},\ }\bibfield  {title}
	{\bibinfo {title} {Probing fundamental physics with spin-based quantum
			sensors},\ }\href {https://doi.org/10.1103/PhysRevA.108.010101} {\bibfield
		{journal} {\bibinfo  {journal} {Physical Review A}\ }\textbf {\bibinfo
			{volume} {108}},\ \bibinfo {pages} {010101} (\bibinfo {year}
		{2023})}\BibitemShut {NoStop}%
	\bibitem [{\citenamefont {Bloch}\ and\ \citenamefont
		{Siegert}(1940)}]{Bloch1940}%
	\BibitemOpen
	\bibfield  {author} {\bibinfo {author} {\bibfnamefont {F.}~\bibnamefont
			{Bloch}}\ and\ \bibinfo {author} {\bibfnamefont {A.}~\bibnamefont
			{Siegert}},\ }\bibfield  {title} {\bibinfo {title} {Magnetic resonance for
			nonrotating fields},\ }\href {https://doi.org/10.1103/PhysRev.57.522}
	{\bibfield  {journal} {\bibinfo  {journal} {Physical Review}\ }\textbf
		{\bibinfo {volume} {57}},\ \bibinfo {pages} {522} (\bibinfo {year}
		{1940})}\BibitemShut {NoStop}%
	\bibitem [{\citenamefont {Sleater}\ \emph {et~al.}(1985)\citenamefont
		{Sleater}, \citenamefont {Hahn}, \citenamefont {Hilbert},\ and\ \citenamefont
		{Clarke}}]{Sleator1985}%
	\BibitemOpen
	\bibfield  {author} {\bibinfo {author} {\bibfnamefont {T.}~\bibnamefont
			{Sleater}}, \bibinfo {author} {\bibfnamefont {E.~L.}\ \bibnamefont {Hahn}},
		\bibinfo {author} {\bibfnamefont {C.}~\bibnamefont {Hilbert}},\ and\ \bibinfo
		{author} {\bibfnamefont {J.}~\bibnamefont {Clarke}},\ }\bibfield  {title}
	{\bibinfo {title} {Nuclear-spin noise},\ }\href
	{https://doi.org/10.1103/PhysRevLett.55.1742} {\bibfield  {journal} {\bibinfo
			{journal} {Physical Review Letters}\ }\textbf {\bibinfo {volume} {55}},\
		\bibinfo {pages} {1742} (\bibinfo {year} {1985})}\BibitemShut {NoStop}%
	\bibitem [{\citenamefont {Crooker}\ \emph {et~al.}(2004)\citenamefont
		{Crooker}, \citenamefont {Rickel}, \citenamefont {Balatsky},\ and\
		\citenamefont {Smith}}]{Crooker2004}%
	\BibitemOpen
	\bibfield  {author} {\bibinfo {author} {\bibfnamefont {S.~A.}\ \bibnamefont
			{Crooker}}, \bibinfo {author} {\bibfnamefont {D.~G.}\ \bibnamefont {Rickel}},
		\bibinfo {author} {\bibfnamefont {A.~V.}\ \bibnamefont {Balatsky}},\ and\
		\bibinfo {author} {\bibfnamefont {D.~L.}\ \bibnamefont {Smith}},\ }\bibfield
	{title} {\bibinfo {title} {Spectroscopy of spontaneous spin noise as a probe
			of spin dynamics and magnetic resonance},\ }\href
	{https://doi.org/10.1038/nature02804} {\bibfield  {journal} {\bibinfo
			{journal} {Nature}\ }\textbf {\bibinfo {volume} {431}},\ \bibinfo {pages}
		{49} (\bibinfo {year} {2004})}\BibitemShut {NoStop}%
	\bibitem [{\citenamefont {Shah}\ \emph {et~al.}(2010)\citenamefont {Shah},
		\citenamefont {Vasilakis},\ and\ \citenamefont {Romalis}}]{Shah2010}%
	\BibitemOpen
	\bibfield  {author} {\bibinfo {author} {\bibfnamefont {V.}~\bibnamefont
			{Shah}}, \bibinfo {author} {\bibfnamefont {G.}~\bibnamefont {Vasilakis}},\
		and\ \bibinfo {author} {\bibfnamefont {M.~V.}\ \bibnamefont {Romalis}},\
	}\bibfield  {title} {\bibinfo {title} {High {{Bandwidth Atomic
					Magnetometery}} with {{Continuous Quantum Nondemolition Measurements}}},\
	}\href {https://doi.org/10.1103/PhysRevLett.104.013601} {\bibfield  {journal}
		{\bibinfo  {journal} {Physical Review Letters}\ }\textbf {\bibinfo {volume}
			{104}},\ \bibinfo {pages} {013601} (\bibinfo {year} {2010})}\BibitemShut
	{NoStop}%
	\bibitem [{\citenamefont {Schlagnitweit}\ and\ \citenamefont
		{M{\"u}ller}(2012)}]{Schlagnitweit2012}%
	\BibitemOpen
	\bibfield  {author} {\bibinfo {author} {\bibfnamefont {J.}~\bibnamefont
			{Schlagnitweit}}\ and\ \bibinfo {author} {\bibfnamefont {N.}~\bibnamefont
			{M{\"u}ller}},\ }\bibfield  {title} {\bibinfo {title} {The first observation
			of {{Carbon-13}} spin noise spectra},\ }\href
	{https://doi.org/10.1016/j.jmr.2012.09.002} {\bibfield  {journal} {\bibinfo
			{journal} {Journal of Magnetic Resonance}\ }\textbf {\bibinfo {volume}
			{224}},\ \bibinfo {pages} {78} (\bibinfo {year} {2012})}\BibitemShut
	{NoStop}%
	\bibitem [{\citenamefont {Bollinger}\ \emph {et~al.}(1996)\citenamefont
		{Bollinger}, \citenamefont {Itano}, \citenamefont {Wineland},\ and\
		\citenamefont {Heinzen}}]{Bollinger1996}%
	\BibitemOpen
	\bibfield  {author} {\bibinfo {author} {\bibfnamefont {J.~J.~.}\ \bibnamefont
			{Bollinger}}, \bibinfo {author} {\bibfnamefont {W.~M.}\ \bibnamefont
			{Itano}}, \bibinfo {author} {\bibfnamefont {D.~J.}\ \bibnamefont
			{Wineland}},\ and\ \bibinfo {author} {\bibfnamefont {D.~J.}\ \bibnamefont
			{Heinzen}},\ }\bibfield  {title} {\bibinfo {title} {Optimal frequency
			measurements with maximally correlated states},\ }\href
	{https://doi.org/10.1103/PhysRevA.54.R4649} {\bibfield  {journal} {\bibinfo
			{journal} {Physical Review A}\ }\textbf {\bibinfo {volume} {54}},\ \bibinfo
		{pages} {R4649} (\bibinfo {year} {1996})}\BibitemShut {NoStop}%
	\bibitem [{\citenamefont {Leibfried}\ \emph {et~al.}(2004)\citenamefont
		{Leibfried}, \citenamefont {Barrett}, \citenamefont {Schaetz}, \citenamefont
		{Britton}, \citenamefont {Chiaverini}, \citenamefont {Itano}, \citenamefont
		{Jost}, \citenamefont {Langer},\ and\ \citenamefont
		{Wineland}}]{Leibfried2004}%
	\BibitemOpen
	\bibfield  {author} {\bibinfo {author} {\bibfnamefont {D.}~\bibnamefont
			{Leibfried}}, \bibinfo {author} {\bibfnamefont {M.~D.}\ \bibnamefont
			{Barrett}}, \bibinfo {author} {\bibfnamefont {T.}~\bibnamefont {Schaetz}},
		\bibinfo {author} {\bibfnamefont {J.}~\bibnamefont {Britton}}, \bibinfo
		{author} {\bibfnamefont {J.}~\bibnamefont {Chiaverini}}, \bibinfo {author}
		{\bibfnamefont {W.~M.}\ \bibnamefont {Itano}}, \bibinfo {author}
		{\bibfnamefont {J.~D.}\ \bibnamefont {Jost}}, \bibinfo {author}
		{\bibfnamefont {C.}~\bibnamefont {Langer}},\ and\ \bibinfo {author}
		{\bibfnamefont {D.~J.}\ \bibnamefont {Wineland}},\ }\bibfield  {title}
	{\bibinfo {title} {Toward {{Heisenberg-Limited Spectroscopy}} with
			{{Multiparticle Entangled States}}},\ }\href
	{https://doi.org/10.1126/science.1097576} {\bibfield  {journal} {\bibinfo
			{journal} {Science}\ }\textbf {\bibinfo {volume} {304}},\ \bibinfo {pages}
		{1476} (\bibinfo {year} {2004})}\BibitemShut {NoStop}%
	\bibitem [{\citenamefont {Omran}\ \emph {et~al.}(2019)\citenamefont {Omran},
		\citenamefont {Levine}, \citenamefont {Keesling}, \citenamefont {Semeghini},
		\citenamefont {Wang}, \citenamefont {Ebadi}, \citenamefont {Bernien},
		\citenamefont {Zibrov}, \citenamefont {Pichler}, \citenamefont {Choi},
		\citenamefont {Cui}, \citenamefont {Rossignolo}, \citenamefont {Rembold},
		\citenamefont {Montangero}, \citenamefont {Calarco}, \citenamefont {Endres},
		\citenamefont {Greiner}, \citenamefont {Vuleti{\'c}},\ and\ \citenamefont
		{Lukin}}]{Omran2019}%
	\BibitemOpen
	\bibfield  {author} {\bibinfo {author} {\bibfnamefont {A.}~\bibnamefont
			{Omran}}, \bibinfo {author} {\bibfnamefont {H.}~\bibnamefont {Levine}},
		\bibinfo {author} {\bibfnamefont {A.}~\bibnamefont {Keesling}}, \bibinfo
		{author} {\bibfnamefont {G.}~\bibnamefont {Semeghini}}, \bibinfo {author}
		{\bibfnamefont {T.~T.}\ \bibnamefont {Wang}}, \bibinfo {author}
		{\bibfnamefont {S.}~\bibnamefont {Ebadi}}, \bibinfo {author} {\bibfnamefont
			{H.}~\bibnamefont {Bernien}}, \bibinfo {author} {\bibfnamefont {A.~S.}\
			\bibnamefont {Zibrov}}, \bibinfo {author} {\bibfnamefont {H.}~\bibnamefont
			{Pichler}}, \bibinfo {author} {\bibfnamefont {S.}~\bibnamefont {Choi}},
		\bibinfo {author} {\bibfnamefont {J.}~\bibnamefont {Cui}}, \bibinfo {author}
		{\bibfnamefont {M.}~\bibnamefont {Rossignolo}}, \bibinfo {author}
		{\bibfnamefont {P.}~\bibnamefont {Rembold}}, \bibinfo {author} {\bibfnamefont
			{S.}~\bibnamefont {Montangero}}, \bibinfo {author} {\bibfnamefont
			{T.}~\bibnamefont {Calarco}}, \bibinfo {author} {\bibfnamefont
			{M.}~\bibnamefont {Endres}}, \bibinfo {author} {\bibfnamefont
			{M.}~\bibnamefont {Greiner}}, \bibinfo {author} {\bibfnamefont
			{V.}~\bibnamefont {Vuleti{\'c}}},\ and\ \bibinfo {author} {\bibfnamefont
			{M.~D.}\ \bibnamefont {Lukin}},\ }\bibfield  {title} {\bibinfo {title}
		{Generation and manipulation of {{Schr{\"o}dinger}} cat states in {{Rydberg}}
			atom arrays},\ }\href {https://doi.org/10.1126/science.aax9743} {\bibfield
		{journal} {\bibinfo  {journal} {Science}\ }\textbf {\bibinfo {volume}
			{365}},\ \bibinfo {pages} {570} (\bibinfo {year} {2019})}\BibitemShut
	{NoStop}%
	\bibitem [{\citenamefont {Kitagawa}\ and\ \citenamefont
		{Ueda}(1993)}]{Kitagawa1993}%
	\BibitemOpen
	\bibfield  {author} {\bibinfo {author} {\bibfnamefont {M.}~\bibnamefont
			{Kitagawa}}\ and\ \bibinfo {author} {\bibfnamefont {M.}~\bibnamefont
			{Ueda}},\ }\bibfield  {title} {\bibinfo {title} {Squeezed spin states},\
	}\href {https://doi.org/10.1103/PhysRevA.47.5138} {\bibfield  {journal}
		{\bibinfo  {journal} {Physical Review A}\ }\textbf {\bibinfo {volume} {47}},\
		\bibinfo {pages} {5138} (\bibinfo {year} {1993})}\BibitemShut {NoStop}%
	\bibitem [{\citenamefont {Wineland}\ \emph {et~al.}(1992)\citenamefont
		{Wineland}, \citenamefont {Bollinger}, \citenamefont {Itano}, \citenamefont
		{Moore},\ and\ \citenamefont {Heinzen}}]{Wineland1992}%
	\BibitemOpen
	\bibfield  {author} {\bibinfo {author} {\bibfnamefont {D.~J.}\ \bibnamefont
			{Wineland}}, \bibinfo {author} {\bibfnamefont {J.~J.}\ \bibnamefont
			{Bollinger}}, \bibinfo {author} {\bibfnamefont {W.~M.}\ \bibnamefont
			{Itano}}, \bibinfo {author} {\bibfnamefont {F.~L.}\ \bibnamefont {Moore}},\
		and\ \bibinfo {author} {\bibfnamefont {D.~J.}\ \bibnamefont {Heinzen}},\
	}\bibfield  {title} {\bibinfo {title} {Spin squeezing and reduced quantum
			noise in spectroscopy},\ }\href {https://doi.org/10.1103/PhysRevA.46.R6797}
	{\bibfield  {journal} {\bibinfo  {journal} {Physical Review A}\ }\textbf
		{\bibinfo {volume} {46}},\ \bibinfo {pages} {R6797} (\bibinfo {year}
		{1992})}\BibitemShut {NoStop}%
	\bibitem [{\citenamefont {Jackson~Kimball}\ \emph {et~al.}(2016)\citenamefont
		{Jackson~Kimball}, \citenamefont {Sushkov},\ and\ \citenamefont
		{Budker}}]{JacksonKimball2016}%
	\BibitemOpen
	\bibfield  {author} {\bibinfo {author} {\bibfnamefont {D.~F.}\ \bibnamefont
			{Jackson~Kimball}}, \bibinfo {author} {\bibfnamefont {A.~O.}\ \bibnamefont
			{Sushkov}},\ and\ \bibinfo {author} {\bibfnamefont {D.}~\bibnamefont
			{Budker}},\ }\bibfield  {title} {\bibinfo {title} {Precessing {{Ferromagnetic
					Needle Magnetometer}}},\ }\href
	{https://doi.org/10.1103/PhysRevLett.116.190801} {\bibfield  {journal}
		{\bibinfo  {journal} {Physical Review Letters}\ }\textbf {\bibinfo {volume}
			{116}},\ \bibinfo {pages} {190801} (\bibinfo {year} {2016})},\ \Eprint
	{https://arxiv.org/abs/1602.02818} {arXiv:1602.02818} \BibitemShut {NoStop}%
	\bibitem [{\citenamefont {Vinante}\ \emph {et~al.}(2021)\citenamefont
		{Vinante}, \citenamefont {Timberlake}, \citenamefont {Budker}, \citenamefont
		{Kimball}, \citenamefont {Sushkov},\ and\ \citenamefont
		{Ulbricht}}]{Vinante2021a}%
	\BibitemOpen
	\bibfield  {author} {\bibinfo {author} {\bibfnamefont {A.}~\bibnamefont
			{Vinante}}, \bibinfo {author} {\bibfnamefont {C.}~\bibnamefont {Timberlake}},
		\bibinfo {author} {\bibfnamefont {D.}~\bibnamefont {Budker}}, \bibinfo
		{author} {\bibfnamefont {D.~F.}\ \bibnamefont {Kimball}}, \bibinfo {author}
		{\bibfnamefont {A.~O.}\ \bibnamefont {Sushkov}},\ and\ \bibinfo {author}
		{\bibfnamefont {H.}~\bibnamefont {Ulbricht}},\ }\bibfield  {title} {\bibinfo
		{title} {Surpassing the {{Energy Resolution Limit}} with {{Ferromagnetic
					Torque Sensors}}},\ }\href {https://doi.org/10.1103/PhysRevLett.127.070801}
	{\bibfield  {journal} {\bibinfo  {journal} {Physical Review Letters}\
		}\textbf {\bibinfo {volume} {127}},\ \bibinfo {pages} {070801} (\bibinfo
		{year} {2021})},\ \Eprint {https://arxiv.org/abs/2104.14425}
	{arXiv:2104.14425} \BibitemShut {NoStop}%
	\bibitem [{\citenamefont {{Schleier-Smith}}\ \emph {et~al.}(2010)\citenamefont
		{{Schleier-Smith}}, \citenamefont {Leroux},\ and\ \citenamefont
		{Vuleti{\'c}}}]{Schleier-Smith2010}%
	\BibitemOpen
	\bibfield  {author} {\bibinfo {author} {\bibfnamefont {M.~H.}\ \bibnamefont
			{{Schleier-Smith}}}, \bibinfo {author} {\bibfnamefont {I.~D.}\ \bibnamefont
			{Leroux}},\ and\ \bibinfo {author} {\bibfnamefont {V.}~\bibnamefont
			{Vuleti{\'c}}},\ }\bibfield  {title} {\bibinfo {title} {Squeezing the
			collective spin of a dilute atomic ensemble by cavity feedback},\ }\href
	{https://doi.org/10.1103/PhysRevA.81.021804} {\bibfield  {journal} {\bibinfo
			{journal} {Physical Review A - Atomic, Molecular, and Optical Physics}\
		}\textbf {\bibinfo {volume} {81}},\ \bibinfo {pages} {021804} (\bibinfo
		{year} {2010})}\BibitemShut {NoStop}%
	\bibitem [{\citenamefont {Leroux}\ \emph {et~al.}(2010)\citenamefont {Leroux},
		\citenamefont {{Schleier-Smith}},\ and\ \citenamefont
		{Vuleti{\'c}}}]{Leroux2010}%
	\BibitemOpen
	\bibfield  {author} {\bibinfo {author} {\bibfnamefont {I.~D.}\ \bibnamefont
			{Leroux}}, \bibinfo {author} {\bibfnamefont {M.~H.}\ \bibnamefont
			{{Schleier-Smith}}},\ and\ \bibinfo {author} {\bibfnamefont {V.}~\bibnamefont
			{Vuleti{\'c}}},\ }\bibfield  {title} {\bibinfo {title} {Implementation of
			cavity squeezing of a collective atomic {{Spin}}},\ }\href
	{https://doi.org/10.1103/PhysRevLett.104.073602} {\bibfield  {journal}
		{\bibinfo  {journal} {Physical Review Letters}\ }\textbf {\bibinfo {volume}
			{104}},\ \bibinfo {pages} {073602} (\bibinfo {year} {2010})}\BibitemShut
	{NoStop}%
	\bibitem [{\citenamefont {Vasilakis}\ \emph
		{et~al.}(2015{\natexlab{b}})\citenamefont {Vasilakis}, \citenamefont {Shen},
		\citenamefont {Jensen}, \citenamefont {Balabas}, \citenamefont {Salart},
		\citenamefont {Chen},\ and\ \citenamefont {Polzik}}]{Vasilakis2015}%
	\BibitemOpen
	\bibfield  {author} {\bibinfo {author} {\bibfnamefont {G.}~\bibnamefont
			{Vasilakis}}, \bibinfo {author} {\bibfnamefont {H.}~\bibnamefont {Shen}},
		\bibinfo {author} {\bibfnamefont {K.}~\bibnamefont {Jensen}}, \bibinfo
		{author} {\bibfnamefont {M.}~\bibnamefont {Balabas}}, \bibinfo {author}
		{\bibfnamefont {D.}~\bibnamefont {Salart}}, \bibinfo {author} {\bibfnamefont
			{B.}~\bibnamefont {Chen}},\ and\ \bibinfo {author} {\bibfnamefont {E.~S.}\
			\bibnamefont {Polzik}},\ }\bibfield  {title} {\bibinfo {title} {Generation of
			a squeezed state of an oscillator by stroboscopic back-action-evading
			measurement},\ }\href {https://doi.org/10.1038/nphys3280} {\bibfield
		{journal} {\bibinfo  {journal} {Nature Physics}\ }\textbf {\bibinfo {volume}
			{11}},\ \bibinfo {pages} {389} (\bibinfo {year} {2015}{\natexlab{b}})},\
	\Eprint {https://arxiv.org/abs/1411.6289} {arXiv:1411.6289} \BibitemShut
	{NoStop}%
	\bibitem [{\citenamefont {Bao}\ \emph {et~al.}(2020)\citenamefont {Bao},
		\citenamefont {Duan}, \citenamefont {Jin}, \citenamefont {Lu}, \citenamefont
		{Li}, \citenamefont {Qu}, \citenamefont {Wang}, \citenamefont {Novikova},
		\citenamefont {Mikhailov}, \citenamefont {Zhao}, \citenamefont {M{\o}lmer},
		\citenamefont {Shen},\ and\ \citenamefont {Xiao}}]{Bao2020}%
	\BibitemOpen
	\bibfield  {author} {\bibinfo {author} {\bibfnamefont {H.}~\bibnamefont
			{Bao}}, \bibinfo {author} {\bibfnamefont {J.}~\bibnamefont {Duan}}, \bibinfo
		{author} {\bibfnamefont {S.}~\bibnamefont {Jin}}, \bibinfo {author}
		{\bibfnamefont {X.}~\bibnamefont {Lu}}, \bibinfo {author} {\bibfnamefont
			{P.}~\bibnamefont {Li}}, \bibinfo {author} {\bibfnamefont {W.}~\bibnamefont
			{Qu}}, \bibinfo {author} {\bibfnamefont {M.}~\bibnamefont {Wang}}, \bibinfo
		{author} {\bibfnamefont {I.}~\bibnamefont {Novikova}}, \bibinfo {author}
		{\bibfnamefont {E.~E.}\ \bibnamefont {Mikhailov}}, \bibinfo {author}
		{\bibfnamefont {K.~F.}\ \bibnamefont {Zhao}}, \bibinfo {author}
		{\bibfnamefont {K.}~\bibnamefont {M{\o}lmer}}, \bibinfo {author}
		{\bibfnamefont {H.}~\bibnamefont {Shen}},\ and\ \bibinfo {author}
		{\bibfnamefont {Y.}~\bibnamefont {Xiao}},\ }\bibfield  {title} {\bibinfo
		{title} {Spin squeezing of 1011 atoms by prediction and retrodiction
			measurements},\ }\href {https://doi.org/10.1038/s41586-020-2243-7} {\bibfield
		{journal} {\bibinfo  {journal} {Nature}\ }\textbf {\bibinfo {volume}
			{581}},\ \bibinfo {pages} {159} (\bibinfo {year} {2020})}\BibitemShut
	{NoStop}%
	\bibitem [{\citenamefont {Colombo}\ \emph {et~al.}(2022)\citenamefont
		{Colombo}, \citenamefont {{Pedrozo-Pe{\~n}afiel}}, \citenamefont
		{Adiyatullin}, \citenamefont {Li}, \citenamefont {Mendez}, \citenamefont
		{Shu},\ and\ \citenamefont {Vuleti{\'c}}}]{Colombo2022a}%
	\BibitemOpen
	\bibfield  {author} {\bibinfo {author} {\bibfnamefont {S.}~\bibnamefont
			{Colombo}}, \bibinfo {author} {\bibfnamefont {E.}~\bibnamefont
			{{Pedrozo-Pe{\~n}afiel}}}, \bibinfo {author} {\bibfnamefont {A.~F.}\
			\bibnamefont {Adiyatullin}}, \bibinfo {author} {\bibfnamefont
			{Z.}~\bibnamefont {Li}}, \bibinfo {author} {\bibfnamefont {E.}~\bibnamefont
			{Mendez}}, \bibinfo {author} {\bibfnamefont {C.}~\bibnamefont {Shu}},\ and\
		\bibinfo {author} {\bibfnamefont {V.}~\bibnamefont {Vuleti{\'c}}},\
	}\bibfield  {title} {\bibinfo {title} {Time-reversal-based quantum metrology
			with many-body entangled states},\ }\href
	{https://doi.org/10.1038/s41567-022-01653-5} {\bibfield  {journal} {\bibinfo
			{journal} {Nature Physics}\ }\textbf {\bibinfo {volume} {18}},\ \bibinfo
		{pages} {925} (\bibinfo {year} {2022})}\BibitemShut {NoStop}%
	\bibitem [{\citenamefont {Serafin}\ \emph {et~al.}(2021)\citenamefont
		{Serafin}, \citenamefont {Fadel}, \citenamefont {Treutlein},\ and\
		\citenamefont {Sinatra}}]{Serafin2021}%
	\BibitemOpen
	\bibfield  {author} {\bibinfo {author} {\bibfnamefont {A.}~\bibnamefont
			{Serafin}}, \bibinfo {author} {\bibfnamefont {M.}~\bibnamefont {Fadel}},
		\bibinfo {author} {\bibfnamefont {P.}~\bibnamefont {Treutlein}},\ and\
		\bibinfo {author} {\bibfnamefont {A.}~\bibnamefont {Sinatra}},\ }\bibfield
	{title} {\bibinfo {title} {Nuclear {{Spin Squeezing}} in {{Helium-3}} by
			{{Continuous Quantum Nondemolition Measurement}}},\ }\href
	{https://doi.org/10.1103/PhysRevLett.127.013601} {\bibfield  {journal}
		{\bibinfo  {journal} {Physical Review Letters}\ }\textbf {\bibinfo {volume}
			{127}},\ \bibinfo {pages} {013601} (\bibinfo {year} {2021})}\BibitemShut
	{NoStop}%
	\bibitem [{\citenamefont {{Arrowsmith-Kron}}\ \emph {et~al.}(2024)\citenamefont
		{{Arrowsmith-Kron}}, \citenamefont {{Athanasakis-Kaklamanakis}},
		\citenamefont {Au}, \citenamefont {Ballof}, \citenamefont {Berger},
		\citenamefont {Borschevsky}, \citenamefont {Breier}, \citenamefont
		{Buchinger}, \citenamefont {Budker}, \citenamefont {Caldwell}, \citenamefont
		{Charles}, \citenamefont {Dattani}, \citenamefont {de~Groote}, \citenamefont
		{DeMille}, \citenamefont {Dickel}, \citenamefont {Dobaczewski}, \citenamefont
		{D{\"u}llmann}, \citenamefont {Eliav}, \citenamefont {Engel}, \citenamefont
		{Fan}, \citenamefont {Flambaum}, \citenamefont {Flanagan}, \citenamefont
		{Gaiser}, \citenamefont {Ruiz}, \citenamefont {Gaul}, \citenamefont {Giesen},
		\citenamefont {Ginges}, \citenamefont {Gottberg}, \citenamefont {Gwinner},
		\citenamefont {Heinke}, \citenamefont {Hoekstra}, \citenamefont {Holt},
		\citenamefont {Hutzler}, \citenamefont {Jayich}, \citenamefont {Karthein},
		\citenamefont {Leach}, \citenamefont {Madison}, \citenamefont
		{{Malbrunot-Ettenauer}}, \citenamefont {Miyagi}, \citenamefont {Moore},
		\citenamefont {Moroch}, \citenamefont {Navratil}, \citenamefont {Nazarewicz},
		\citenamefont {Neyens}, \citenamefont {Norrgard}, \citenamefont {Nusgart},
		\citenamefont {Pa{\v s}teka}, \citenamefont {Petrov}, \citenamefont
		{Pla{\ss}}, \citenamefont {Ready}, \citenamefont {Reiter}, \citenamefont
		{Reponen}, \citenamefont {Rothe}, \citenamefont {Safronova}, \citenamefont
		{Scheidenerger}, \citenamefont {Shindler}, \citenamefont {Singh},
		\citenamefont {Skripnikov}, \citenamefont {Titov}, \citenamefont {Udrescu},
		\citenamefont {Wilkins},\ and\ \citenamefont {Yang}}]{Arrowsmith-Kron2024}%
	\BibitemOpen
	\bibfield  {author} {\bibinfo {author} {\bibfnamefont {G.}~\bibnamefont
			{{Arrowsmith-Kron}}}, \bibinfo {author} {\bibfnamefont {M.}~\bibnamefont
			{{Athanasakis-Kaklamanakis}}}, \bibinfo {author} {\bibfnamefont
			{M.}~\bibnamefont {Au}}, \bibinfo {author} {\bibfnamefont {J.}~\bibnamefont
			{Ballof}}, \bibinfo {author} {\bibfnamefont {R.}~\bibnamefont {Berger}},
		\bibinfo {author} {\bibfnamefont {A.}~\bibnamefont {Borschevsky}}, \bibinfo
		{author} {\bibfnamefont {A.~A.}\ \bibnamefont {Breier}}, \bibinfo {author}
		{\bibfnamefont {F.}~\bibnamefont {Buchinger}}, \bibinfo {author}
		{\bibfnamefont {D.}~\bibnamefont {Budker}}, \bibinfo {author} {\bibfnamefont
			{L.}~\bibnamefont {Caldwell}}, \bibinfo {author} {\bibfnamefont
			{C.}~\bibnamefont {Charles}}, \bibinfo {author} {\bibfnamefont
			{N.}~\bibnamefont {Dattani}}, \bibinfo {author} {\bibfnamefont {R.~P.}\
			\bibnamefont {de~Groote}}, \bibinfo {author} {\bibfnamefont {D.}~\bibnamefont
			{DeMille}}, \bibinfo {author} {\bibfnamefont {T.}~\bibnamefont {Dickel}},
		\bibinfo {author} {\bibfnamefont {J.}~\bibnamefont {Dobaczewski}}, \bibinfo
		{author} {\bibfnamefont {C.~E.}\ \bibnamefont {D{\"u}llmann}}, \bibinfo
		{author} {\bibfnamefont {E.}~\bibnamefont {Eliav}}, \bibinfo {author}
		{\bibfnamefont {J.}~\bibnamefont {Engel}}, \bibinfo {author} {\bibfnamefont
			{M.}~\bibnamefont {Fan}}, \bibinfo {author} {\bibfnamefont {V.}~\bibnamefont
			{Flambaum}}, \bibinfo {author} {\bibfnamefont {K.~T.}\ \bibnamefont
			{Flanagan}}, \bibinfo {author} {\bibfnamefont {A.~N.}\ \bibnamefont
			{Gaiser}}, \bibinfo {author} {\bibfnamefont {R.~F.~G.}\ \bibnamefont {Ruiz}},
		\bibinfo {author} {\bibfnamefont {K.}~\bibnamefont {Gaul}}, \bibinfo {author}
		{\bibfnamefont {T.~F.}\ \bibnamefont {Giesen}}, \bibinfo {author}
		{\bibfnamefont {J.~S.~M.}\ \bibnamefont {Ginges}}, \bibinfo {author}
		{\bibfnamefont {A.}~\bibnamefont {Gottberg}}, \bibinfo {author}
		{\bibfnamefont {G.}~\bibnamefont {Gwinner}}, \bibinfo {author} {\bibfnamefont
			{R.}~\bibnamefont {Heinke}}, \bibinfo {author} {\bibfnamefont
			{S.}~\bibnamefont {Hoekstra}}, \bibinfo {author} {\bibfnamefont {J.~D.}\
			\bibnamefont {Holt}}, \bibinfo {author} {\bibfnamefont {N.~R.}\ \bibnamefont
			{Hutzler}}, \bibinfo {author} {\bibfnamefont {A.}~\bibnamefont {Jayich}},
		\bibinfo {author} {\bibfnamefont {J.}~\bibnamefont {Karthein}}, \bibinfo
		{author} {\bibfnamefont {K.~G.}\ \bibnamefont {Leach}}, \bibinfo {author}
		{\bibfnamefont {K.~W.}\ \bibnamefont {Madison}}, \bibinfo {author}
		{\bibfnamefont {S.}~\bibnamefont {{Malbrunot-Ettenauer}}}, \bibinfo {author}
		{\bibfnamefont {T.}~\bibnamefont {Miyagi}}, \bibinfo {author} {\bibfnamefont
			{I.~D.}\ \bibnamefont {Moore}}, \bibinfo {author} {\bibfnamefont
			{S.}~\bibnamefont {Moroch}}, \bibinfo {author} {\bibfnamefont
			{P.}~\bibnamefont {Navratil}}, \bibinfo {author} {\bibfnamefont
			{W.}~\bibnamefont {Nazarewicz}}, \bibinfo {author} {\bibfnamefont
			{G.}~\bibnamefont {Neyens}}, \bibinfo {author} {\bibfnamefont {E.~B.}\
			\bibnamefont {Norrgard}}, \bibinfo {author} {\bibfnamefont {N.}~\bibnamefont
			{Nusgart}}, \bibinfo {author} {\bibfnamefont {L.~F.}\ \bibnamefont {Pa{\v
					s}teka}}, \bibinfo {author} {\bibfnamefont {A.~N.}\ \bibnamefont {Petrov}},
		\bibinfo {author} {\bibfnamefont {W.~R.}\ \bibnamefont {Pla{\ss}}}, \bibinfo
		{author} {\bibfnamefont {R.~A.}\ \bibnamefont {Ready}}, \bibinfo {author}
		{\bibfnamefont {M.~P.}\ \bibnamefont {Reiter}}, \bibinfo {author}
		{\bibfnamefont {M.}~\bibnamefont {Reponen}}, \bibinfo {author} {\bibfnamefont
			{S.}~\bibnamefont {Rothe}}, \bibinfo {author} {\bibfnamefont {M.~S.}\
			\bibnamefont {Safronova}}, \bibinfo {author} {\bibfnamefont {C.}~\bibnamefont
			{Scheidenerger}}, \bibinfo {author} {\bibfnamefont {A.}~\bibnamefont
			{Shindler}}, \bibinfo {author} {\bibfnamefont {J.~T.}\ \bibnamefont {Singh}},
		\bibinfo {author} {\bibfnamefont {L.~V.}\ \bibnamefont {Skripnikov}},
		\bibinfo {author} {\bibfnamefont {A.~V.}\ \bibnamefont {Titov}}, \bibinfo
		{author} {\bibfnamefont {S.-M.}\ \bibnamefont {Udrescu}}, \bibinfo {author}
		{\bibfnamefont {S.~G.}\ \bibnamefont {Wilkins}},\ and\ \bibinfo {author}
		{\bibfnamefont {X.}~\bibnamefont {Yang}},\ }\bibfield  {title} {\bibinfo
		{title} {Opportunities for fundamental physics research with radioactive
			molecules},\ }\href {https://doi.org/10.1088/1361-6633/ad1e39} {\bibfield
		{journal} {\bibinfo  {journal} {Reports on Progress in Physics}\ }\textbf
		{\bibinfo {volume} {87}},\ \bibinfo {pages} {084301} (\bibinfo {year}
		{2024})}\BibitemShut {NoStop}%
	\bibitem [{\citenamefont {Budker}\ \emph {et~al.}(2006)\citenamefont {Budker},
		\citenamefont {Lamoreaux}, \citenamefont {Sushkov},\ and\ \citenamefont
		{Sushkov}}]{Budker2006}%
	\BibitemOpen
	\bibfield  {author} {\bibinfo {author} {\bibfnamefont {D.}~\bibnamefont
			{Budker}}, \bibinfo {author} {\bibfnamefont {S.~K.}\ \bibnamefont
			{Lamoreaux}}, \bibinfo {author} {\bibfnamefont {A.~O.}\ \bibnamefont
			{Sushkov}},\ and\ \bibinfo {author} {\bibfnamefont {O.~P.}\ \bibnamefont
			{Sushkov}},\ }\bibfield  {title} {\bibinfo {title} {Sensitivity of
			condensed-matter {{P}} - {{And T}} -violation experiments},\ }\href
	{https://doi.org/10.1103/PhysRevA.73.022107} {\bibfield  {journal} {\bibinfo
			{journal} {Physical Review A}\ }\textbf {\bibinfo {volume} {73}},\ \bibinfo
		{pages} {022107} (\bibinfo {year} {2006})}\BibitemShut {NoStop}%
	\bibitem [{\citenamefont {Eckel}\ \emph {et~al.}(2012)\citenamefont {Eckel},
		\citenamefont {Sushkov},\ and\ \citenamefont {Lamoreaux}}]{Eckel2012}%
	\BibitemOpen
	\bibfield  {author} {\bibinfo {author} {\bibfnamefont {S.}~\bibnamefont
			{Eckel}}, \bibinfo {author} {\bibfnamefont {A.~O.}\ \bibnamefont {Sushkov}},\
		and\ \bibinfo {author} {\bibfnamefont {S.~K.}\ \bibnamefont {Lamoreaux}},\
	}\bibfield  {title} {\bibinfo {title} {Limit on the electron electric dipole
			moment using paramagnetic ferroelectric {{Eu}} 0.{{5Ba}} 0.{{5TiO}} 3},\
	}\href {https://doi.org/10.1103/PhysRevLett.109.193003} {\bibfield  {journal}
		{\bibinfo  {journal} {Physical Review Letters}\ }\textbf {\bibinfo {volume}
			{109}},\ \bibinfo {pages} {193003} (\bibinfo {year} {2012})},\ \Eprint
	{https://arxiv.org/abs/1208.4420} {arXiv:1208.4420} \BibitemShut {NoStop}%
	\bibitem [{\citenamefont {Budker}\ \emph {et~al.}(2014)\citenamefont {Budker},
		\citenamefont {Graham}, \citenamefont {Ledbetter}, \citenamefont
		{Rajendran},\ and\ \citenamefont {Sushkov}}]{Budker2014}%
	\BibitemOpen
	\bibfield  {author} {\bibinfo {author} {\bibfnamefont {D.}~\bibnamefont
			{Budker}}, \bibinfo {author} {\bibfnamefont {P.~W.}\ \bibnamefont {Graham}},
		\bibinfo {author} {\bibfnamefont {M.}~\bibnamefont {Ledbetter}}, \bibinfo
		{author} {\bibfnamefont {S.}~\bibnamefont {Rajendran}},\ and\ \bibinfo
		{author} {\bibfnamefont {A.~O.}\ \bibnamefont {Sushkov}},\ }\bibfield
	{title} {\bibinfo {title} {Proposal for a cosmic axion spin precession
			experiment ({{CASPEr}})},\ }\href {https://doi.org/10.1103/PhysRevX.4.021030}
	{\bibfield  {journal} {\bibinfo  {journal} {Physical Review X}\ }\textbf
		{\bibinfo {volume} {4}},\ \bibinfo {pages} {021030} (\bibinfo {year}
		{2014})}\BibitemShut {NoStop}%
	\bibitem [{\citenamefont {Aybas}\ \emph
		{et~al.}(2021{\natexlab{a}})\citenamefont {Aybas}, \citenamefont {Bekker},
		\citenamefont {Blanchard}, \citenamefont {Budker}, \citenamefont {Centers},
		\citenamefont {Figueroa}, \citenamefont {Gramolin}, \citenamefont
		{Jackson~Kimball}, \citenamefont {Wickenbrock},\ and\ \citenamefont
		{Sushkov}}]{Aybas2021b}%
	\BibitemOpen
	\bibfield  {author} {\bibinfo {author} {\bibfnamefont {D.}~\bibnamefont
			{Aybas}}, \bibinfo {author} {\bibfnamefont {H.}~\bibnamefont {Bekker}},
		\bibinfo {author} {\bibfnamefont {J.~W.}\ \bibnamefont {Blanchard}}, \bibinfo
		{author} {\bibfnamefont {D.}~\bibnamefont {Budker}}, \bibinfo {author}
		{\bibfnamefont {G.~P.}\ \bibnamefont {Centers}}, \bibinfo {author}
		{\bibfnamefont {N.~L.}\ \bibnamefont {Figueroa}}, \bibinfo {author}
		{\bibfnamefont {A.~V.}\ \bibnamefont {Gramolin}}, \bibinfo {author}
		{\bibfnamefont {D.~F.}\ \bibnamefont {Jackson~Kimball}}, \bibinfo {author}
		{\bibfnamefont {A.}~\bibnamefont {Wickenbrock}},\ and\ \bibinfo {author}
		{\bibfnamefont {A.~O.}\ \bibnamefont {Sushkov}},\ }\bibfield  {title}
	{\bibinfo {title} {Quantum sensitivity limits of nuclear magnetic resonance
			experiments searching for new fundamental physics},\ }\href
	{https://doi.org/10.1088/2058-9565/abfbbc} {\bibfield  {journal} {\bibinfo
			{journal} {Quantum Science and Technology}\ }\textbf {\bibinfo {volume}
			{6}},\ \bibinfo {pages} {034007} (\bibinfo {year} {2021}{\natexlab{a}})},\
	\Eprint {https://arxiv.org/abs/2103.06284} {arXiv:2103.06284} \BibitemShut
	{NoStop}%
	\bibitem [{\citenamefont {Eichhorn}\ \emph {et~al.}(2022)\citenamefont
		{Eichhorn}, \citenamefont {Parker}, \citenamefont {Josten}, \citenamefont
		{M{\"u}ller}, \citenamefont {Scheuer}, \citenamefont {Steiner}, \citenamefont
		{Gierse}, \citenamefont {Handwerker}, \citenamefont {Keim}, \citenamefont
		{Lucas}, \citenamefont {Qureshi}, \citenamefont {Marshall}, \citenamefont
		{Salhov}, \citenamefont {Quan}, \citenamefont {Binder}, \citenamefont
		{Jahnke}, \citenamefont {Neumann}, \citenamefont {Knecht}, \citenamefont
		{Blanchard}, \citenamefont {Plenio}, \citenamefont {Jelezko}, \citenamefont
		{Emsley}, \citenamefont {Vassiliou}, \citenamefont {Hautle},\ and\
		\citenamefont {Schwartz}}]{Eichhorn2022}%
	\BibitemOpen
	\bibfield  {author} {\bibinfo {author} {\bibfnamefont {T.~R.}\ \bibnamefont
			{Eichhorn}}, \bibinfo {author} {\bibfnamefont {A.~J.}\ \bibnamefont
			{Parker}}, \bibinfo {author} {\bibfnamefont {F.}~\bibnamefont {Josten}},
		\bibinfo {author} {\bibfnamefont {C.}~\bibnamefont {M{\"u}ller}}, \bibinfo
		{author} {\bibfnamefont {J.}~\bibnamefont {Scheuer}}, \bibinfo {author}
		{\bibfnamefont {J.~M.}\ \bibnamefont {Steiner}}, \bibinfo {author}
		{\bibfnamefont {M.}~\bibnamefont {Gierse}}, \bibinfo {author} {\bibfnamefont
			{J.}~\bibnamefont {Handwerker}}, \bibinfo {author} {\bibfnamefont
			{M.}~\bibnamefont {Keim}}, \bibinfo {author} {\bibfnamefont {S.}~\bibnamefont
			{Lucas}}, \bibinfo {author} {\bibfnamefont {M.~U.}\ \bibnamefont {Qureshi}},
		\bibinfo {author} {\bibfnamefont {A.}~\bibnamefont {Marshall}}, \bibinfo
		{author} {\bibfnamefont {A.}~\bibnamefont {Salhov}}, \bibinfo {author}
		{\bibfnamefont {Y.}~\bibnamefont {Quan}}, \bibinfo {author} {\bibfnamefont
			{J.}~\bibnamefont {Binder}}, \bibinfo {author} {\bibfnamefont {K.~D.}\
			\bibnamefont {Jahnke}}, \bibinfo {author} {\bibfnamefont {P.}~\bibnamefont
			{Neumann}}, \bibinfo {author} {\bibfnamefont {S.}~\bibnamefont {Knecht}},
		\bibinfo {author} {\bibfnamefont {J.~W.}\ \bibnamefont {Blanchard}}, \bibinfo
		{author} {\bibfnamefont {M.~B.}\ \bibnamefont {Plenio}}, \bibinfo {author}
		{\bibfnamefont {F.}~\bibnamefont {Jelezko}}, \bibinfo {author} {\bibfnamefont
			{L.}~\bibnamefont {Emsley}}, \bibinfo {author} {\bibfnamefont {C.~C.}\
			\bibnamefont {Vassiliou}}, \bibinfo {author} {\bibfnamefont {P.}~\bibnamefont
			{Hautle}},\ and\ \bibinfo {author} {\bibfnamefont {I.}~\bibnamefont
			{Schwartz}},\ }\bibfield  {title} {\bibinfo {title} {Hyperpolarized
			{{Solution-State NMR Spectroscopy}} with {{Optically Polarized Crystals}}},\
	}\href {https://doi.org/10.1021/jacs.1c09119} {\bibfield  {journal} {\bibinfo
			{journal} {Journal of the American Chemical Society}\ }\textbf {\bibinfo
			{volume} {144}},\ \bibinfo {pages} {2511} (\bibinfo {year}
		{2022})}\BibitemShut {NoStop}%
	\bibitem [{\citenamefont {Sinha}\ \emph {et~al.}(2003)\citenamefont {Sinha},
		\citenamefont {Emerson}, \citenamefont {Boulant}, \citenamefont {Fortunato},
		\citenamefont {Havel},\ and\ \citenamefont {Cory}}]{Sinha2003}%
	\BibitemOpen
	\bibfield  {author} {\bibinfo {author} {\bibfnamefont {S.}~\bibnamefont
			{Sinha}}, \bibinfo {author} {\bibfnamefont {J.}~\bibnamefont {Emerson}},
		\bibinfo {author} {\bibfnamefont {N.}~\bibnamefont {Boulant}}, \bibinfo
		{author} {\bibfnamefont {E.~M.}\ \bibnamefont {Fortunato}}, \bibinfo {author}
		{\bibfnamefont {T.~F.}\ \bibnamefont {Havel}},\ and\ \bibinfo {author}
		{\bibfnamefont {D.~G.}\ \bibnamefont {Cory}},\ }\bibfield  {title} {\bibinfo
		{title} {Experimental {{Simulation}} of {{Spin Squeezing}} by {{Nuclear
					Magnetic Resonance}}},\ }\href
	{https://doi.org/10.1023/B:QINP.0000042202.87144.cb} {\bibfield  {journal}
		{\bibinfo  {journal} {Quantum Information Processing}\ }\textbf {\bibinfo
			{volume} {2}},\ \bibinfo {pages} {433} (\bibinfo {year} {2003})}\BibitemShut
	{NoStop}%
	\bibitem [{\citenamefont {Auccaise}\ \emph {et~al.}(2015)\citenamefont
		{Auccaise}, \citenamefont {{Araujo-Ferreira}}, \citenamefont {Sarthour},
		\citenamefont {Oliveira}, \citenamefont {Bonagamba},\ and\ \citenamefont
		{Roditi}}]{Auccaise2015}%
	\BibitemOpen
	\bibfield  {author} {\bibinfo {author} {\bibfnamefont {R.}~\bibnamefont
			{Auccaise}}, \bibinfo {author} {\bibfnamefont {A.~G.}\ \bibnamefont
			{{Araujo-Ferreira}}}, \bibinfo {author} {\bibfnamefont {R.~S.}\ \bibnamefont
			{Sarthour}}, \bibinfo {author} {\bibfnamefont {I.~S.}\ \bibnamefont
			{Oliveira}}, \bibinfo {author} {\bibfnamefont {T.~J.}\ \bibnamefont
			{Bonagamba}},\ and\ \bibinfo {author} {\bibfnamefont {I.}~\bibnamefont
			{Roditi}},\ }\bibfield  {title} {\bibinfo {title} {Spin squeezing in a
			quadrupolar nuclei {{NMR}} system},\ }\href
	{https://doi.org/10.1103/PhysRevLett.114.043604} {\bibfield  {journal}
		{\bibinfo  {journal} {Physical Review Letters}\ }\textbf {\bibinfo {volume}
			{114}},\ \bibinfo {pages} {043604} (\bibinfo {year} {2015})}\BibitemShut
	{NoStop}%
	\bibitem [{\citenamefont {Aksu~Korkmaz}\ and\ \citenamefont
		{Bulutay}(2016)}]{AksuKorkmaz2016}%
	\BibitemOpen
	\bibfield  {author} {\bibinfo {author} {\bibfnamefont {Y.}~\bibnamefont
			{Aksu~Korkmaz}}\ and\ \bibinfo {author} {\bibfnamefont {C.}~\bibnamefont
			{Bulutay}},\ }\bibfield  {title} {\bibinfo {title} {Nuclear spin squeezing
			via electric quadrupole interaction},\ }\href
	{https://doi.org/10.1103/PhysRevA.93.013812} {\bibfield  {journal} {\bibinfo
			{journal} {Physical Review A}\ }\textbf {\bibinfo {volume} {93}},\ \bibinfo
		{pages} {013812} (\bibinfo {year} {2016})}\BibitemShut {NoStop}%
	\bibitem [{\citenamefont {Rudner}\ \emph {et~al.}(2011)\citenamefont {Rudner},
		\citenamefont {Vandersypen}, \citenamefont {Vuleti{\'c}},\ and\ \citenamefont
		{Levitov}}]{Rudner2011}%
	\BibitemOpen
	\bibfield  {author} {\bibinfo {author} {\bibfnamefont {M.~S.}\ \bibnamefont
			{Rudner}}, \bibinfo {author} {\bibfnamefont {L.~M.}\ \bibnamefont
			{Vandersypen}}, \bibinfo {author} {\bibfnamefont {V.}~\bibnamefont
			{Vuleti{\'c}}},\ and\ \bibinfo {author} {\bibfnamefont {L.~S.}\ \bibnamefont
			{Levitov}},\ }\bibfield  {title} {\bibinfo {title} {Generating entanglement
			and squeezed states of nuclear spins in quantum dots},\ }\href
	{https://doi.org/10.1103/PhysRevLett.107.206806} {\bibfield  {journal}
		{\bibinfo  {journal} {Physical Review Letters}\ }\textbf {\bibinfo {volume}
			{107}},\ \bibinfo {pages} {206806} (\bibinfo {year} {2011})}\BibitemShut
	{NoStop}%
	\bibitem [{\citenamefont {Schuetz}\ \emph {et~al.}(2013)\citenamefont
		{Schuetz}, \citenamefont {Kessler}, \citenamefont {Vandersypen},
		\citenamefont {Cirac},\ and\ \citenamefont {Giedke}}]{Schuetz2013}%
	\BibitemOpen
	\bibfield  {author} {\bibinfo {author} {\bibfnamefont {M.~J.~A.}\
			\bibnamefont {Schuetz}}, \bibinfo {author} {\bibfnamefont {E.~M.}\
			\bibnamefont {Kessler}}, \bibinfo {author} {\bibfnamefont {L.~M.~K.}\
			\bibnamefont {Vandersypen}}, \bibinfo {author} {\bibfnamefont {J.~I.}\
			\bibnamefont {Cirac}},\ and\ \bibinfo {author} {\bibfnamefont
			{G.}~\bibnamefont {Giedke}},\ }\bibfield  {title} {\bibinfo {title}
		{Steady-{{State Entanglement}} in the {{Nuclear Spin Dynamics}} of a {{Double
					Quantum Dot}}},\ }\href {https://doi.org/10.1103/PhysRevLett.111.246802}
	{\bibfield  {journal} {\bibinfo  {journal} {Physical Review Letters}\
		}\textbf {\bibinfo {volume} {111}},\ \bibinfo {pages} {246802} (\bibinfo
		{year} {2013})}\BibitemShut {NoStop}%
	\bibitem [{\citenamefont {Sinatra}(2022)}]{Sinatra2022}%
	\BibitemOpen
	\bibfield  {author} {\bibinfo {author} {\bibfnamefont {A.}~\bibnamefont
			{Sinatra}},\ }\bibfield  {title} {\bibinfo {title} {Spin-squeezed states for
			metrology},\ }\href {https://doi.org/10.1063/5.0084096} {\bibfield  {journal}
		{\bibinfo  {journal} {Applied Physics Letters}\ }\textbf {\bibinfo {volume}
			{120}},\ \bibinfo {pages} {120501} (\bibinfo {year} {2022})}\BibitemShut
	{NoStop}%
	\bibitem [{\citenamefont {Groszkowski}\ \emph {et~al.}(2020)\citenamefont
		{Groszkowski}, \citenamefont {Lau}, \citenamefont {Leroux}, \citenamefont
		{Govia},\ and\ \citenamefont {Clerk}}]{Groszkowski2020}%
	\BibitemOpen
	\bibfield  {author} {\bibinfo {author} {\bibfnamefont {P.}~\bibnamefont
			{Groszkowski}}, \bibinfo {author} {\bibfnamefont {H.-K.}\ \bibnamefont
			{Lau}}, \bibinfo {author} {\bibfnamefont {C.}~\bibnamefont {Leroux}},
		\bibinfo {author} {\bibfnamefont {L.~C.~G.}\ \bibnamefont {Govia}},\ and\
		\bibinfo {author} {\bibfnamefont {A.~A.}\ \bibnamefont {Clerk}},\ }\bibfield
	{title} {\bibinfo {title} {Heisenberg-{{Limited Spin Squeezing}} via
			{{Bosonic Parametric Driving}}},\ }\href
	{https://doi.org/10.1103/PhysRevLett.125.203601} {\bibfield  {journal}
		{\bibinfo  {journal} {Physical Review Letters}\ }\textbf {\bibinfo {volume}
			{125}},\ \bibinfo {pages} {203601} (\bibinfo {year} {2020})}\BibitemShut
	{NoStop}%
	\bibitem [{\citenamefont {Auzinsh}\ \emph {et~al.}(2004)\citenamefont
		{Auzinsh}, \citenamefont {Budker}, \citenamefont {Kimball}, \citenamefont
		{Rochester}, \citenamefont {Stalnaker}, \citenamefont {Sushkov},\ and\
		\citenamefont {Yashchuk}}]{Auzinsh2004}%
	\BibitemOpen
	\bibfield  {author} {\bibinfo {author} {\bibfnamefont {M.}~\bibnamefont
			{Auzinsh}}, \bibinfo {author} {\bibfnamefont {D.}~\bibnamefont {Budker}},
		\bibinfo {author} {\bibfnamefont {D.~F.}\ \bibnamefont {Kimball}}, \bibinfo
		{author} {\bibfnamefont {S.~M.}\ \bibnamefont {Rochester}}, \bibinfo {author}
		{\bibfnamefont {J.~E.}\ \bibnamefont {Stalnaker}}, \bibinfo {author}
		{\bibfnamefont {A.~O.}\ \bibnamefont {Sushkov}},\ and\ \bibinfo {author}
		{\bibfnamefont {V.~V.}\ \bibnamefont {Yashchuk}},\ }\bibfield  {title}
	{\bibinfo {title} {Can a quantum nondemolition measurement improve the
			sensitivity of an atomic magnetometer?},\ }\bibfield  {journal} {\bibinfo
		{journal} {Physical Review Letters}\ }\textbf {\bibinfo {volume} {93}},\
	\href {https://doi.org/10.1103/PhysRevLett.93.173002}
	{10.1103/PhysRevLett.93.173002} (\bibinfo {year} {2004}),\ \Eprint
	{https://arxiv.org/abs/physics/0403097} {arXiv:physics/0403097} \BibitemShut
	{NoStop}%
	\bibitem [{\citenamefont {Ma}\ \emph {et~al.}(2011)\citenamefont {Ma},
		\citenamefont {Wang}, \citenamefont {Sun},\ and\ \citenamefont
		{Nori}}]{Ma2011}%
	\BibitemOpen
	\bibfield  {author} {\bibinfo {author} {\bibfnamefont {J.}~\bibnamefont
			{Ma}}, \bibinfo {author} {\bibfnamefont {X.}~\bibnamefont {Wang}}, \bibinfo
		{author} {\bibfnamefont {C.~P.}\ \bibnamefont {Sun}},\ and\ \bibinfo {author}
		{\bibfnamefont {F.}~\bibnamefont {Nori}},\ }\bibfield  {title} {\bibinfo
		{title} {Quantum spin squeezing},\ }\href
	{https://doi.org/10.1016/j.physrep.2011.08.003} {\bibfield  {journal}
		{\bibinfo  {journal} {Physics Reports}\ }\textbf {\bibinfo {volume} {509}},\
		\bibinfo {pages} {89} (\bibinfo {year} {2011})},\ \Eprint
	{https://arxiv.org/abs/1011.2978} {arXiv:1011.2978} \BibitemShut {NoStop}%
	\bibitem [{\citenamefont {Chin}\ \emph {et~al.}(2012)\citenamefont {Chin},
		\citenamefont {Huelga},\ and\ \citenamefont {Plenio}}]{Chin2012}%
	\BibitemOpen
	\bibfield  {author} {\bibinfo {author} {\bibfnamefont {A.~W.}\ \bibnamefont
			{Chin}}, \bibinfo {author} {\bibfnamefont {S.~F.}\ \bibnamefont {Huelga}},\
		and\ \bibinfo {author} {\bibfnamefont {M.~B.}\ \bibnamefont {Plenio}},\
	}\bibfield  {title} {\bibinfo {title} {Quantum metrology in non-markovian
			environments},\ }\href {https://doi.org/10.1103/PhysRevLett.109.233601}
	{\bibfield  {journal} {\bibinfo  {journal} {Physical Review Letters}\
		}\textbf {\bibinfo {volume} {109}},\ \bibinfo {pages} {233601} (\bibinfo
		{year} {2012})},\ \Eprint {https://arxiv.org/abs/1103.1219} {arXiv:1103.1219}
	\BibitemShut {NoStop}%
	\bibitem [{\citenamefont {Sushkov}(2023)}]{Sushkov2023d}%
	\BibitemOpen
	\bibfield  {author} {\bibinfo {author} {\bibfnamefont {A.~O.}\ \bibnamefont
			{Sushkov}},\ }\bibfield  {title} {\bibinfo {title} {{{EuCl3}}.{{6H2O}} as the
			solid-state platform for searches for new {{P}},{{T-odd}} physics},\ }\href
	{http://arxiv.org/abs/2304.12105} {\bibfield  {journal} {\bibinfo  {journal}
			{arXiv:2304.12105}\ } (\bibinfo {year} {2023})},\ \Eprint
	{https://arxiv.org/abs/2304.12105} {arXiv:2304.12105 [cond-mat]} \BibitemShut
	{NoStop}%
	\bibitem [{\citenamefont {Hosten}\ \emph
		{et~al.}(2016{\natexlab{b}})\citenamefont {Hosten}, \citenamefont
		{Krishnakumar}, \citenamefont {Engelsen},\ and\ \citenamefont
		{Kasevich}}]{Hosten2016a}%
	\BibitemOpen
	\bibfield  {author} {\bibinfo {author} {\bibfnamefont {O.}~\bibnamefont
			{Hosten}}, \bibinfo {author} {\bibfnamefont {R.}~\bibnamefont
			{Krishnakumar}}, \bibinfo {author} {\bibfnamefont {N.~J.}\ \bibnamefont
			{Engelsen}},\ and\ \bibinfo {author} {\bibfnamefont {M.~A.}\ \bibnamefont
			{Kasevich}},\ }\bibfield  {title} {\bibinfo {title} {Quantum phase
			magnification},\ }\href {https://doi.org/10.1126/science.aaf3397} {\bibfield
		{journal} {\bibinfo  {journal} {Science}\ }\textbf {\bibinfo {volume}
			{352}},\ \bibinfo {pages} {1552} (\bibinfo {year}
		{2016}{\natexlab{b}})}\BibitemShut {NoStop}%
	\bibitem [{\citenamefont {Davis}\ \emph {et~al.}(2016)\citenamefont {Davis},
		\citenamefont {Bentsen},\ and\ \citenamefont {{Schleier-Smith}}}]{Davis2016}%
	\BibitemOpen
	\bibfield  {author} {\bibinfo {author} {\bibfnamefont {E.}~\bibnamefont
			{Davis}}, \bibinfo {author} {\bibfnamefont {G.}~\bibnamefont {Bentsen}},\
		and\ \bibinfo {author} {\bibfnamefont {M.}~\bibnamefont {{Schleier-Smith}}},\
	}\bibfield  {title} {\bibinfo {title} {Approaching the {{Heisenberg Limit}}
			without {{Single-Particle Detection}}},\ }\href
	{https://doi.org/10.1103/PhysRevLett.116.053601} {\bibfield  {journal}
		{\bibinfo  {journal} {Physical Review Letters}\ }\textbf {\bibinfo {volume}
			{116}},\ \bibinfo {pages} {053601} (\bibinfo {year} {2016})},\ \Eprint
	{https://arxiv.org/abs/1508.04110} {arXiv:1508.04110} \BibitemShut {NoStop}%
	\bibitem [{\citenamefont {Nolan}\ \emph {et~al.}(2017)\citenamefont {Nolan},
		\citenamefont {Szigeti},\ and\ \citenamefont {Haine}}]{Nolan2017}%
	\BibitemOpen
	\bibfield  {author} {\bibinfo {author} {\bibfnamefont {S.~P.}\ \bibnamefont
			{Nolan}}, \bibinfo {author} {\bibfnamefont {S.~S.}\ \bibnamefont {Szigeti}},\
		and\ \bibinfo {author} {\bibfnamefont {S.~A.}\ \bibnamefont {Haine}},\
	}\bibfield  {title} {\bibinfo {title} {Optimal and {{Robust Quantum Metrology
					Using Interaction-Based Readouts}}},\ }\href
	{https://doi.org/10.1103/PhysRevLett.119.193601} {\bibfield  {journal}
		{\bibinfo  {journal} {Physical Review Letters}\ }\textbf {\bibinfo {volume}
			{119}},\ \bibinfo {pages} {193601} (\bibinfo {year} {2017})},\ \Eprint
	{https://arxiv.org/abs/1703.10417} {arXiv:1703.10417} \BibitemShut {NoStop}%
	\bibitem [{\citenamefont {Leroux}\ \emph {et~al.}(2018)\citenamefont {Leroux},
		\citenamefont {Govia},\ and\ \citenamefont {Clerk}}]{Leroux2018}%
	\BibitemOpen
	\bibfield  {author} {\bibinfo {author} {\bibfnamefont {C.}~\bibnamefont
			{Leroux}}, \bibinfo {author} {\bibfnamefont {L.~C.}\ \bibnamefont {Govia}},\
		and\ \bibinfo {author} {\bibfnamefont {A.~A.}\ \bibnamefont {Clerk}},\
	}\bibfield  {title} {\bibinfo {title} {Enhancing {{Cavity Quantum
					Electrodynamics}} via {{Antisqueezing}}: {{Synthetic Ultrastrong
					Coupling}}},\ }\href {https://doi.org/10.1103/PhysRevLett.120.093602}
	{\bibfield  {journal} {\bibinfo  {journal} {Physical Review Letters}\
		}\textbf {\bibinfo {volume} {120}},\ \bibinfo {pages} {093602} (\bibinfo
		{year} {2018})}\BibitemShut {NoStop}%
	\bibitem [{\citenamefont {Guarrera}\ \emph {et~al.}(2019)\citenamefont
		{Guarrera}, \citenamefont {Gartman}, \citenamefont {Bevilacqua},
		\citenamefont {Barontini},\ and\ \citenamefont {Chalupczak}}]{Guarrera2019}%
	\BibitemOpen
	\bibfield  {author} {\bibinfo {author} {\bibfnamefont {V.}~\bibnamefont
			{Guarrera}}, \bibinfo {author} {\bibfnamefont {R.}~\bibnamefont {Gartman}},
		\bibinfo {author} {\bibfnamefont {G.}~\bibnamefont {Bevilacqua}}, \bibinfo
		{author} {\bibfnamefont {G.}~\bibnamefont {Barontini}},\ and\ \bibinfo
		{author} {\bibfnamefont {W.}~\bibnamefont {Chalupczak}},\ }\bibfield  {title}
	{\bibinfo {title} {Parametric {{Amplification}} and {{Noise Squeezing}} in
			{{Room Temperature Atomic Vapors}}},\ }\href
	{https://doi.org/10.1103/PhysRevLett.123.033601} {\bibfield  {journal}
		{\bibinfo  {journal} {Physical Review Letters}\ }\textbf {\bibinfo {volume}
			{123}},\ \bibinfo {pages} {033601} (\bibinfo {year} {2019})},\ \Eprint
	{https://arxiv.org/abs/1903.00212} {arXiv:1903.00212} \BibitemShut {NoStop}%
	\bibitem [{\citenamefont {Koppenh{\"o}fer}\ \emph {et~al.}(2022)\citenamefont
		{Koppenh{\"o}fer}, \citenamefont {Groszkowski}, \citenamefont {Lau},\ and\
		\citenamefont {Clerk}}]{Koppenhofer2022}%
	\BibitemOpen
	\bibfield  {author} {\bibinfo {author} {\bibfnamefont {M.}~\bibnamefont
			{Koppenh{\"o}fer}}, \bibinfo {author} {\bibfnamefont {P.}~\bibnamefont
			{Groszkowski}}, \bibinfo {author} {\bibfnamefont {H.~K.}\ \bibnamefont
			{Lau}},\ and\ \bibinfo {author} {\bibfnamefont {A.~A.}\ \bibnamefont
			{Clerk}},\ }\bibfield  {title} {\bibinfo {title} {Dissipative {{Superradiant
					Spin Amplifier}} for {{Enhanced Quantum Sensing}}},\ }\href
	{https://doi.org/10.1103/PRXQuantum.3.030330} {\bibfield  {journal} {\bibinfo
			{journal} {PRX Quantum}\ }\textbf {\bibinfo {volume} {3}},\ \bibinfo {pages}
		{030330} (\bibinfo {year} {2022})},\ \Eprint
	{https://arxiv.org/abs/2111.15647} {arXiv:2111.15647} \BibitemShut {NoStop}%
	\bibitem [{\citenamefont {Koppenh{\"o}fer}\ \emph {et~al.}(2023)\citenamefont
		{Koppenh{\"o}fer}, \citenamefont {Groszkowski},\ and\ \citenamefont
		{Clerk}}]{Koppenhofer2023}%
	\BibitemOpen
	\bibfield  {author} {\bibinfo {author} {\bibfnamefont {M.}~\bibnamefont
			{Koppenh{\"o}fer}}, \bibinfo {author} {\bibfnamefont {P.}~\bibnamefont
			{Groszkowski}},\ and\ \bibinfo {author} {\bibfnamefont {A.~A.}\ \bibnamefont
			{Clerk}},\ }\bibfield  {title} {\bibinfo {title} {Squeezed {{Superradiance
					Enables Robust Entanglement-Enhanced Metrology Even}} with {{Highly Imperfect
					Readout}}},\ }\href {https://doi.org/10.1103/PhysRevLett.131.060802}
	{\bibfield  {journal} {\bibinfo  {journal} {Physical Review Letters}\
		}\textbf {\bibinfo {volume} {131}},\ \bibinfo {pages} {060802} (\bibinfo
		{year} {2023})}\BibitemShut {NoStop}%
	\bibitem [{\citenamefont {Malnou}\ \emph
		{et~al.}(2019{\natexlab{b}})\citenamefont {Malnou}, \citenamefont {Palken},
		\citenamefont {Brubaker}, \citenamefont {Vale}, \citenamefont {Hilton},\ and\
		\citenamefont {Lehnert}}]{Malnou2019}%
	\BibitemOpen
	\bibfield  {author} {\bibinfo {author} {\bibfnamefont {M.}~\bibnamefont
			{Malnou}}, \bibinfo {author} {\bibfnamefont {D.~A.}\ \bibnamefont {Palken}},
		\bibinfo {author} {\bibfnamefont {B.~M.}\ \bibnamefont {Brubaker}}, \bibinfo
		{author} {\bibfnamefont {L.~R.}\ \bibnamefont {Vale}}, \bibinfo {author}
		{\bibfnamefont {G.~C.}\ \bibnamefont {Hilton}},\ and\ \bibinfo {author}
		{\bibfnamefont {K.~W.}\ \bibnamefont {Lehnert}},\ }\bibfield  {title}
	{\bibinfo {title} {Squeezed {{Vacuum Used}} to {{Accelerate}} the {{Search}}
			for a {{Weak Classical Signal}}},\ }\href
	{https://doi.org/10.1103/PhysRevX.9.021023} {\bibfield  {journal} {\bibinfo
			{journal} {Physical Review X}\ }\textbf {\bibinfo {volume} {9}},\ \bibinfo
		{pages} {021023} (\bibinfo {year} {2019}{\natexlab{b}})},\ \Eprint
	{https://arxiv.org/abs/1809.06470} {arXiv:1809.06470} \BibitemShut {NoStop}%
	\bibitem [{\citenamefont {Backes}\ \emph
		{et~al.}(2021{\natexlab{b}})\citenamefont {Backes}, \citenamefont {Palken},
		\citenamefont {Kenany}, \citenamefont {Brubaker}, \citenamefont {Cahn},
		\citenamefont {Droster}, \citenamefont {Hilton}, \citenamefont {Ghosh},
		\citenamefont {Jackson}, \citenamefont {Lamoreaux}, \citenamefont {Leder},
		\citenamefont {Lehnert}, \citenamefont {Lewis}, \citenamefont {Malnou},
		\citenamefont {Maruyama}, \citenamefont {Rapidis}, \citenamefont
		{Simanovskaia}, \citenamefont {Singh}, \citenamefont {Speller}, \citenamefont
		{Urdinaran}, \citenamefont {Vale}, \citenamefont {{van Assendelft}},
		\citenamefont {{van Bibber}},\ and\ \citenamefont {Wang}}]{Backes2021}%
	\BibitemOpen
	\bibfield  {author} {\bibinfo {author} {\bibfnamefont {K.~M.}\ \bibnamefont
			{Backes}}, \bibinfo {author} {\bibfnamefont {D.~A.}\ \bibnamefont {Palken}},
		\bibinfo {author} {\bibfnamefont {S.~A.}\ \bibnamefont {Kenany}}, \bibinfo
		{author} {\bibfnamefont {B.~M.}\ \bibnamefont {Brubaker}}, \bibinfo {author}
		{\bibfnamefont {S.~B.}\ \bibnamefont {Cahn}}, \bibinfo {author}
		{\bibfnamefont {A.}~\bibnamefont {Droster}}, \bibinfo {author} {\bibfnamefont
			{G.~C.}\ \bibnamefont {Hilton}}, \bibinfo {author} {\bibfnamefont
			{S.}~\bibnamefont {Ghosh}}, \bibinfo {author} {\bibfnamefont
			{H.}~\bibnamefont {Jackson}}, \bibinfo {author} {\bibfnamefont {S.~K.}\
			\bibnamefont {Lamoreaux}}, \bibinfo {author} {\bibfnamefont {A.~F.}\
			\bibnamefont {Leder}}, \bibinfo {author} {\bibfnamefont {K.~W.}\ \bibnamefont
			{Lehnert}}, \bibinfo {author} {\bibfnamefont {S.~M.}\ \bibnamefont {Lewis}},
		\bibinfo {author} {\bibfnamefont {M.}~\bibnamefont {Malnou}}, \bibinfo
		{author} {\bibfnamefont {R.~H.}\ \bibnamefont {Maruyama}}, \bibinfo {author}
		{\bibfnamefont {N.~M.}\ \bibnamefont {Rapidis}}, \bibinfo {author}
		{\bibfnamefont {M.}~\bibnamefont {Simanovskaia}}, \bibinfo {author}
		{\bibfnamefont {S.}~\bibnamefont {Singh}}, \bibinfo {author} {\bibfnamefont
			{D.~H.}\ \bibnamefont {Speller}}, \bibinfo {author} {\bibfnamefont
			{I.}~\bibnamefont {Urdinaran}}, \bibinfo {author} {\bibfnamefont {L.~R.}\
			\bibnamefont {Vale}}, \bibinfo {author} {\bibfnamefont {E.~C.}\ \bibnamefont
			{{van Assendelft}}}, \bibinfo {author} {\bibfnamefont {K.}~\bibnamefont {{van
					Bibber}}},\ and\ \bibinfo {author} {\bibfnamefont {H.}~\bibnamefont {Wang}},\
	}\bibfield  {title} {\bibinfo {title} {A quantum enhanced search for dark
			matter axions},\ }\href {https://doi.org/10.1038/s41586-021-03226-7}
	{\bibfield  {journal} {\bibinfo  {journal} {Nature}\ }\textbf {\bibinfo
			{volume} {590}},\ \bibinfo {pages} {238} (\bibinfo {year}
		{2021}{\natexlab{b}})},\ \Eprint {https://arxiv.org/abs/2008.01853}
	{arXiv:2008.01853} \BibitemShut {NoStop}%
	\bibitem [{\citenamefont {Aybas}\ \emph
		{et~al.}(2021{\natexlab{b}})\citenamefont {Aybas}, \citenamefont {Adam},
		\citenamefont {Blumenthal}, \citenamefont {Gramolin}, \citenamefont
		{Johnson}, \citenamefont {Kleyheeg}, \citenamefont {Afach}, \citenamefont
		{Blanchard}, \citenamefont {Centers}, \citenamefont {Garcon}, \citenamefont
		{Engler}, \citenamefont {Figueroa}, \citenamefont {Sendra}, \citenamefont
		{Wickenbrock}, \citenamefont {Lawson}, \citenamefont {Wang}, \citenamefont
		{Wu}, \citenamefont {Luo}, \citenamefont {Mani}, \citenamefont {Mauskopf},
		\citenamefont {Graham}, \citenamefont {Rajendran}, \citenamefont {Kimball},
		\citenamefont {Budker},\ and\ \citenamefont {Sushkov}}]{Aybas2021a}%
	\BibitemOpen
	\bibfield  {author} {\bibinfo {author} {\bibfnamefont {D.}~\bibnamefont
			{Aybas}}, \bibinfo {author} {\bibfnamefont {J.}~\bibnamefont {Adam}},
		\bibinfo {author} {\bibfnamefont {E.}~\bibnamefont {Blumenthal}}, \bibinfo
		{author} {\bibfnamefont {A.~V.}\ \bibnamefont {Gramolin}}, \bibinfo {author}
		{\bibfnamefont {D.}~\bibnamefont {Johnson}}, \bibinfo {author} {\bibfnamefont
			{A.}~\bibnamefont {Kleyheeg}}, \bibinfo {author} {\bibfnamefont
			{S.}~\bibnamefont {Afach}}, \bibinfo {author} {\bibfnamefont {J.~W.}\
			\bibnamefont {Blanchard}}, \bibinfo {author} {\bibfnamefont {G.~P.}\
			\bibnamefont {Centers}}, \bibinfo {author} {\bibfnamefont {A.}~\bibnamefont
			{Garcon}}, \bibinfo {author} {\bibfnamefont {M.}~\bibnamefont {Engler}},
		\bibinfo {author} {\bibfnamefont {N.~L.}\ \bibnamefont {Figueroa}}, \bibinfo
		{author} {\bibfnamefont {M.~G.}\ \bibnamefont {Sendra}}, \bibinfo {author}
		{\bibfnamefont {A.}~\bibnamefont {Wickenbrock}}, \bibinfo {author}
		{\bibfnamefont {M.}~\bibnamefont {Lawson}}, \bibinfo {author} {\bibfnamefont
			{T.}~\bibnamefont {Wang}}, \bibinfo {author} {\bibfnamefont {T.}~\bibnamefont
			{Wu}}, \bibinfo {author} {\bibfnamefont {H.}~\bibnamefont {Luo}}, \bibinfo
		{author} {\bibfnamefont {H.}~\bibnamefont {Mani}}, \bibinfo {author}
		{\bibfnamefont {P.}~\bibnamefont {Mauskopf}}, \bibinfo {author}
		{\bibfnamefont {P.~W.}\ \bibnamefont {Graham}}, \bibinfo {author}
		{\bibfnamefont {S.}~\bibnamefont {Rajendran}}, \bibinfo {author}
		{\bibfnamefont {D.~F.}\ \bibnamefont {Kimball}}, \bibinfo {author}
		{\bibfnamefont {D.}~\bibnamefont {Budker}},\ and\ \bibinfo {author}
		{\bibfnamefont {A.~O.}\ \bibnamefont {Sushkov}},\ }\bibfield  {title}
	{\bibinfo {title} {Search for {{Axionlike Dark Matter Using Solid-State
					Nuclear Magnetic Resonance}}},\ }\href
	{https://doi.org/10.1103/PhysRevLett.126.141802} {\bibfield  {journal}
		{\bibinfo  {journal} {Physical Review Letters}\ }\textbf {\bibinfo {volume}
			{126}},\ \bibinfo {pages} {141802} (\bibinfo {year} {2021}{\natexlab{b}})},\
	\Eprint {https://arxiv.org/abs/2101.01241} {arXiv:2101.01241} \BibitemShut
	{NoStop}%
	\bibitem [{\citenamefont {Garcon}\ \emph {et~al.}(2019)\citenamefont {Garcon},
		\citenamefont {Blanchard}, \citenamefont {Centers}, \citenamefont {Figueroa},
		\citenamefont {Graham}, \citenamefont {Jackson~Kimball}, \citenamefont
		{Rajendran}, \citenamefont {Sushkov}, \citenamefont {Stadnik}, \citenamefont
		{Wickenbrock}, \citenamefont {Wu},\ and\ \citenamefont
		{Budker}}]{Garcon2019}%
	\BibitemOpen
	\bibfield  {author} {\bibinfo {author} {\bibfnamefont {A.}~\bibnamefont
			{Garcon}}, \bibinfo {author} {\bibfnamefont {J.~W.}\ \bibnamefont
			{Blanchard}}, \bibinfo {author} {\bibfnamefont {G.~P.}\ \bibnamefont
			{Centers}}, \bibinfo {author} {\bibfnamefont {N.~L.}\ \bibnamefont
			{Figueroa}}, \bibinfo {author} {\bibfnamefont {P.~W.}\ \bibnamefont
			{Graham}}, \bibinfo {author} {\bibfnamefont {D.~F.}\ \bibnamefont
			{Jackson~Kimball}}, \bibinfo {author} {\bibfnamefont {S.}~\bibnamefont
			{Rajendran}}, \bibinfo {author} {\bibfnamefont {A.~O.}\ \bibnamefont
			{Sushkov}}, \bibinfo {author} {\bibfnamefont {Y.~V.}\ \bibnamefont
			{Stadnik}}, \bibinfo {author} {\bibfnamefont {A.}~\bibnamefont
			{Wickenbrock}}, \bibinfo {author} {\bibfnamefont {T.}~\bibnamefont {Wu}},\
		and\ \bibinfo {author} {\bibfnamefont {D.}~\bibnamefont {Budker}},\
	}\bibfield  {title} {\bibinfo {title} {Constraints on bosonic dark matter
			from ultralow-field nuclear magnetic resonance},\ }\href
	{https://doi.org/10.1126/sciadv.aax4539} {\bibfield  {journal} {\bibinfo
			{journal} {Science Advances}\ }\textbf {\bibinfo {volume} {5}},\ \bibinfo
		{pages} {eaax4539} (\bibinfo {year} {2019})},\ \Eprint
	{https://arxiv.org/abs/1902.04644} {arXiv:1902.04644} \BibitemShut {NoStop}%
	\bibitem [{\citenamefont {Blais}\ \emph {et~al.}(2021)\citenamefont {Blais},
		\citenamefont {Grimsmo}, \citenamefont {Girvin},\ and\ \citenamefont
		{Wallraff}}]{Blais2021}%
	\BibitemOpen
	\bibfield  {author} {\bibinfo {author} {\bibfnamefont {A.}~\bibnamefont
			{Blais}}, \bibinfo {author} {\bibfnamefont {A.~L.}\ \bibnamefont {Grimsmo}},
		\bibinfo {author} {\bibfnamefont {S.~M.}\ \bibnamefont {Girvin}},\ and\
		\bibinfo {author} {\bibfnamefont {A.}~\bibnamefont {Wallraff}},\ }\bibfield
	{title} {\bibinfo {title} {Circuit quantum electrodynamics},\ }\href
	{https://doi.org/10.1103/RevModPhys.93.025005} {\bibfield  {journal}
		{\bibinfo  {journal} {Reviews of Modern Physics}\ }\textbf {\bibinfo {volume}
			{93}},\ \bibinfo {pages} {025005} (\bibinfo {year} {2021})}\BibitemShut
	{NoStop}%
	\bibitem [{\citenamefont {Dicke}(1954)}]{Dicke1954}%
	\BibitemOpen
	\bibfield  {author} {\bibinfo {author} {\bibfnamefont {R.~H.}\ \bibnamefont
			{Dicke}},\ }\bibfield  {title} {\bibinfo {title} {Coherence in {{Spontaneous
					Radiation Processes}}},\ }\href {https://doi.org/10.1103/PhysRev.93.99}
	{\bibfield  {journal} {\bibinfo  {journal} {Physical Review}\ }\textbf
		{\bibinfo {volume} {93}},\ \bibinfo {pages} {99} (\bibinfo {year}
		{1954})}\BibitemShut {NoStop}%
	\bibitem [{\citenamefont {Kohler}\ \emph {et~al.}(2017)\citenamefont {Kohler},
		\citenamefont {Spethmann}, \citenamefont {Schreppler},\ and\ \citenamefont
		{{Stamper-Kurn}}}]{Kohler2017}%
	\BibitemOpen
	\bibfield  {author} {\bibinfo {author} {\bibfnamefont {J.}~\bibnamefont
			{Kohler}}, \bibinfo {author} {\bibfnamefont {N.}~\bibnamefont {Spethmann}},
		\bibinfo {author} {\bibfnamefont {S.}~\bibnamefont {Schreppler}},\ and\
		\bibinfo {author} {\bibfnamefont {D.~M.}\ \bibnamefont {{Stamper-Kurn}}},\
	}\bibfield  {title} {\bibinfo {title} {Cavity-{{Assisted Measurement}} and
			{{Coherent Control}} of {{Collective Atomic Spin Oscillators}}},\ }\href
	{https://doi.org/10.1103/PhysRevLett.118.063604} {\bibfield  {journal}
		{\bibinfo  {journal} {Physical Review Letters}\ }\textbf {\bibinfo {volume}
			{118}},\ \bibinfo {pages} {063604} (\bibinfo {year} {2017})},\ \Eprint
	{https://arxiv.org/abs/1607.07941} {arXiv:1607.07941} \BibitemShut {NoStop}%
	\bibitem [{\citenamefont {Jin}\ \emph {et~al.}(2009)\citenamefont {Jin},
		\citenamefont {Liu},\ and\ \citenamefont {Liu}}]{Jin2009}%
	\BibitemOpen
	\bibfield  {author} {\bibinfo {author} {\bibfnamefont {G.-R.}\ \bibnamefont
			{Jin}}, \bibinfo {author} {\bibfnamefont {Y.-C.}\ \bibnamefont {Liu}},\ and\
		\bibinfo {author} {\bibfnamefont {W.-M.}\ \bibnamefont {Liu}},\ }\bibfield
	{title} {\bibinfo {title} {Spin squeezing in a generalized one-axis twisting
			model},\ }\href {https://doi.org/10.1088/1367-2630/11/7/073049} {\bibfield
		{journal} {\bibinfo  {journal} {New Journal of Physics}\ }\textbf {\bibinfo
			{volume} {11}},\ \bibinfo {pages} {073049} (\bibinfo {year}
		{2009})}\BibitemShut {NoStop}%
	\bibitem [{\citenamefont {Pezz{\`e}}\ \emph {et~al.}(2018)\citenamefont
		{Pezz{\`e}}, \citenamefont {Smerzi}, \citenamefont {Oberthaler},
		\citenamefont {Schmied},\ and\ \citenamefont {Treutlein}}]{Pezze2018}%
	\BibitemOpen
	\bibfield  {author} {\bibinfo {author} {\bibfnamefont {L.}~\bibnamefont
			{Pezz{\`e}}}, \bibinfo {author} {\bibfnamefont {A.}~\bibnamefont {Smerzi}},
		\bibinfo {author} {\bibfnamefont {M.~K.}\ \bibnamefont {Oberthaler}},
		\bibinfo {author} {\bibfnamefont {R.}~\bibnamefont {Schmied}},\ and\ \bibinfo
		{author} {\bibfnamefont {P.}~\bibnamefont {Treutlein}},\ }\bibfield  {title}
	{\bibinfo {title} {Quantum metrology with nonclassical states of atomic
			ensembles},\ }\href {https://doi.org/10.1103/RevModPhys.90.035005} {\bibfield
		{journal} {\bibinfo  {journal} {Reviews of Modern Physics}\ }\textbf
		{\bibinfo {volume} {90}},\ \bibinfo {pages} {035005} (\bibinfo {year}
		{2018})},\ \Eprint {https://arxiv.org/abs/1609.01609} {arXiv:1609.01609}
	\BibitemShut {NoStop}%
	\bibitem [{\citenamefont {Hu}\ \emph {et~al.}(2017)\citenamefont {Hu},
		\citenamefont {Chen}, \citenamefont {Vendeiro}, \citenamefont {Urvoy},
		\citenamefont {Braverman},\ and\ \citenamefont {Vuleti{\'c}}}]{Hu2017}%
	\BibitemOpen
	\bibfield  {author} {\bibinfo {author} {\bibfnamefont {J.}~\bibnamefont
			{Hu}}, \bibinfo {author} {\bibfnamefont {W.}~\bibnamefont {Chen}}, \bibinfo
		{author} {\bibfnamefont {Z.}~\bibnamefont {Vendeiro}}, \bibinfo {author}
		{\bibfnamefont {A.}~\bibnamefont {Urvoy}}, \bibinfo {author} {\bibfnamefont
			{B.}~\bibnamefont {Braverman}},\ and\ \bibinfo {author} {\bibfnamefont
			{V.}~\bibnamefont {Vuleti{\'c}}},\ }\bibfield  {title} {\bibinfo {title}
		{Vacuum spin squeezing},\ }\href {https://doi.org/10.1103/PhysRevA.96.050301}
	{\bibfield  {journal} {\bibinfo  {journal} {Physical Review A}\ }\textbf
		{\bibinfo {volume} {96}},\ \bibinfo {pages} {050301} (\bibinfo {year}
		{2017})},\ \Eprint {https://arxiv.org/abs/1703.02439} {arXiv:1703.02439}
	\BibitemShut {NoStop}%
	\bibitem [{\citenamefont {S{\o}rensen}\ \emph {et~al.}(2001)\citenamefont
		{S{\o}rensen}, \citenamefont {Duan}, \citenamefont {Cirac},\ and\
		\citenamefont {Zoller}}]{Sorensen2001}%
	\BibitemOpen
	\bibfield  {author} {\bibinfo {author} {\bibfnamefont {A.}~\bibnamefont
			{S{\o}rensen}}, \bibinfo {author} {\bibfnamefont {L.-M.}\ \bibnamefont
			{Duan}}, \bibinfo {author} {\bibfnamefont {J.~I.}\ \bibnamefont {Cirac}},\
		and\ \bibinfo {author} {\bibfnamefont {P.}~\bibnamefont {Zoller}},\
	}\bibfield  {title} {\bibinfo {title} {Many-particle entanglement with
			{{Bose}}--{{Einstein}} condensates},\ }\href
	{https://doi.org/10.1038/35051038} {\bibfield  {journal} {\bibinfo  {journal}
			{Nature}\ }\textbf {\bibinfo {volume} {409}},\ \bibinfo {pages} {63}
		(\bibinfo {year} {2001})}\BibitemShut {NoStop}%
	\bibitem [{\citenamefont {Bracker}\ \emph {et~al.}(2005)\citenamefont
		{Bracker}, \citenamefont {Stinaff}, \citenamefont {Gammon}, \citenamefont
		{Ware}, \citenamefont {Tischler}, \citenamefont {Shabaev}, \citenamefont
		{Efros}, \citenamefont {Park}, \citenamefont {Gershoni}, \citenamefont
		{Korenev},\ and\ \citenamefont {Merkulov}}]{Bracker2005}%
	\BibitemOpen
	\bibfield  {author} {\bibinfo {author} {\bibfnamefont {A.~S.}\ \bibnamefont
			{Bracker}}, \bibinfo {author} {\bibfnamefont {E.~A.}\ \bibnamefont
			{Stinaff}}, \bibinfo {author} {\bibfnamefont {D.}~\bibnamefont {Gammon}},
		\bibinfo {author} {\bibfnamefont {M.~E.}\ \bibnamefont {Ware}}, \bibinfo
		{author} {\bibfnamefont {J.~G.}\ \bibnamefont {Tischler}}, \bibinfo {author}
		{\bibfnamefont {A.}~\bibnamefont {Shabaev}}, \bibinfo {author} {\bibfnamefont
			{{\relax Al}.~L.}\ \bibnamefont {Efros}}, \bibinfo {author} {\bibfnamefont
			{D.}~\bibnamefont {Park}}, \bibinfo {author} {\bibfnamefont {D.}~\bibnamefont
			{Gershoni}}, \bibinfo {author} {\bibfnamefont {V.~L.}\ \bibnamefont
			{Korenev}},\ and\ \bibinfo {author} {\bibfnamefont {I.~A.}\ \bibnamefont
			{Merkulov}},\ }\bibfield  {title} {\bibinfo {title} {Optical {{Pumping}} of
			the {{Electronic}} and {{Nuclear Spin}} of {{Single Charge-Tunable Quantum
					Dots}}},\ }\href {https://doi.org/10.1103/PhysRevLett.94.047402} {\bibfield
		{journal} {\bibinfo  {journal} {Physical Review Letters}\ }\textbf {\bibinfo
			{volume} {94}},\ \bibinfo {pages} {047402} (\bibinfo {year}
		{2005})}\BibitemShut {NoStop}%
	\bibitem [{\citenamefont {Matsuzaki}\ \emph {et~al.}(2011)\citenamefont
		{Matsuzaki}, \citenamefont {Benjamin},\ and\ \citenamefont
		{Fitzsimons}}]{Matsuzaki2011}%
	\BibitemOpen
	\bibfield  {author} {\bibinfo {author} {\bibfnamefont {Y.}~\bibnamefont
			{Matsuzaki}}, \bibinfo {author} {\bibfnamefont {S.~C.}\ \bibnamefont
			{Benjamin}},\ and\ \bibinfo {author} {\bibfnamefont {J.}~\bibnamefont
			{Fitzsimons}},\ }\bibfield  {title} {\bibinfo {title} {Magnetic field sensing
			beyond the standard quantum limit under the effect of decoherence},\ }\href
	{https://doi.org/10.1103/PhysRevA.84.012103} {\bibfield  {journal} {\bibinfo
			{journal} {Physical Review A}\ }\textbf {\bibinfo {volume} {84}},\ \bibinfo
		{pages} {012103} (\bibinfo {year} {2011})}\BibitemShut {NoStop}%
	\bibitem [{\citenamefont {Wang}\ \emph {et~al.}(2010)\citenamefont {Wang},
		\citenamefont {Miranowicz}, \citenamefont {Liu}, \citenamefont {Sun},\ and\
		\citenamefont {Nori}}]{Wang2010}%
	\BibitemOpen
	\bibfield  {author} {\bibinfo {author} {\bibfnamefont {X.}~\bibnamefont
			{Wang}}, \bibinfo {author} {\bibfnamefont {A.}~\bibnamefont {Miranowicz}},
		\bibinfo {author} {\bibfnamefont {Y.-x.}\ \bibnamefont {Liu}}, \bibinfo
		{author} {\bibfnamefont {C.~P.}\ \bibnamefont {Sun}},\ and\ \bibinfo {author}
		{\bibfnamefont {F.}~\bibnamefont {Nori}},\ }\bibfield  {title} {\bibinfo
		{title} {Sudden vanishing of spin squeezing under decoherence},\ }\href
	{https://doi.org/10.1103/PhysRevA.81.022106} {\bibfield  {journal} {\bibinfo
			{journal} {Physical Review A}\ }\textbf {\bibinfo {volume} {81}},\ \bibinfo
		{pages} {022106} (\bibinfo {year} {2010})}\BibitemShut {NoStop}%
	\bibitem [{\citenamefont {{Lewis-Swan}}\ \emph {et~al.}(2018)\citenamefont
		{{Lewis-Swan}}, \citenamefont {Norcia}, \citenamefont {Cline}, \citenamefont
		{Thompson},\ and\ \citenamefont {Rey}}]{Lewis-Swan2018}%
	\BibitemOpen
	\bibfield  {author} {\bibinfo {author} {\bibfnamefont {R.~J.}\ \bibnamefont
			{{Lewis-Swan}}}, \bibinfo {author} {\bibfnamefont {M.~A.}\ \bibnamefont
			{Norcia}}, \bibinfo {author} {\bibfnamefont {J.~R.}\ \bibnamefont {Cline}},
		\bibinfo {author} {\bibfnamefont {J.~K.}\ \bibnamefont {Thompson}},\ and\
		\bibinfo {author} {\bibfnamefont {A.~M.}\ \bibnamefont {Rey}},\ }\bibfield
	{title} {\bibinfo {title} {Robust {{Spin Squeezing}} via {{Photon-Mediated
					Interactions}} on an {{Optical Clock Transition}}},\ }\href
	{https://doi.org/10.1103/PhysRevLett.121.070403} {\bibfield  {journal}
		{\bibinfo  {journal} {Physical Review Letters}\ }\textbf {\bibinfo {volume}
			{121}},\ \bibinfo {pages} {070403} (\bibinfo {year} {2018})},\ \Eprint
	{https://arxiv.org/abs/1804.06784} {arXiv:1804.06784} \BibitemShut {NoStop}%
	\bibitem [{\citenamefont {Huelga}\ \emph {et~al.}(1997)\citenamefont {Huelga},
		\citenamefont {Macchiavello}, \citenamefont {Pellizzari}, \citenamefont
		{Ekert}, \citenamefont {Plenio},\ and\ \citenamefont {Cirac}}]{Huelga1997}%
	\BibitemOpen
	\bibfield  {author} {\bibinfo {author} {\bibfnamefont {S.~F.}\ \bibnamefont
			{Huelga}}, \bibinfo {author} {\bibfnamefont {C.}~\bibnamefont
			{Macchiavello}}, \bibinfo {author} {\bibfnamefont {T.}~\bibnamefont
			{Pellizzari}}, \bibinfo {author} {\bibfnamefont {A.~K.}\ \bibnamefont
			{Ekert}}, \bibinfo {author} {\bibfnamefont {M.~B.}\ \bibnamefont {Plenio}},\
		and\ \bibinfo {author} {\bibfnamefont {J.~I.}\ \bibnamefont {Cirac}},\
	}\bibfield  {title} {\bibinfo {title} {Improvement of frequency standards
			with quantum entanglement},\ }\href
	{https://doi.org/10.1103/PhysRevLett.79.3865} {\bibfield  {journal} {\bibinfo
			{journal} {Physical Review Letters}\ }\textbf {\bibinfo {volume} {79}},\
		\bibinfo {pages} {3865} (\bibinfo {year} {1997})}\BibitemShut {NoStop}%
	\bibitem [{\citenamefont {Agarwal}\ \emph {et~al.}(1997)\citenamefont
		{Agarwal}, \citenamefont {Puri},\ and\ \citenamefont {Singh}}]{Agarwal1997}%
	\BibitemOpen
	\bibfield  {author} {\bibinfo {author} {\bibfnamefont {G.~S.}\ \bibnamefont
			{Agarwal}}, \bibinfo {author} {\bibfnamefont {R.~R.}\ \bibnamefont {Puri}},\
		and\ \bibinfo {author} {\bibfnamefont {R.~P.}\ \bibnamefont {Singh}},\
	}\bibfield  {title} {\bibinfo {title} {Atomic
			{{Schr}}{\textbackslash}"odinger cat states},\ }\href
	{https://doi.org/10.1103/PhysRevA.56.2249} {\bibfield  {journal} {\bibinfo
			{journal} {Physical Review A}\ }\textbf {\bibinfo {volume} {56}},\ \bibinfo
		{pages} {2249} (\bibinfo {year} {1997})}\BibitemShut {NoStop}%
	\bibitem [{\citenamefont {Andr{\'e}}\ \emph {et~al.}(2004)\citenamefont
		{Andr{\'e}}, \citenamefont {S{\o}rensen},\ and\ \citenamefont
		{Lukin}}]{Andre2004}%
	\BibitemOpen
	\bibfield  {author} {\bibinfo {author} {\bibfnamefont {A.}~\bibnamefont
			{Andr{\'e}}}, \bibinfo {author} {\bibfnamefont {A.~S.}\ \bibnamefont
			{S{\o}rensen}},\ and\ \bibinfo {author} {\bibfnamefont {M.~D.}\ \bibnamefont
			{Lukin}},\ }\bibfield  {title} {\bibinfo {title} {Stability of {{Atomic
					Clocks Based}} on {{Entangled Atoms}}},\ }\href
	{https://doi.org/10.1103/PhysRevLett.92.230801} {\bibfield  {journal}
		{\bibinfo  {journal} {Physical Review Letters}\ }\textbf {\bibinfo {volume}
			{92}},\ \bibinfo {pages} {230801} (\bibinfo {year} {2004})}\BibitemShut
	{NoStop}%
\end{thebibliography}

%

\clearpage
\onecolumngrid

\appendix

\section{Derivation of the Rotating Frame Hamiltonian}
\label{app:Sx2_derivation}
Here we present a brief derivation of the rotating frame Hamiltonian via the Magnus expansion, and the effective spin Hamiltonian in the dispersive regime via a Schrieffer–Wolff transformation.

We begin with the Dicke model in the lab frame:
\begin{equation}
	H_\mathrm{lab}\left(t\right)=\left[\omega_{s}+\Delta\left(t\right)\right]a^{\dagger}a+\omega_{s}S_{z}+g\left(t\right)\left(a+a^{\dagger}\right)\left(S_{+}+S_{-}\right)\label{eq:Hamiltonian-2}
\end{equation}
The Larmor frequency is $\omega_s$, the circuit frequency is $\omega_s+\Delta\left(t\right)$ where $\Delta\left(t\right)$ is a time-dependent detuning, and the time-dependent single spin bare coupling is $g\left(t\right)$. We transform to the rotating frame defined by the operator:
\begin{align}
	U=\exp\left(-i\left[\omega_{s}S_{z}+\omega_{s}a^{\dagger}a\right]t\right)
\end{align}
Which gives a rotating frame Hamiltonian of:
\begin{equation}
	H_{u}\left(t\right)=\Delta\left(t\right)a^{\dagger}a+g\left(t\right)\left(a\exp\left(i\omega_{s}t\right)+a^{\dagger}\exp\left(-i\omega_{s}t\right)\right)\left(S_{+}\exp\left(-i\omega_{s}t\right)+S_{-}\exp\left(i\omega_{s}t\right)\right)\label{eq:Rotating_Hamiltonian}
\end{equation}

If the time dependent terms $\Delta\left(t\right)$ and $g\left(t\right)$ are periodic with period $\omega_s$, then this problem can be well approximated with a Magnus expansion which rewrites the time dependent Hamiltonian as a power series in $1/\omega_s$ of time independent Hamiltonians. Thus, we assume $\Delta\left(t\right)$ and $g\left(t\right)$ are periodic in $\omega_s$ and define their Fourier expansion:
\begin{align}
	g\left(t\right) & =\sum_{l=-\infty}^{\infty}g_{l}\exp\left(il\omega_{s}t\right)\nonumber \\
	\Delta\left(t\right) & =\sum_{l=-\infty}^{\infty}\Delta_{l}\exp\left(il\omega_{s}t\right)\label{eq:Fourier_transforms}
\end{align}
The Magnus expansion gives:
\begin{equation}
	H_{\mathrm{eff}}=\sum_{n=0}^{\infty}\mathcal{H}_{n},\:with\:\mathcal{H}_{n+1}\ll\mathcal{H}_{n}\label{eq:H_eff}
\end{equation}
Where each $\mathcal{H}_{n}$ is time independent. The first few terms are given by:
\begin{align}
	\mathcal{H}_{0} & =H_{0}\nonumber \\
	\mathcal{H}_{1} & =\frac{1}{\omega_{s}}\sum_{l\geq1}\frac{1}{l}\left[H_{l},H_{-l}\right]\nonumber \\
	\vdots\label{eq:Magnus_series}
\end{align}
With the Hamiltonian at each order:
\begin{equation}
	H_{l}=\Delta_{l}a^{\dagger}a+g_{l}\left[aS_{+}+a^{\dagger}S_{-}\right]+g_{l+2}a^{\dagger}S_{+}+g_{l-2}aS_{-}\label{eq:H_l}
\end{equation}

To get the leading order Hamiltonian, we set the phase of the signal such that $g_2 = g_{-2}$ to give:
\begin{align}
	H_{\mathrm{eff}} = \Delta_{0}a^{\dagger}a+g_{0}\left(aS_{+}+a^{\dagger}S_{-}\right)+g_{2}\left(a^{\dagger}S_{+}+aS_{-} \right)\label{eq:Main}
\end{align}
Which gives Eq.~(\ref{eq:H_Dicke_rotating_eff}) in the main text for $g_2 = 0$, and Eq.~(\ref{eq:H_Dicke_rotating_modulated}) when $g_0=g_2=g$.

Next, we want to derive an effective spin Hamiltonian in the dispersive regime, $g_0\sqrt{N \bar{n}}, g_2\sqrt{N \bar{n}} \ll \Delta_0$. When the detuning is large, the spin and photon modes mostly decouple, allowing for an effective Hamiltonian on just the spins that arises from spin-spin interactions mediated by the circuit. We start with the leading order Hamiltonian:
\begin{equation}
	H_{eff}\cong H_{0}=\Delta_{0}a^{\dagger}a+g_{0}\left[aS_{+}+a^{\dagger}S_{-}\right]+g_{2}a^{\dagger}S_{+}+g_{-2}aS_{-}
\end{equation}

Now we perform a Shrieffer-Wolff transformation to derive an effective Hamiltonian $H_{SW}$
\begin{equation}
	H_{0}\rightarrow\exp\left(P\right)H_{0}\exp\left(-P\right)=H_{SW},\label{eq:Schrefer_Wolf}
\end{equation}
with the transformation operator:
\begin{equation}
	P=\frac{1}{\Delta_{0}}\left(g_{0}\left[a^{\dagger}S_{-}-aS_{+}\right]+g_{2}a^{\dagger}S_{+}-g_{-2}aS_{-}\right).\label{eq:Screiffer_wolf}
\end{equation}
Let us drop constant terms and assume the following form for the Fourier coefficients of $g(t)$:
\begin{align}
	g_{2} & =g_{-2},\:g_{0}=\left(1+\varepsilon\right)g_{2}.
\end{align}
We introduce the small parameter $\epsilon$ to allow for small errors in the modulation scheme and explore how such errors affect the effective Hamiltonian.
We find that the Shrieffer-Wolff effective Hamiltonian is:
\begin{align}
	H_{SW} & =\frac{g_{2}^{2}}{\Delta_{0}}\left[\varepsilon^{2}S_{z}^{2} - 4\left(1+\varepsilon\right)S_{x}^{2}\right]
\end{align}

For the case when $g_0 = g_2$, then $\varepsilon = 0$ and we recover the Hamiltonian in Eq.~(\ref{eq:H_modulated_Sx2}). If there are small errors in the modulation scheme so that $0 < \left| \varepsilon \right| \ll 1$, then the Hamiltonian is still nearly a OAT Hamiltonian since the other term scales as $\varepsilon^2$ and it can be shown that a maximum squeezing of $\xi^2 \sim \varepsilon^2$ can still be achieved. Thus, the scheme is robust against small errors in the modulation, provided $0 < \left| g_0 - g_2 \right| \ll g_0$.

\section{Square Wave Modulation Scheme}
\label{app:square_modulation}
In the main text we presented a modulation function $g(t) = g_0 (1 + 2 \cos(2 \omega_s t)) $ that achieves $g_0 = g_2$, giving the Hamiltonian proportional to $S_x^2$. In practice, this modulation may be difficult to achieve since it requires continuous variation of the coupling.

Experimentally, it may be easier to implement square wave modulation where the coupling $g(t)$ modulates between two constant values, $g$ and $-\alpha g$, with duty cycle $d$ at frequency $2\omega_s$, where $0 \leq \alpha \leq 1$ is a dimensionless constant. The Fourier components are given by:
\begin{align}
	g_0 &= g( d(1 + \alpha ) - \alpha ) \\
	g_2 &= g\frac{1 + \alpha}{\pi} \sin ( \pi d)
\end{align}
For a given choice of $\alpha$, the value of $d$ can be determined to satisfy $g_0 = g_2$, giving a constraint on these parameters.

When the degree of squeezing is limited by dephasing, we want to maximize the squeezing rate to achieve as much squeezing as possible in as short a time as possible. So, we want to maximize $g_0$ while meeting the constraint $g_0 = g_2$. Performing this constrained optimization numerically, we find optimal values of $\alpha = .95$ and $d = .73$ giving $g_0 = 0.47g$

When the degree of squeezing is instead limited by circuit noise, the optimum parameters will change. This is because the circuit noise rate is proportional to the average squared coupling: $\bar{\gamma}_c \propto \overline{g(t)^2}$. Since the squeezing strength is proportional to $g_0^2$, we now want to maximize $g_0^2/\overline{g(t)^2}$ to achieve a large squeezing strength but weak circuit noise strength, while still satisfying the constraint $g_0 = g_2$. Performing this constrained optimization numerically, we find the optimal parameter values are $\alpha=.22$, $d=.50$, corresponding to $g_0 \approx 0.39 g$ and $\overline{g(t)^2}\approx 0.52 g^2$.

Although the parameters are quite different in each regime, the values of $g_0$ only differ by about $20\%$. Thus, for simplicity we use the latter parameters of $\alpha=.22$, $d=.50$ in the main text.

\section{The effect of inhomogeneous spin coupling}
\label{app:inhomogeneous}

Throughout this work we have worked in the collective spin basis which assumes all spin couplings are equal. In practice, there will be inhomogeneous variation of the spin coupling for different spins in the ensemble. Here, we examine how this variation in the spin coupling affects squeezing and show that in the large $N$ limit, reasonable variation in the coupling has negligible effect on the creation of the squeezed state.

In deriving the spin-circuit coupling, Eq.~(\ref{eq:240}), we assumed that the magnetic field generated in the coil is constant across the entire sample. However, for a finite inductor, the magnetic field will vary radially and axially across the sample. For example, modeling the inductor as a cylindrical sheet of current, the on-axis field at the ends of the sample will be smaller than in the middle of the sample by a factor of $\approx \sqrt{5/2} \approx 1.6$. Thus, spins at the edges of the sample will couple less strongly to the circuit, and hence less strongly to other spins in the dispersive regime.

We model this with the following effective spin Hamiltonian in the rotating frame:
\begin{align}
	\label{eq:H_vary_coupling}
	H = -\sum_{i \neq j} g_{ij} \sigma_x^{i} \sigma_x^{j}
\end{align}
Where the couplings $g_{ij}$ can vary, though we require $g_{ij}=g_{ji}$. When $g_{ij}=g_0$, then up to constant terms and factors of order unity we recover the Hamiltonian $H= -g_e S_x^2$ in Eq.~(\ref{eq:H_modulated_Sx2}).

To determine how variation in the coupling constants will affect squeezing, we will calculate the squeezing parameter under the Hamiltonian eq.~(\ref{eq:H_vary_coupling}) and then assume for simplicity a Gaussian distribution of the couplings $g_{ij}$ and compare to the case of uniform coupling. We assume the initial spin state is along the z-axis with unity polarization, so that the expectations of the operators are:
\begin{equation}
	\left\langle \sigma_{x}^{k}\left(t=0\right)\right\rangle =\left\langle \sigma_{y}^{k}\left(t=0\right)\right\rangle =0,\quad\left\langle \sigma_{z}^{k}\left(t=0\right)\right\rangle =1\label{eq:Polarization}
\end{equation}
The time evolution operator for the Hamiltonian eq.~(\ref{eq:H_vary_coupling}) is:
\begin{align}
	U\left(t\right) &= \prod_{i\neq j}\exp\left(-i\theta_{ij}\sigma_{x}^{i}\sigma_{x}^{j}\right)\label{eq:Unitary} \\
	\theta_{ij} &= -g_{ij} t
\end{align}

The squeezing parameter along an angle $\phi$ in the transverse plane is given by:
\begin{equation}
	\xi_{\phi}^{2}\equiv\frac{\left\langle \left[\sum_{i}\left(\cos\left(\phi\right)\sigma_{x}^{i}+\sin\left(\phi\right)\sigma_{y}^{i}\right)\right]^{2}\right\rangle }{N/4}=\frac{\mathcal{A}}{N/4},\label{eq:Ratio}
\end{equation}
where the time dependence is implicit in the state. Calculating the expectations of the operators on the time evolved state, we find:
\begin{align}
	\mathcal{A} & =N+\sum_{k\neq l}P_{l}P_{k}\sin^{2}\left(\phi\right)\prod_{j\neq k,l}\left(\cos\left[4\theta_{jk}\right]\cos\left[4\theta_{jl}\right]-\sin\left[4\theta_{jk}\right]\sin\left[4\theta_{jl}\right]\right)-\nonumber \\
	& -\sum_{k\neq l}P_{l}P_{k}\sin^{2}\left(\phi\right)\prod_{j\neq k,l}\left(\cos\left[4\theta_{jk}\right]\cos\left[4\theta_{jl}\right]+\sin\left[4\theta_{jk}\right]\sin\left[4\theta_{jl}\right]\right) \\
	& +2\sin\left(2\phi\right)\sum_{k\neq l}P_{l}\sin\left[4\theta_{kl}\right]\prod_{i\neq l,k}\cos\left[4\theta_{il}\right].\nonumber \label{eq:a_b}
\end{align}
The optimal squeezing occurs at the angle $\phi$ which minimizes the squeezing parameter, but we can leave it general for now.

We now assume that $g_{ij}$ is Gaussian distributed, giving a distribution of the phases $\theta_{ij}$ with mean $\theta_0 = g_0 t$ and standard deviation $\kappa g_0 t$. This corresponds to the couplings have a typical variation by a factor $\kappa$. Additionally, we assume that the collective spin does not decrease substantially due squeezing~\cite{Kitagawa1993}. Thus, we calculate the squeezing parameter averaged over the distribution of couplings:
\begin{align}
	\bar{\xi_{\phi}^{2}} & \equiv\int\prod_{i<j}d\theta_{ij}p\left(\left\{ \theta_{ij}\right\} \right)\xi_{\phi}^{2} \\
	p\left(\left\{ \theta_{ij}\right\} \right) & =\mathcal{N}\prod_{i<j}\exp\left(-\alpha\left(\theta_{ij}-\theta_{0}\right)^{2}\right)\label{eq:Gaussian} \\ 
	\mathcal{N} &=\left(\frac{\pi}{\alpha}\right)^{\frac{N\left(N-1\right)}{4}},
\end{align}
where $1/\alpha = \kappa^2 \theta_0^2$. After much algebra, we find that the squeezing is parameter is given by:
\begin{align}
	\bar{\xi_{\phi}^{2}} = 1&+\sin^{2}\left(\phi\right)\left(N-1\right)\exp\left(-\frac{8\left(N-2\right)}{\alpha}\right)\left[\cos^{N-2}\left(8\theta_{0}\right)-1\right] \\ 
	&+2\sin\left(2\phi\right)\left(N-1\right)\exp\left(-\frac{4\left(N-1\right)}{\alpha}\right)\sin\left(4\theta_{0}\right)\cos^{N-2}\left(4\theta_{0}\right)\label{eq:Final}
\end{align}

The functional form of squeezing is identical to the case of uniform coupling, except for factors that scale as:
\begin{align}
	\exp\left( -\frac{N}{\alpha} \right) = \exp\left( -N \kappa^2 \theta_0^2 \right)
\end{align}
If the argument of these factors is large, then the degree of squeezing will be exponentially suppressed. However, for optimal squeezing 
$\theta_0 =g_et_o$, where $t_o\propto (Ng_e)^{-2/3}$ is given by Eq.~\eqref{eq:optimal}.
Therefore these factors scale as:
\begin{align}
	\exp\left( - \frac{\kappa^2 (Ng_e)^{2/3}}{N} \right) \approx 1 + O\left(\frac{1}{N^{1/3}}\right)\approx 1
\end{align}

Thus, variation in the coupling has a negligible affect on the generation of squeezing in the large $N$ limit, at least in the approximation where the distribution is Gaussian. This ultimately occurs because squeezing occurs on time scales $t \leq 1/(Ng_e)^{2/3}$, so the relative differences in the phases is small over the squeezing time. For our case, where the couplings will vary by factors of order unity over the sample due to the non-uniform magnetic field, this variation can be well ignored. A slightly better estimate of squeezing would use the average coupling over the sample rather than the max as we used in the main text, but this would change our results by a factor of order unity and hence we ignore it here for simplicity.

\section{\label{sec:Decay-of-squeezed}Decay of squeezed states due to dephasing}

For nuclear spin ensembles in solids, the most relevant source
of decoherence is spin dephasing $\gamma_s$. We model this dephasing as a Krauss
process with Krauss operators:
\begin{align}
\mathcal{K}_{i,\mathbb{I}} & =\sqrt{s_{i}}\mathbb{I}\nonumber \\
\mathcal{K}_{i,0} & =\sqrt{p_{i}}\left|0_{i}\right\rangle \left\langle 0_{i}\right|\nonumber \\
\mathcal{K}_{i,1} & =\sqrt{p_{i}}\left|1_{i}\right\rangle \left\langle 1_{i}\right|\label{eq:Krauss}
\end{align}
with $s_{i}=1-p_{i}=\exp\left(-\gamma_st\right)$. Now the
total Krauss operator for the density matrix with $N$ spins is given
by: 
\begin{equation}
\mathcal{K}_{\left\{ \alpha\right\} }^{N}=\prod_{i=1}^{N}\mathcal{K}_{i,\alpha_{i}}\label{eq:Product}
\end{equation}
Where $\alpha_{i}=\mathbb{I},0,1$. Now we wish to compute the expectation
value of an arbitrary operator $O_{N}$ under Krauss evolution, we
have that: 
\begin{align}
\left\langle O_{N}\left(t\right)\right\rangle  & =\sum_{\left\{ \alpha\right\} }Tr\left[O_{N}\prod_{i=1}^{N}\mathcal{K}_{i,\alpha_{i}}\rho_{N}\left(t=0\right)\prod_{i=1}^{N}\mathcal{K}_{i,\alpha_{i}}^{\dagger}\right]\nonumber \\
 & =\sum_{\left\{ \alpha\right\} }Tr\left[\prod_{i=1}^{N}\mathcal{K}_{i,\alpha_{i}}^{\dagger}O_{N}\prod_{i=1}^{N}\mathcal{K}_{i,\alpha_{i}}\rho_{N}\left(t=0\right)\right]\label{eq:Krauss-1}
\end{align}
Now we have that 
\begin{align}
\mathcal{K}_{i,\mathbb{I}}^{\dagger}\sigma_{x}^{i}\mathcal{K}_{i,\mathbb{I}} & =s_{i}\sigma_{x}^{i},\:\mathcal{K}_{i,0}^{\dagger}\sigma_{x}^{i}\mathcal{K}_{i,0}=\mathcal{K}_{i,1}^{\dagger}\sigma_{x}^{i}\mathcal{K}_{i,1}=0\nonumber \\
\mathcal{K}_{i,\mathbb{I}}^{\dagger}\sigma_{y}^{i}\mathcal{K}_{i,\mathbb{I}}^{\dagger} & =s_{i}\sigma_{y}^{i},\:\mathcal{K}_{i,0}^{\dagger}\sigma_{y}^{i}\mathcal{K}_{i,0}=\mathcal{K}_{i,1}^{\dagger}\sigma_{y}^{i}\mathcal{K}_{i,1}=0\label{eq:Zero}
\end{align}
Now for initial states close to z-axis we have that 
\begin{align}
\left\langle \sigma_{x}^{i}\left(t\right)\right\rangle  & =Tr\left[s_{i}\sigma_{x}^{i}\rho_{N}\left(t=0\right)\right]=0\nonumber \\
\left\langle \sigma_{y}^{i}\left(t\right)\right\rangle  & =Tr\left[s_{i}\sigma_{y}^{i}\rho_{N}\left(t=0\right)\right]=0\label{eq:Zero-1}
\end{align}
Furthermore we have that: 
\begin{equation}
\left\langle \sigma_{\alpha}^{i}\left(t\right)\sigma_{\beta}^{j}\left(t\right)\right\rangle =s_{i}s_{j}Tr\left[\sigma_{\alpha}^{i}\sigma_{\beta}^{j}\rho_{N}\left(t=0\right)\right]=s_{i}s_{j}\left\langle \sigma_{\alpha}^{i}\left(t=0\right)\sigma_{\beta}^{j}\left(t=0\right)\right\rangle \label{eq:Zero-2}
\end{equation}
With $\alpha,\beta=x/y$ and $i\neq j$. Now we have that the expectation
value of 
\begin{align}
\left\langle \left[\sum_{i=1}^{N}\left[\sigma_{x}^{i}\left(t\right)\cos\left(\theta\right)+\sigma_{y}^{i}\left(t\right)\sin\left(\theta\right)\right]\right]^{2}\right\rangle  & =N+\sum_{i\neq j}\left\langle \left[\sigma_{x}^{i}\left(t\right)\cos\left(\theta\right)+\sigma_{y}^{i}\left(t\right)\sin\left(\theta\right)\right]\left[\sigma_{x}^{j}\left(t\right)\cos\left(\theta\right)+\sigma_{y}^{j}\left(t\right)\sin\left(\theta\right)\right]\right\rangle \nonumber \\
 & =N+\sum_{i\neq j}s_{i}s_{j}\left\langle \left[\sigma_{x}^{i}\cos\left(\theta\right)+\sigma_{y}^{i}\sin\left(\theta\right)\right]\left[\sigma_{x}^{j}\cos\left(\theta\right)+\sigma_{y}^{j}\sin\left(\theta\right)\right]\right\rangle \label{eq:Time}
\end{align}
Now we know that at $t=0$ we must have that:
\begin{align}
&N\xi^{2} =N+\sum_{i\neq j}\left\langle \left[\sigma_{x}^{i}\cos\left(\theta\right)+\sigma_{y}^{i}\sin\left(\theta\right)\right]\left[\sigma_{x}^{j}\cos\left(\theta\right)+\sigma_{y}^{j}\sin\left(\theta\right)\right]\right\rangle \nonumber \\
&\Rightarrow\sum_{i\neq j}\left\langle \left[\sigma_{x}^{i}\cos\left(\theta\right)+\sigma_{y}^{i}\sin\left(\theta\right)\right]\left[\sigma_{x}^{j}\cos\left(\theta\right)+\sigma_{y}^{j}\sin\left(\theta\right)\right]\right\rangle  =-N\left(1-\xi^{2}\right)\label{eq:Squeezing}
\end{align}
Furthermore we will assume that $s_{i}=s_{j}=s=\exp\left(-\gamma_st\right)$,
then we have that: 
\begin{align}
N\xi^{2}\left(t\right) & =N-N\left(1-\xi^{2}\right)s^{2}\nonumber \\
\xi^{2}\left(t\right) & =1-\left(1-\xi^{2}\right)s^{2}\nonumber \\
\frac{\xi^{2}\left(t\right)}{\xi^{2}} & =s^{2}+\frac{1-s^{2}}{\xi^{2}}.\label{eq:Coherence}
\end{align}
We see that for $t\ll1/\gamma_s$ the squeezing parameter degrades at the rate $2\gamma_s$.

\section{\label{sec:Finite-polarization-effects}Finite polarization effects}

Consider the following initial density matrix. 
\begin{equation}
\rho=\prod_{i}\left(\mathbb{I}+p\sigma_{z}^{i}\right),\label{eq:Density_matrix}
\end{equation}
where $p$ is ensemble polarization.
Now consider acting on the intial state by the following squeezing Hamiltonian
\begin{equation}
H=-g_e\left(\sum_{i}\sigma_{x}^{i}\right)^{2}\label{eq:Squeezing}
\end{equation}
for a time $t$ and write $\theta_{0}=g_et\ll1$. Therefore we have
that:
\begin{align}
\xi_{\theta}^{2} & \equiv\frac{\left\langle \left[\sum_{i}\left(\cos\left(\theta\right)\sigma_{x}^{i}+\sin\left(\theta\right)\sigma_{y}^{i}\right)\right]^{2}\right\rangle }{N/4}=\label{eq:Ratio-1}\\
 & =1-\left(N-1\right)p^{2}\sin^{2}\left(\theta\right)\left[\cos^{N-2}\left(8\theta_{0}\right)-1\right]+2p\left(N-1\right)\sin\left(2\theta\right)\sin\left(4\theta_{0}\right)\cos^{N-2}\left(4\theta_{0}\right).
\end{align}
Now we write $\sin^{2}\left(\theta\right)=\frac{1}{2}\left(1-\cos\left(2\theta\right)\right)$.
As such we have that: 
\begin{align}
\xi_{\theta}^{2} & =1-\frac{\left(N-1\right)}{2}p^2\left[\cos^{N-2}\left(8\theta_{0}\right)-1\right]+\nonumber \\
 & +\left(N-1\right)p\left[\frac{1}{2}p\left[1-\cos^{N-2}\left(8\theta_{0}\right)\right]\cos\left(2\theta\right)+2\sin\left(2\theta\right)\sin\left(4\theta_{0}\right)\cos^{N-2}\left(4\theta_{0}\right)\right]\label{eq:Squeezing_angle}
\end{align}
Now use the trig identity that 
\begin{equation}
A\cos\left(2\theta\right)+B\sin\left(2\theta\right)=\sqrt{A^{2}+B^{2}}\cos\left(2\theta-\varphi\right),\quad s.t.\quad\tan\left(\varphi\right)=\frac{B}{A}\label{eq:Trig_idenity}
\end{equation}
Furthermore $\min_{\varphi}[\cos\left(2\theta-\varphi\right)]=-1$, so
that:
\begin{align}
\Rightarrow\xi_{min}^{2} & =1-\frac{\left(N-1\right)}{2}p^2\left[\cos^{N-2}\left(8\theta_{0}\right)-1\right]-\nonumber \\
 & -\left(N-1\right)p\sqrt{\frac{P^{2}}{4}\left(1-\cos^{N-2}\left(8\theta_{0}\right)\right)^{2}+4\sin^{2}\left(4\theta_{0}\right)\cos^{2\left(N-2\right)}\left(4\theta_{0}\right)}\nonumber \\
 & \cong 1+\frac{\left(N-1\right)}{2}p^2\left[1-\cos^{N-2}\left(8\theta_{0}\right)\right]-\nonumber \\
 & -\frac{\left(N-1\right)}{2}p^2\left[1-\cos^{N-2}\left(8\theta_{0}\right)\right]\left(1+\frac{1}{2}\left(\frac{4\sin^{2}\left(4\theta_{0}\right)\cos^{2\left(N-2\right)}\left(4\theta_{0}\right)}{\frac{p^{2}}{4}\left(1-\cos^{N-2}\left(8\theta_{0}\right)\right)^{2}}\right)-\frac{1}{8}\left(\frac{4\sin^{2}\left(4\theta_{0}\right)\cos^{2\left(N-2\right)}\left(4\theta_{0}\right)}{\frac{p^{2}}{4}\left(1-\cos^{N-2}\left(8\theta_{0}\right)\right)^{2}}\right)^{2}+...\right)\nonumber \\
 & =1-\frac{N-1}{2}\left(4\left(\frac{\sin^{2}\left(4\theta_{0}\right)\cos^{2\left(N-2\right)}\left(4\theta_{0}\right)}{\left(1-\cos^{N-2}\left(8\theta_{0}\right)\right)}\right)-32p^{-2}\left(\frac{\sin^{4}\left(4\theta_{0}\right)\cos^{4\left(N-2\right)}\left(4\theta_{0}\right)}{\left(1-\cos^{N-2}\left(8\theta_{0}\right)\right)^{3}}\right)+...\right)\nonumber \\
 & =1-2\left(N-1\right)\left(\frac{\sin^{2}\left(4\theta_{0}\right)\cos^{2\left(N-2\right)}\left(4\theta_{0}\right)}{\left(1-\cos^{N-2}\left(8\theta_{0}\right)\right)}\right)\left[1-8p^{-2}\left(\frac{\sin^{2}\left(4\theta_{0}\right)\cos^{2\left(N-2\right)}\left(4\theta_{0}\right)}{\left(1-\cos^{N-2}\left(8\theta_{0}\right)\right)^{2}}\right)\right],\label{eq:Min_variance}
\end{align}
where we have used that $\theta_{0}\ll1$ to make the Taylor expansion.
Now we have that: 
\begin{align}
\cos^{2\left(N-2\right)}\left(4\theta_{0}\right) & =\exp\left(2\left(N-2\right)\ln\left(\cos\left(4\theta_{0}\right)\right)\right)\nonumber \\
 & \cong\exp\left(2\left(N-2\right)\ln\left[1-8\theta_{0}^{2}\right]\right)\nonumber \\
 & \cong\exp\left(2\left(N-2\right)\left[\left(-8\theta_{0}^{2}\right)\right]\right)\nonumber \\
 & \cong1+2\left(N-2\right)\left[\left(-8\theta_{0}^{2}\right)\right]+32\left(N-2\right)^{2}\theta_{0}^{4}\nonumber \\
\cos^{N-2}\left(8\theta_{0}\right) & =\exp\left(\left(N-2\right)\ln\left(\cos\left(8\theta_{0}\right)\right)\right)\nonumber \\
 & \cong\exp\left(\left(N-2\right)\ln\left[1-32\theta_{0}^{2}\right]\right)\nonumber \\
 & \cong\exp\left(\left(N-2\right)\left[-32\theta_{0}^{2}\right]\right)\nonumber \\
 & \cong1-32\left(N-2\right)\theta_{0}^{2}+512\left(N-2\right)^{2}\theta_{0}^{4}+\frac{2^{12}}{3}\left(N-2\right)^{3}\theta_{0}^{6}\label{eq:Expansions}
\end{align}
This means that: 

\begin{align}
\frac{\sin^{2}\left(4\theta_{0}\right)\cos^{2\left(N-2\right)}\left(4\theta_{0}\right)}{\left(1-\cos^{N-2}\left(8\theta_{0}\right)\right)} & \cong\frac{1}{2\left(N-2\right)}\frac{\left[1-\left(N-2\right)16\theta_{0}^{2}+32\left(N-2\right)^{2}\theta_{0}^{4}\right]}{\left[1-\left(N-2\right)16\theta_{0}^{2}+\frac{128}{3}\left(N-2\right)^{2}\theta_{0}^{4}\right]}\cong\frac{1}{2\left(N-1\right)}\left[\left(1-\left(N-1\right)^{2}\frac{32}{3}\theta_{0}^{4}\right)\right]\nonumber \\
\frac{\sin^{2}\left(4\theta_{0}\right)\cos^{2\left(N-2\right)}\left(4\theta_{0}\right)}{\left(1-\cos^{N-2}\left(8\theta_{0}\right)\right)^{2}} & =\frac{1}{128\left(N-2\right)^{2}\theta_{0}^{2}}\cong\frac{1}{128\left(N-1\right)^{2}\theta_{0}^{2}}\label{eq:Simplifications}
\end{align}
This means that: 
\begin{align}
\xi_{min}^{2} & \cong 1-\left(1-\left(N-1\right)^{2}\frac{32}{3}\theta_{0}^{4}\right)\left[1-\frac{p^{-2}}{16\left(N-1\right)^{2}\theta_{0}^{2}}\right]\nonumber \\
 & \cong \frac{1}{16p^2N^{2}\theta_{0}^{2}}+\frac{32}{3}N^{2}\theta_{0}^{4}+....\nonumber \\
 & =\frac{1}{16p^2N^{2}g_e^{2}t^{2}}+\frac{32}{3}Ng_e^{2}t^{2}\label{eq:Final-1}
\end{align}

\end{document}